\documentclass[12pt,preprint]{aastex}

  \shorttitle{The delayed injection of PAHs into the ISM of galaxies}
  \shortauthors{Galliano et al.}

  \usepackage{ulem}

\newcommand{\ncode}[1]{{\tt #1}}

\newcommand{\peg}{\ncode{P\'EGASE}}

\usepackage{color}
\definecolor{grey}{rgb}{0.5,0.5,0.5}

\newcommand{\sms}[1]{{\mbox{{\scriptsize #1}}}}
\newcommand{\dd}{{\rm d}}
\newcommand{\ddiff}{{\;\rm d}}
\newcommand{\E}[1]{\times10^{#1}}
\newcommand{\petm}[2]{^{+#1}_{-#2}}
\newcommand{\unc}{$^{(?)}$}

\newcommand{\refeq}[1]{Eq.~(\ref{#1})}
\newcommand{\refeqs}[1]{Eqs.~(\ref{#1})}
\newcommand{\refeqp}[1]{(Eq.~\ref{#1})}
\newcommand{\reftab}[1]{Table~\ref{#1}}
\newcommand{\reffig}[1]{Fig.~\ref{#1}}
\newcommand{\reffigs}[1]{Figs.~\ref{#1}}

\newcommand{\refS}[1]{\S\ref{#1}}

\newcounter{textlistctr}
\newcommand{\thetextlist}{, }
\newcommand{\textlist}[1]
            {\setcounter{textlistctr}{1}
             \renewcommand{\thetextlist}
             {{\it (\roman{textlistctr})}\stepcounter{textlistctr}}#1
              }

\newcounter{obsrefctr}
\newcommand{\obsref}[1]
           {\refstepcounter{obsrefctr}[\arabic{obsrefctr}]\label{#1}}
\newcommand{\refobs}[1]{[\ref{#1}]}

\newcommand{\iras}{{\it IRAS}}
\newcommand{\iso}{{\it ISO}}
\newcommand{\isoST}{{\it Infrared Space Observatory}}
\newcommand{\spitz}{{\it Spitzer}}
\newcommand{\spitzST}{{\it Spitzer Space Telescope}}

\newcommand{\irac}{{\it Spitzer}/IRAC}
\newcommand{\mips}{{\it Spitzer}/MIPS}
\newcommand{\irs}{{\it Spitzer}/IRS}
\newcommand{\isocam}{{\it ISO}/CAM}
\newcommand{\isosws}{{\it ISO}/SWS}

\newcommand{\irasi}{\iras$_\sms{12$\mu m$}$}
\newcommand{\irasii}{\iras$_\sms{25$\mu m$}$}

\newcommand{\iraciv}{\irac$_\sms{8$\mu m$}$}
\newcommand{\IRACiv}{IRAC$_\sms{8$\mu m$}$}

\newcommand{\mipsi}{\mips$_\sms{24$\mu m$}$}
\newcommand{\MIPSi}{MIPS$_\sms{24$\mu m$}$}

\newcommand{\arp}[1]{Arp$\;$#1}
\newcommand{\haro}[1]{Haro$\;$#1}
\newcommand{\IC}[1]{IC$\;$#1}

\newcommand{\IRAS}[1]{IRAS$\;$#1}
\newcommand{\M}[1]{M$\;$#1}
\newcommand{\mrk}[1]{Mrk$\;$#1}
\newcommand{\ngc}[1]{NGC$\;$#1}
\newcommand{\tol}[1]{Tol$\;$#1}
\newcommand{\um}[1]{UM$\;$#1}

\newcommand{\smcb}{SMC$\;$B1$\#$1}

\newcommand{\cenA}{Centaurus$\;$A}
\newcommand{\hen}{He$\;$2-10}

\newcommand{\hstruc}{HS$\;$0822+3542}
\newcommand{\izw}{I$\;$Zw$\;$18}
\newcommand{\iizw}{II$\;$Zw$\;$40}
\newcommand{\sbs}{SBS$\;$0335-052}
\newcommand{\viizw}{VII$\;$Zw$\;$403}

\newcommand{\snii}{SN$\;${\sc ii}}

\newcommand{\msun}{\;M_\odot}
\newcommand{\zsun}{\;Z_\odot}
\newcommand{\mic}{\;\mu {\rm m}}

\newcommand{\upfov}{''{\rm pixel}^{-1}}

\newcommand{\ariii}{Ar$\,${\sc iii}}
\newcommand{\arii}{Ar$\,${\sc ii}}

\newcommand{\feii}{Fe$\,${\sc ii}}

\newcommand{\hi}{H$\,${\sc i}}
\newcommand{\hii}{H$\,${\sc ii}}

\newcommand{\neiii}{Ne$\,${\sc iii}}
\newcommand{\neii}{Ne$\,${\sc ii}}

\newcommand{\siii}{S$\,${\sc iii}}
\newcommand{\siv}{S$\,${\sc iv}}
\newcommand{\siII}{Si$\,${\sc ii}}
\newcommand{\hmol}{H$_2$}

\newcommand{\ariiiline}{[\ariii]$_{8.99\mu m}$}
\newcommand{\ariiline}{[\arii]$_{6.98\mu m}$}

\newcommand{\neiiiline}{[\neiii]$_{15.56\mu m}$}
\newcommand{\neiiline}{[\neii]$_{12.81\mu m}$}

\newcommand{\feiiline}{[\feii]$_{25.99\mu m}$}

\newcommand{\siiiline}{[\siii]$_{18.68\mu m}$}
\newcommand{\siiilineb}{[\siii]$_{33.48\mu m}$}
\newcommand{\sivline}{[\siv]$_{10.51\mu m}$}
\newcommand{\siIIline}{[\siII]$_{34.82\mu m}$}

  \graphicspath{{.}}
  \DeclareGraphicsExtensions{.eps}

  \slugcomment{\today\ (submitted to \apj\ November 10, 2006)}

\begin{document}

\title{STELLAR EVOLUTIONARY EFFECTS ON THE ABUNDANCES OF PAH AND SN-CONDENSED
       DUST IN GALAXIES}

\author{Fr\'ed\'eric~Galliano,
        Eli~Dwek}
  \affil{Observational Cosmology Lab., Code 665,
         NASA Goddard Space Flight Center,
         Greenbelt MD 20771, USA}
  \email{galliano@astro.umd.edu}
\and
\author{Pierre~Chanial}
  \affil{Astrophysics Group, Blackett Laboratory, Imperial College, 
         Prince Consort Road, London SW7 2AZ, UK}

\begin{abstract}
Spectral and photometric observations of nearby galaxies show a correlation 
between the strength of their mid-IR aromatic features, attributed to PAH molecules, and their metal abundance, leading to a deficiency of these features
in low-metallicity galaxies. 
In this paper, we suggest that the observed correlation represents a trend of 
PAH abundance with galactic age, reflecting the delayed injection of 
carbon dust into the ISM by AGB stars in the final post-AGB phase of their evolution.
AGB stars are the primary sources of PAHs and carbon dust in galaxies, 
and recycle their ejecta back to the interstellar medium only after a few 
hundred million years of evolution on the main sequence.
In contrast, more massive stars that explode as Type~II supernovae inject 
their metals and dust almost instantaneously after their formation. 
We first determined the PAH abundance in galaxies by constructing detailed 
models of UV-to-radio SED of galaxies that estimate the contribution of dust in 
PAH-free \hii\ regions, and PAHs and dust from photodissociation regions, 
to the IR emission. 
All model components: the galaxies' stellar content, properties of their \hii\ 
regions, and their ionizing and non-ionizing radiation fields and dust 
abundances, are constrained by their observed multiwavelength spectrum. 
After determining the PAH and dust abundances in 35 nearby galaxies using
our SED model, we use a chemical evolution model to show that the delayed 
injection of carbon dust by AGB stars provides a natural explanation to the 
dependence of the PAH content in galaxies with metallicity.
We also show that larger dust particles giving rise to the far-IR emission 
follow a distinct evolutionary trend closely related to the injection of 
dust by massive stars into the ISM.
\end{abstract}



\section{INTRODUCTION}
\label{sec:intro}

Spectral and photometric observations of nearby galaxies with the \isoST\ 
(\iso) and the \spitzST\ have provided the opportunity to investigate the 
inter-relations between global galactic properties, such as morphology, star 
formation activity, spectral energy distribution (SED), metallicity, and dust abundance and composition. 
In particular, these observations have enabled detailed studies of the 
correlation of dust abundances and composition with metal enrichment in 
galaxies spanning a wide range of metallicities. 
Since the metallicity of galaxies evolves monotonically with time, galaxies 
with different metallicities provide snapshots of the evolutionary history 
of galaxies.

An exciting result provided by \iso\ spectral observations of nearby galaxies 
was the discovery of a striking correlation between the strength of their 
mid-IR aromatic features and their metallicity \citep{madden06}. 
Low-metallicity galaxies exhibited very weak or no aromatic features. Observations of the 8-to-24$\mic$ bands flux ratio obtained with \irac\ and 
\mips\ instruments showed a correlation of this flux ratio with the galaxies'
oxygen abundance \citep{engelbracht05}. 
Since the \iraciv\ band is supposed to trace the strength of the aromatic 
features, and the \mipsi\ that of the continuum emission from the hot 
non-aromatic dust component, this correlation seemed to confirm the trends 
discovered by \iso. 
This correlation was independently confirmed by \citet{wu06} and 
\citet{ohalloran06} from spectral observations of low-metallicity 
blue compact dwarf galaxies with the \irs\ instrument.

The aromatic features are most commonly attributed to the vibrational modes of 
Polycyclic Aromatic Hydrocarbons \citep[PAHs;][]{leger84,allamandola85}, which 
are large planar molecules made of 50 to 1000 carbon atoms. 
Their ubiquity makes them an important component of dust models 
\citep{desert90,dwek97,zubko04,draine07a}, locking up $\sim15-20\%$ of 
the total amount of interstellar carbon
\citep[][with solar abundance constraints]{zubko04}. 
They mostly reside in galactic photodissociation regions (PDRs), where they 
play an important role in the heating of the gas by providing photo-electrons 
\citep[{e.g.}][]{tielens85,bakes94}, and in interstellar chemistry by 
providing surfaces for chemical reactions. 
Because of their small sizes, PAHs are stochastically heated by the 
interstellar radiation fields. 
The relative strength of some of the aromatic features depends on their 
ionization state, and varies therefore significantly with the physical 
condition of the environment \citep{hony01,vermeij02,galliano07}. 
Understanding the evolution of PAHs and their relation to the global properties 
of galaxies is therefore extremely important because of the complex mutual 
influences between PAHs, and their ambient radiative and gaseous surroundings.

Several explanations have been offered for the correlation of the intensity 
of the PAH features with metallicity. 
The first suggests that the trend reflects an increase in the destruction efficiency of PAHs in low metallicity environments 
\citep{galliano03,galliano05,madden06}. 
Low-metallicity environments are bathed with harder photons than our
Galaxy, due to the higher effective temperature of their stars, and their
young age.
Moreover, the paucity of dust allows this hard radiation field to penetrate 
deeper into the ISM, compared to high-metallicity systems, selectively 
destroying the PAH molecules by photoevaporation or photodissociation. 
This explanation is consistent with the models of population synthesis
and dust evolution of \citet{dwek00}.
Assuming that PAHs are efficiently destroyed by UV photons in \hii\ regions, 
their models showed an evolutionary trend of PAH features with time, as the
relative contribution of ionizing OB stars to the galaxy's SED decreases
with time.
A second explanation has been proposed by \citet{ohalloran06}, who suggested
that PAHs are destroyed by the numerous shocks observed in low metallicity 
systems.
To support their proposition, they showed an anti-correlation between the 
PAH-to-continuum ratio and the \feiiline/\neiiline\ line ratio, the latter 
being supposedly a shock tracer.
The problem with this explanation is that there is no observational evidence
that PAHs are selectively destroyed in shocks. 
On the contrary, \citet{reach02} showed that, in the shocked medium of 
3C$\,$391, both the PAH features and the underlying continuum disappear.

All previous explanations attribute the paucity of PAHs to destructive
processes that are more efficient in the early stages of galaxy evolution.
In contrast, \citet{dwek05} suggested that the observed correlation reflects an
evolutionary trend of the sources of interstellar PAHs with metallicity. 
PAHs and carbon dust are mostly produced in asymptotic giant  branch (AGB) 
stars which, unlike massive stars, recycle their ejecta into the ISM after a
significantly longer time of main sequence evolution. 
The observed correlation of PAH line intensities with metallicity is therefore 
a trend of PAH abundance with galactic age, reflecting the delayed injection 
of PAHs and carbon dust into the ISM by AGB stars in their final, post-AGB, 
phase of their evolution.

Such distinct evolutionary trends of SN and AGB produced dust with time 
were predicted by \citet{dwek98} and more recently by \citet{morgan03}.
In particular, the latter show the evolutionary trend of AGB dust with time.
This trend can be translated as a trend with metallicity if galaxies 
approximately share a global ``cosmic'' star formation history.

The goal of this paper is to present a detailed evolutionary model to examine 
whether the observed trend of PAH line intensity with metallicity reflects an 
evolutionary trend of PAH abundance with metallicity. 
This requires the determination of PAH abundance from the strength of their 
aromatic features in the galaxies for which this trend has been observed, and 
the use of a chemical evolution model to follow the change of PAH abundance with 
galactic metallicity \citep{dwek98}. 
The paper is organized as follows. 
We first present in \refS{sec:sample} the sample of nearby galaxies that 
were considered in our analysis. 
In \refS{sec:method}, we describe the method we used to separate the contributions of \hii\ regions and photodissociation regions to the global SED, in order to determine the abundances of PAHs and larger grains in these 
galaxies in \refS{sec:results}. 
In \refS{sec:dustvol}, we briefly describe the chemical evolution model 
used in the calculations, and compare its results to the abundances derived
from our SED modeling. 
The results of the paper are briefly summarized in \refS{sec:conclusion}.

Throughout this paper we will refer to the solar abundances by 
\citet{grevesse98}, the oxygen number abundance being 
$12+\log(\rm O/H)_\odot=8.83$, the Helium and heavy elements to gas mass 
ratios $Y_\odot=0.248$ and $Z_\odot=0.017$, respectively.
Besides, we assume that the helium abundance is independent of the metallicity.


\section{THE SAMPLE OF NEARBY GALAXIES}
\label{sec:sample}

  \subsection{Source Selection}

In order to properly characterize the PAH emission, we considered galaxies
whose mid-IR spectrum has been observed, either with one of the spectrographs
onboard the \iso\ satellite, or with the \irs.
We combined ISO samples of starbursts and AGNs 
\citep{laurent00}, spirals \citep{roussel01}, ellipticals \citep{xilouris04}, 
dwarf galaxies \citep{madden06}, and the low-metallicity sources of 
the \spitz\ sample presented by \citet{engelbracht05} which were observed by 
\irs.
Among the ellipticals, only \ngc{1399} has a mid-IR spectrum.
Consequently, our sample includes various types of galaxies and covers a wide 
range of metallicities and star formation activity (\reftab{tab:source}).

The modeling that will be presented in \refS{sec:method} requires the
assembly of data covering the stellar as well as the dust emission components, 
for each galaxy.
First, we need most of the U, B, V, R, I, J, H, K fluxes, to constrain the
stellar spectrum shape.
Photometric J, H and K bands are available for almost all of our sources, 
thanks to the 2MASS survey \citep{jarrett03}.
We rejected the galaxies \arp{118}, \arp{236}, \arp{299} and \ngc{4038}, 
for which no B and V bands were reported.
Second, the far-IR SED is used to constrain the intensity of the interstellar
radiation field (ISRF).
Hence, we rejected galaxies which have not been detected by IRAS or MIPS,
such as \hstruc, \tol{1214-277} and \tol{65}.
In addition, our mass estimates are normalised to the Hydrogen mass.
We therefore rejected the galaxy \IRAS{23128-5919}, for which no \hi\ 
observation was reported.
Finally, we removed \M{31} from our sample, since its angular diameter is
too large to build a consistent observed total SED.

The global properties of the selected sources are presented in 
\reftab{tab:source}.
If relevant, the distances were homogenised to 
$H_\sms{0}=71\; \rm km\,s^{-1}\,Mpc^{-1}$.
The masses have been scaled to the adopted distance.
\begin{deluxetable}{llllrrrrrrrl}
  \tabletypesize{\footnotesize}
  \rotate
  \tablecolumns{12}
  \tablewidth{0pc}
  \tablecaption{Select Properties of the Sample}
  \tablehead{
    \colhead{Name} & \colhead{R.A.}    & \colhead{Dec.}                 & 
      \colhead{Mid-IR} & \colhead{Distance}       &
      \multicolumn{2}{c}{$12+\log({\rm O/H})$} &
      \multicolumn{2}{c}{$M_\sms{\hi}$}        & 
      \multicolumn{2}{c}{$M_\sms{\hmol}$}      & \colhead{Notes}   \\
    \colhead{}     & \colhead{(J2000)} & \colhead{(J2000)}              & 
      \colhead{spectrograph}   & \colhead{(Mpc)}          & 
                            & \colhead{[ref.]} &
      \colhead{$(10^8\;M_\odot)$} & \colhead{[ref.]} & 
      \colhead{$(10^8\;M_\odot)$} & \colhead{[ref.]} & }
  \startdata
    \haro{11}      & $00^h36^m52\fs5$  & $-33\degr33\arcmin19\arcsec$   &
      \irs                  & 92                       &
      7.9           & \obsref{ZHaro11}            &
      $\lesssim1$\unc       & \refobs{ZHaro11} &
      $\lesssim1$           & \refobs{ZHaro11} & Pec \hii \\
    \ngc{253}      & $00^h47^m32\fs9$  & $-25\degr17\arcmin18\arcsec$   &
      \isocam                      & 3.3                      & 
      9.0           & \obsref{ZN253}              &
      $18$                  & \obsref{HIN253}  &
      $17$                  & \obsref{H2N253}  & \hii \\
    \ngc{520}      & $01^h24^m34\fs9$  & $+03\degr47\arcmin31\arcsec$   &
      \isocam                      & 27                       & 
      \unc          &                             &
      $35$                  & \obsref{HIN520}  &
      $35$                  & \obsref{H2N520}  & Pec \hii \\
    \ngc{613}      & $01^h34^m17\fs5$  & $-29\degr24\arcmin58\arcsec$   &
      \isocam                      & 19                       &
      9.2           & \obsref{ZN613}              &
      $37$                  & \obsref{bettoni} &
      \nodata               &                  & Sy \\
    \ngc{891}      & $02^h22^m33\fs4$  & $+42\degr20\arcmin57\arcsec$   &
      \isocam                      & 9.6                      & 
      8.9           & \obsref{ZN891}              &
      $76$                  & \refobs{bettoni} &
      $49$                  & \refobs{bettoni} & Edge-on \\
    \ngc{1068}     & $02^h42^m40\fs6$  & $-00\degr00\arcmin47\arcsec$   &
      \isocam                      & 15                       &
      9.0           & \obsref{ZN1068}             &
      $22$                  & \obsref{HIN1068} &
      $74$                  & \obsref{H2N1068} & Sy \\
    \ngc{1097}     & $02^h46^m19\fs1$  & $-30\degr16\arcmin28\arcsec$   &
      \isocam                      & 12                       &
      9.0           & \obsref{ZN1097}             &
      $51$                  & \obsref{HIN1097} & 
      $\gtrsim7.3$          & \obsref{H2N1097} & Sy \\
    \ngc{1140}     & $02^h54^m33\fs5$  & $-10\degr01\arcmin44\arcsec$   &
      \isocam                      & 25                       & 
      8.0           & \obsref{heckman}            &
      $52$                  & \obsref{HIN1140} & 
      \nodata               &                  & Irr \hii \\
    \ngc{1365}     & $03^h33^m35\fs6$  & $-36\degr08\arcmin23\arcsec$   &
      \isocam                      & 19                       &
      9.1           & \obsref{ZN1365}             &
      $130$                 & \obsref{HIN1365} &
      $170$                 & \obsref{H2N1365} & Sy \\
    \sbs           & $03^h37^m44\fs0$  & $-05\degr02\arcmin38\arcsec$   &
      \irs                  & 54                       & 
      7.3           & \obsref{Zsbs}               &
      $9.9$                 & \obsref{HIsbs}   &
      \nodata               &                  & BCD \\
    \ngc{1399}     & $03^h38^m29\fs1$  & $-35\degr27\arcmin03\arcsec$   &
      \isocam                      & 21                       & 
      \unc          &                             &
      $\lesssim2$           & \refobs{bettoni} &
      \nodata               &                  & cD \\
    \IC{342}       & $03^h46^m49\fs7$  & $+68\degr05\arcmin45\arcsec$   &
      \isocam                      & 3.8                      & 
      8.9           & \obsref{Zic}                &
      $130$                 & \refobs{bettoni} &
      $50$                  & \refobs{bettoni} & \hii \\
    \ngc{1569}     & $04^h30^m49\fs1$  & $+64\degr50\arcmin52\arcsec$   &
      \isocam                      & 2.2                      & 
      8.2           & \obsref{ZN1569}             &
      $1.3$                 & \obsref{HIN1569} &
      $0.50$                & \obsref{H2N1569} & Irr \hii \\
    \ngc{1808}     & $05^h07^m42\fs3$  & $-37\degr30\arcmin47\arcsec$   &
      \isocam                      & 11                       &
      9.1           & \obsref{ZN1808}             &
      $18$                  & \obsref{HIN1808} &
      $20$                  & \obsref{H2N1808} & Sy \\
    \iizw          & $05^h55^m42\fs7$  & $+03\degr23\arcmin29\arcsec$   &
      \isocam                      & 10                       & 
      8.1           & \obsref{Ziizw}              &
      $4.2$                 & \refobs{bettoni} &
      $0.23$                & \refobs{bettoni} & BCD \\
    \hen           & $08^h36^m15\fs2$  & $-26\degr24\arcmin34\arcsec$   &
      \irs                  & 8.7                      & 
      8.9           & \refobs{ZN1569}             &
      $3.1$                 & \obsref{HIhen}   &
      $1.6$                 & \refobs{HIhen}   & Irr \hii \\
    \izw           & $09^h34^m02\fs0$  & $+55\degr14\arcmin28\arcsec$   &
      \irs                  & 13                       & 
      7.2           & \refobs{Zsbs}               &
      $1.2$                 & \obsref{HIIZw18} &
      \nodata               &                  & BCD \\
    \M{82}         & $09^h55^m51\fs8$  & $+69\degr40\arcmin46\arcsec$   &
      \isocam                      & 3.6                      & 
      9.0           & \obsref{ZM82}               &
      $9.0$                 & \obsref{HIM82}   & 
      $13$                  & \obsref{H2M82}   & Irr \hii \\
    \ngc{3256}     & $10^h27^m51\fs1$  & $-43\degr54\arcmin17\arcsec$   &
      \isocam                      & 37                       &
      8.9           & \obsref{ZN3256}             &
      $62$                  & \obsref{casasola}&
      $300$                 & \obsref{H2N3256} & Pec \hii \\
    \mrk{33}       & $10^h32^m31\fs9$  & $+54\degr24\arcmin04\arcsec$   &
      \irs                  & 20                        & 
      8.4           & \refobs{ZN3256}             &
      $4.3$                 & \obsref{HIMrk33} &
      $0.70$                & \obsref{H2Mrk33} & Irr \hii \\
    \mrk{153}      & $10^h49^m05\fs0$  & $+52\degr20\arcmin08\arcsec$   &
      \irs                  & 37                        &
      7.8           & \obsref{ZMrk153}            &
      $6.7$                 & \obsref{HIMrk153} &
      \nodata               &                   & BCD \\
    \viizw         & $11^h27^m59\fs9$  & $+78\degr59\arcmin39\arcsec$   &
      \irs                  & 4.5                       &
      7.7           & \refobs{Zsbs}             &
      $0.69$                & \refobs{HIMrk33}  &
      \nodata               &                   & BCD \\
    \um{448}       & $11^h42^m12\fs4$  & $+00\degr20\arcmin03\arcsec$   &
      \irs                  & 70                       &
      8.0           & \obsref{ZUM448}             &
      $47$                  & \obsref{sage92}  &
      $24$                  & \refobs{sage92}  & Pec \hii \\ 
    \ngc{4945}     & $13^h05^m26\fs2$  & $-49\degr28\arcmin15\arcsec$   &
      \isocam                      & 3.9                      &
      \unc          &                             &
      $45$                  & \obsref{huchtmeier}&
      $17$                  & \obsref{H2N4945} & Edge-on Sy \\
    \cenA          & $13^h25^m28\fs0$  & $-43\degr01\arcmin06\arcsec$   &
      \isocam                      & 3.8                      &
      $\sim9$       & \obsref{ZcenA}              &
      $11$                  & \obsref{HIcenA}  &
      $2.0$                 & \obsref{H2cenA}  & Elliptical Sy \\
    \M{51}         & $13^h29^m52\fs7$  & $+47\degr11\arcmin43\arcsec$   &
      \isocam                      & 8.4                      &
      8.7           & \obsref{ZM51}               &
      $50$                  & \obsref{HIM51}   &
      $56$                  & \obsref{H2M51}   & \\
    \M{83}         & $13^h37^m00\fs7$  & $-29\degr51\arcmin58\arcsec$   &
      \isocam                      & 4.5                      &
      9.2           & \obsref{ZM83}               &
      $51$                  & \obsref{HIM83}   &
      $49$                  & \obsref{H2M83}   & \hii \\
    \tol{89}       & $14^h01^m21\fs5$  & $-33\degr03\arcmin50\arcsec$   &
      \irs                  & 15                       &
      8.0           & \obsref{Ztol89}             &
      $13$                  & \obsref{HItol89} &
      \nodata               &                  & \hii \\
    Circinus       & $14^h13^m09\fs6$  & $-65\degr20\arcmin21\arcsec$   &
      \isocam                      & 4.0                      & 
      \unc          &                             &
      $55$                  & \refobs{huchtmeier}&
      $11$                  & \obsref{H2circ}  & Sy \\
    \ngc{5253}     & $13^h39^m55\fs7$  & $-31\degr38\arcmin29\arcsec$   &
      \isosws                      & 3.3                      &
      8.2           & \refobs{heckman}            &
      $0.91$                & \obsref{HIN5253} &
      $\lesssim0.1$         & \obsref{H2N5253} & Irr \hii \\
    \arp{220}      & $15^h34^m57\fs2$  & $+23\degr30\arcmin11\arcsec$   &
      \isocam                      & 73                       & 
      \unc          &                             &
      $280$                 & \refobs{casasola}&   
      $160$                 & \obsref{H2a220}  & ULIRG \\
    \ngc{6240}     & $16^h52^m58\fs8$  & $+02\degr24\arcmin06\arcsec$   &
      \isocam                      & 98                       &
      \unc          &                             &
      $88$                  & \refobs{bettoni} &
      $310$                 & \refobs{bettoni} & LIRG \\
    \ngc{6946}     & $20^h34^m51\fs2$  & $+60\degr09\arcmin17\arcsec$   &
      \isocam                      & 5.5                      &
      9.1           & \obsref{ZN6946}             &
      $56$                  & \obsref{HIN6946} &
      $33$                  & \obsref{H2N6946} & \hii \\
    \mrk{930}      & $23^h31^m58\fs3$  & $+28\degr56\arcmin50\arcsec$   &
      \irs                  & 73                       & 
      8.1           & \refobs{ZUM448}              &
      $28$                  & \obsref{HImrk930}    & 
      \nodata               &                  & \hii \\ 
    \ngc{7714}     & $23^h36^m14\fs1$  & $+02\degr09\arcmin19\arcsec$   &
      \irs                  & 37                       & 
      8.5           & \obsref{ZN7714}             &
      $17$                  & \obsref{HIN7714} & 
      $22$                  & \refobs{HIN7714} & Pec \hii
  \enddata
  \label{tab:source}
  \tablerefs{\refobs{ZHaro11}~\citet{bergvall00};
             \refobs{ZN253}~\citet{zaritsky94};
             \refobs{HIN253}~\citet{boomsma05};
             \refobs{H2N253}~\citet{houghton97};
             \refobs{HIN520}~\citet{bernloehr93};
             \refobs{H2N520}~\citet{yun01};
             \refobs{ZN613}~\citet{alloin79b};
             \refobs{bettoni}~\citet{bettoni03};
             \refobs{ZN891}~\citet{otte01};
             \refobs{ZN1068}~\citet{dutil99};
             \refobs{HIN1068}~\citet{staveley-smith87};
             \refobs{H2N1068}~\citet{helfer03};
             \refobs{ZN1097}~\citet{storchi-bergmann95};
             \refobs{HIN1097}~\citet{ondrechen89};
             \refobs{H2N1097}~\citet{gerin88};
             \refobs{heckman}~\citet{heckman98};
             \refobs{HIN1140}~\citet{hunter94};
             \refobs{ZN1365}~\citet{roy97};
             \refobs{HIN1365}~\citet{jorsater95};
             \refobs{H2N1365}~\citet{sandqvist95};
             \refobs{Zsbs}~\citet{izotov99};
             \refobs{HIsbs}~\citet{thuan99b};
             \refobs{Zic}~\citet{pilyugin04};
             \refobs{ZN1569}~\citet{kobulnicky97};
             \refobs{HIN1569}~\citet{stil02};
             \refobs{H2N1569}~\citet{israel97};
             \refobs{ZN1808}~\citet{ravindranath01};
             \refobs{HIN1808}~\citet{dahlem01};
             \refobs{H2N1808}~\citet{dahlem90};
             \refobs{Ziizw}~\citet{perez-montero03};
             \refobs{HIhen}~\citet{sauvage97};
             \refobs{HIIZw18}~\citet{van-zee98};
             \refobs{ZM82}~\citet{boselli02};
             \refobs{HIM82}~\citet{appleton81};
             \refobs{H2M82}~\citet{walter02};
             \refobs{ZN3256}~\citet{mas-hesse99};
             \refobs{casasola}~\citet{casasola04};
             \refobs{H2N3256}~\citet{sargent89};
             \refobs{HIMrk33}~\citet{thuan04};
             \refobs{H2Mrk33}~\citet{israel05};
             \refobs{ZMrk153}~\citet{kunth85};
             \refobs{HIMrk153}~\citet{thuan81};
             \refobs{ZUM448}~\citet{izotov98};
             \refobs{sage92}~\citet{sage92};
             \refobs{huchtmeier}~\citet{huchtmeier88};
             \refobs{H2N4945}~\citet{dahlem93};
             \refobs{ZcenA}~\citet{schaerer00};
             \refobs{HIcenA}~\citet{richter94};
             \refobs{H2cenA}~\citet{wild97};
             \refobs{ZM51}~\citet{bresolin04};
             \refobs{HIM51}~\citet{dean75};
             \refobs{H2M51}~\citet{helfer03};
             \refobs{ZM83}~\citet{webster83};
             \refobs{HIM83}~\citet{tilanus93};
             \refobs{H2M83}~\citet{lundgren04};
             \refobs{Ztol89}~\citet{durret85};
             \refobs{HItol89}~\citet{paturel03};
             \refobs{H2circ}~\citet{elmouttie98};
             \refobs{HIN5253}~\citet{reif82};
             \refobs{H2N5253}~\citet{meier02};
             \refobs{H2a220}~\citet{sanders91};
             \refobs{ZN6946}~\citet{kobulnicky99};
             \refobs{HIN6946}~\citet{carignan90};
             \refobs{H2N6946}~\citet{tacconi86};
             \refobs{HImrk930}~\citet{hopkins02};
             \refobs{ZN7714}~\citet{gonzalez-delgado95};
             \refobs{HIN7714}~\citet{struck03}.}
  \tablecomments{The sources are ordered according to their right ascension.
                 The symbol \unc\ identifies uncertain values.
                 Entries for the metallicity for which no data exist are marked
                 by \nodata, and assumed to be solar-to-supersolar.
                 The \hmol\ masses are not used to calculate the dust-to-gas 
                 mass ratio, because the IR emission was assumed to originate 
                 entirely from the \hi\ gas.
                 }
\end{deluxetable}

  \subsection{\isocam\ Data Reduction}
  \label{sec:cam}

Most of the sources in \reftab{tab:source} were observed with \isocam\ 
\citep{cesarsky96cam} on board the \iso\ satellite \citep{kessler96}.
These spectra were used by \citet{madden06}, and we refer to this work for a 
detailed description of the data reduction.
The CVF performed spectral imaging using a $32\times 32$ detector 
array, with a sampling of $3\upfov$ or $6\upfov$ in our cases, from 
$\lambda=5\mic$ to $16.5\mic$ with one pointing of two CVFs, from $\lambda=5$ 
to $9.5\mic$ and from $\lambda=9.0$ to $16.5\mic$.
The spectral resolution goes from $\lambda/\Delta\lambda=35$ to 51 across the 
full spectra.

For the data treatment, we used the CAM Interactive Reduction
\citep[CIR, version AUG01;][]{chanial03}. 
The subtraction of the dark currents was performed using the \citet{biviano98}
model which predicts the time evolution for each row of the detector, taking 
into account drifts along each orbit and each revolution. 
We masked the glitches using multi-resolution median filtering 
\citep{starck99} on each block of data after slicing the cube. 
Additional deglitching was performed manually, examining the temporal cut for 
each pixel. 
We corrected the systematic memory effects using the Fouks-Schubert method 
\citep{coulais00}. 
We computed a hybrid flat-field image placing a mask on the source and
computing a flat field outside this mask from the median of the temporal cut 
for each pixel. 
For the pixels which were on-source, the flat-field response was set to the 
corresponding calibration flat-field.
The conversion from Analog Digital Units to mJy/pixel was performed using the 
standard in-flight calibration data base.
To remove the sky contribution, the sources smaller than the array were masked
and, for a given wavelength, the median of the pixels which are off-source 
were subtracted from each pixel. 
For the more extended sources, we subtracted an independently observed 
zodiacal spectrum.
The intensity of this spectrum was a free parameter varied in order to match
the properly sky subtracted fluxes in the LW2 ($6.7\mic$) and LW3 ($14.3\mic$)
broadbands.
The final product is a 3D spectral-image of each galaxy.
We integrated the spectrum into an aperture encompassing the entire galaxy, 
and to obtain the global SED of the galaxy.
When the angular size of the source was larger than the one of the array, we 
scaled-up the spectrum to match the \irasi\ broadband flux.

  \subsection{\irs\ Spectrum Extraction}
  \label{sec:irs}

Several of the low-metallicity sources in \reftab{tab:source} were not 
observed by \isocam, we therefore complemented our database with publically
released mid-IR spectra from the \irs\ spectrometer on board the \spitzST\ 
\citep{houck04irs,werner04}.
Among these galaxies, the spectra of \sbs, \ngc{7714}, \viizw, \haro{11} and
\izw\ are described in details in \citet{houck04}, \citet{brandl04} and
\citet{wu06}.
We considered only low-resolution data, taken with
the SL (Short-Low) module, from $\lambda=5.2\mic$ to $14.5\mic$, 
and the LL (Long-Low) module, from $\lambda=14.0\mic$ to $38.0\mic$, both 
with a spectral resolution of $\lambda/\Delta\lambda\simeq64-128$.

We retrieved the Basic Calibrated Data (BCD) that have been preprocessed by
the Spitzer Science Center (SSC) data reduction pipeline,
and converted to flux density, corrected for stray light and
flatfielded.
The extraction of the spectra from the 2D space/wavelength images was 
performed with the Spectral Modeling, Analysis and Reduction Tool 
\citep[SMART, version 5.5.6;][]{higdon04}.
We first inspected the BCD images and identified the hot pixels which have not 
been masked out by the SSC.
We replaced them by the median of their neighbors.
For each module, an off-source position is observed.
We subtracted this spectrum from the on-source one, in order to remove the 
sky emission.
The extraction of the 1D spectrum was performed inside a column whose width 
vary with the wavelength. 
We have excluded the bonus order.
Then, the various frames, for each nod position, were coadded.
Since the long wavelength end of the SL module and the short wavelength end 
of the LL one were not systematically overlapping, we finally scaled the SL 
module, in order to obtain a continuous spectrum.
This scaling factor can be as large as $50\%$.
At the time when this publication is written, the \irs\ data handbook 
recommends not to derive the signal-to-noise ratio from the uncertainties 
generated by the pipeline.
Instead of that, we adopted the recommended systematic error of $20\%$.
However, the SL module of the noisiest spectra (\viizw, \mrk{153}, \mrk{930}, 
\izw, \tol{89}) exhibits fluctuations larger than this value.
To take into account these statistical variations, we smoothed these spectra
into a $\Delta\lambda\simeq 0.2\mic$ window (4 points), and took the standard 
deviation inside this window as the error.
Similarly to what we did with \isocam\ spectra (\refS{sec:cam}), in order
to compensate the fact that we may be overlooking some extended emission,
we scaled-up the \irs\ SL and LL spectra to match the \irasi\ and \irasii.
In the case of \iizw, the \irs\ spectrum does not exhibit the PAH 
features that \citet{madden06} detected in the extended emission,
thus we prefer to use the \isocam\ data for this galaxy.
For \tol{89}, the \iraciv\ broadband flux is higher than the integrated
spectrum into the same band.
This is probably due to the fact that the slit of \irs\ measures only the 
nucleus emission, which likely has a steeper continuum and weaker features 
than the extended emission contributing to the total broadband.
Therefore, we will consider the mass of PAHs derived for this galaxy to be 
a lower limit.

  \subsection{The Multiwavelength SEDs}

The UV-to-radio SED of each galaxy was built using data from different catalogs.
We extensively used the HYPERLEDA and NED databases.
Most of the SEDs include \iras\ broadband fluxes at 12, 25, 60 and $100\mic$ 
\citep{moshir90,rice88}, and J, H, K 2MASS fluxes \citep{jarrett03}.
For some of the galaxies, we completed the IR SED with the broadband \spitz\ 
data reported by \citet{dale05}.
We kept only data encompassing the entire galaxy.
The optical data were corrected for Galactic extinction, using the 
\citet{schlegel98} extinction maps.

For \ngc{5253}, we used the \isosws\ observations of \citet{crowther99} to 
model the spectrum of this galaxy.


\section{MODELING THE PANCHROMATIC SEDS OF GALAXIES TO DERIVE THE PAH AND
         DUST ABUNDANCES}
\label{sec:method}

  \subsection{Motivations and Approach}

Determining the PAH abundance from the 8 and $24\mic$ fluxes or even from the
mid-IR spectrum 
requires:\textlist{\thetextlist~knowledge of the contribution of other dust 
           species to the mid-IR emission, in order to determine the intrinsic
           emission in the PAH bands; and 
         \thetextlist~knowledge of the ISRF they are subjected to, since PAHs 
           are stochastically heated.}The dust that contributes to the mid-IR emission is hot dust radiating at equilibrium temperature in \hii\ regions, as well as very small grains undergoing temperature fluctuations in PDRs.
Therefore, determining the PAH abundance requires a panchromatic approach,
to model the SEDs of galaxies.
It requires knowledge of star formation history, and processing of the stellar
SED by the gas and dust in the ionized and neutral phases of the galaxy.

The $8/24\mic$ flux ratio provides information on the strength of the PAH
features in the galaxy, although it is relatively limited.
In general, the \iraciv\ flux is dominated by the $6-9\mic$ PAH features, but 
in some cases, such as low metallicity galaxies or deeply embedded sources, the
$8\mic$ flux may be dominated by, respectively, the continuum 
\citep{madden06,wu06} or the $9.7\mic$ silicate feature 
\citep[e.g.][]{spoon04,hao07}.
When there are no PAHs and the \iraciv\ band is dominated by the continuum,
the $8/24\mic$ ratio reflects the color of the mid-IR emitting silicates and
graphites.
The \mipsi\ flux is dominated by the continuum emission from the silicate 
and carbon grains.
However, this part of the SED is very sensitive to the abundance of small
grains \citep[e.g.][]{galliano03,galliano05}, and can originate additionally
from large grains close to strong sources of radiation
\citep[e.g.][]{plante02,vanzi04}.

\begin{figure}[htbp]
  \plotone{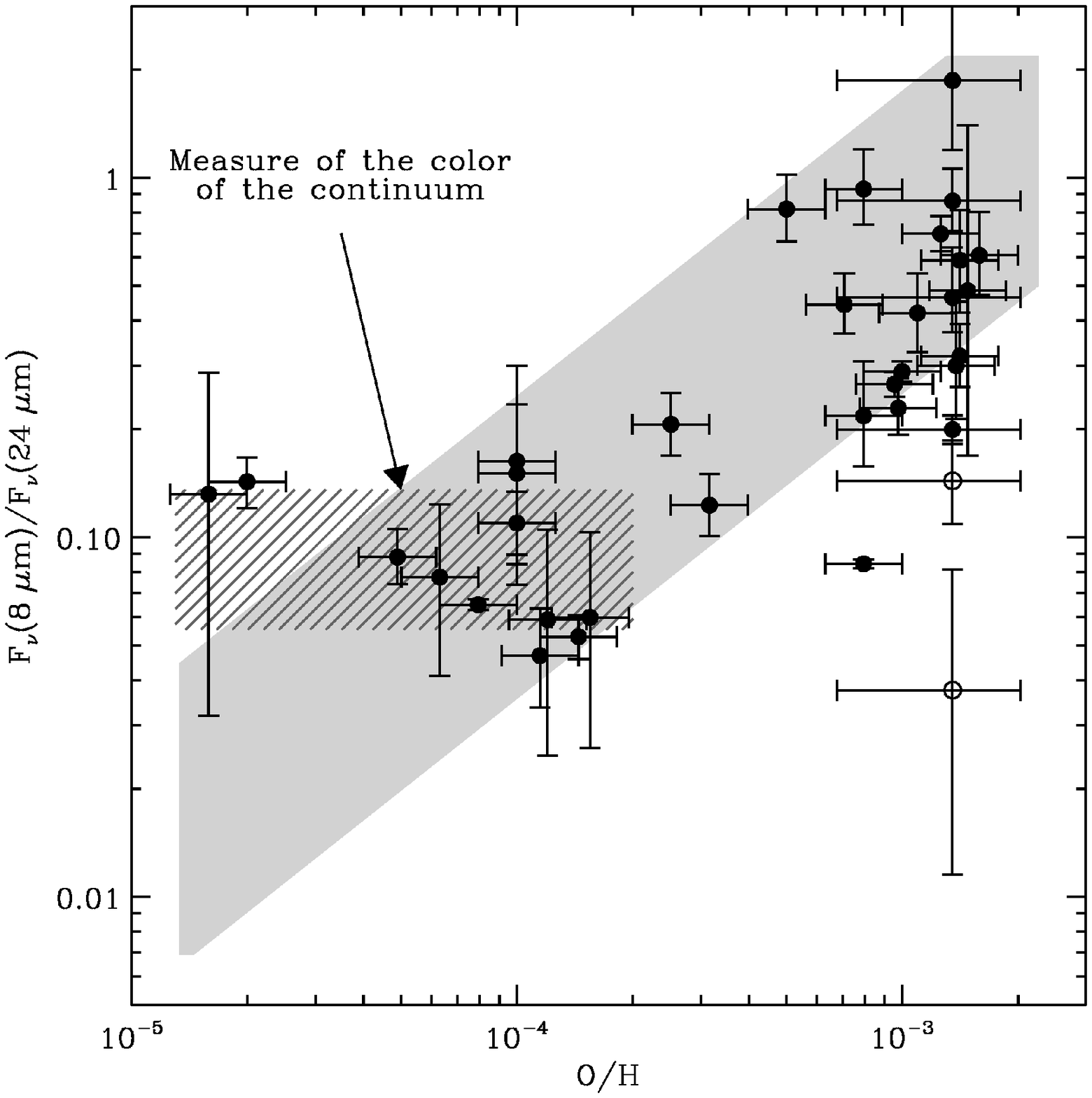}
  \caption{Mid-IR color as a function of the oxygen number abundance, for 
           the sample in \reftab{tab:source}.
           This figure illustrates the importance of knowing the mid-IR 
           continuum emission in order to derive the intensity of the PAH bands.
           Our observed SEDs have been integrated over the \iraciv\ and 
           \mipsi\ bandpasses, in order to produce this plot.
           The two open circles are the (U)LIRGs of our sample.
           The grey stripe is the $\pm1\sigma$ linear correlation between
           $\log(\rm O/H)$ and $\log\left(F_\nu(8\mic)/F_\nu(24\mic)\right)$.
           The hatched stripe shows the range of the ratio 
           ($0.06\lesssim F_\nu(8\mic)/F_\nu(24\mic)\lesssim0.13$), which is
           a measure of the color of the silicate and graphite continuum,
           when the PAH features are weak.}
  \label{fig:band_ratios}
\end{figure}

\reffig{fig:band_ratios} shows the ratio of the two bands, for 
the sources presented in \reftab{tab:source}.
The figure shows that when the PAH-to-continuum ratio is small, the $8/24\mic$
ratio plateaus around a value of $\simeq0.1$, which roughly represents the
color of the mid-IR continuum.
The figure illustrates that the separation between the PAH bands and the
continuum is important to study the evolution of their strengths with the
metallicity.
The error bars on the ratio, in \reffig{fig:band_ratios}, come from the 
propagation of the observational errors.
The error bars on the metallicity were not systematically given by the authors 
who published them.
An error of 0.1~dex in the O/H number abundance accounts for the typical 
dispersion between independent measurements.
There were 6 sources, in \reftab{tab:source}, for which no metallicity 
measurements were reported.
However, all of them have the morphology, optical colors, IR emission and 
\hmol\ content of solar or supersolar systems, with \ngc{1399} probably 
having the largest metallicity.
We assign an arbitrary $Z=(2\pm1)\:Z_\odot$ to these galaxies, and consider 
them to be uncertain.

\begin{figure}[htbp]
  \plotone{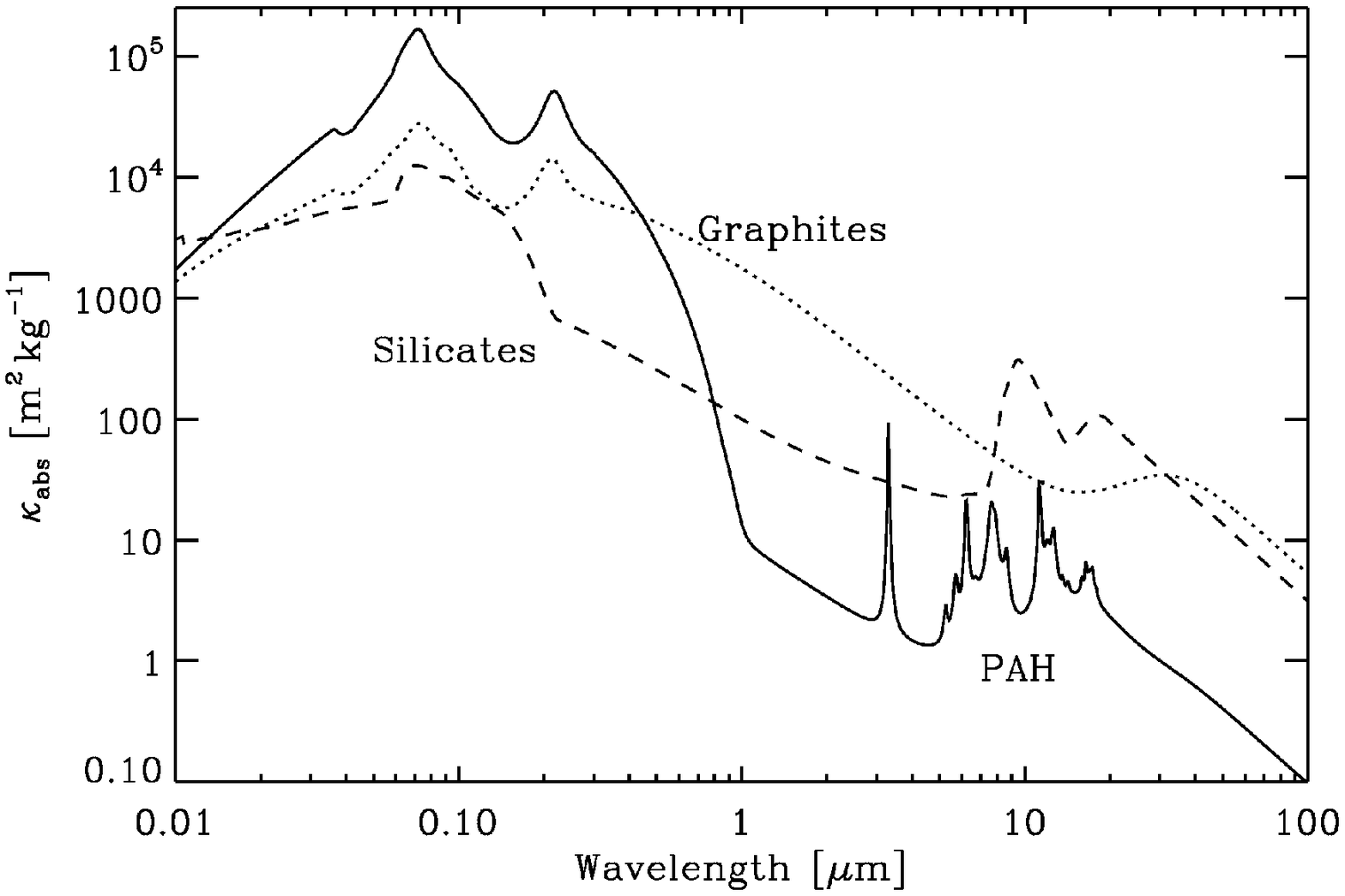}
  \caption{The wavelength-dependence of the mass absorption coefficient of the 
           PAHs \citep{draine07a}, graphites \citep{laor93} and 
           silicates \citep{weingartner01}, integrated over the \citet{zubko04}
           size distribution, for the bare grain model with solar abundance
           constraints (BARE-GR-S).}
  \label{fig:kappabs}
\end{figure}

\begin{figure}[htbp]
  \plotone{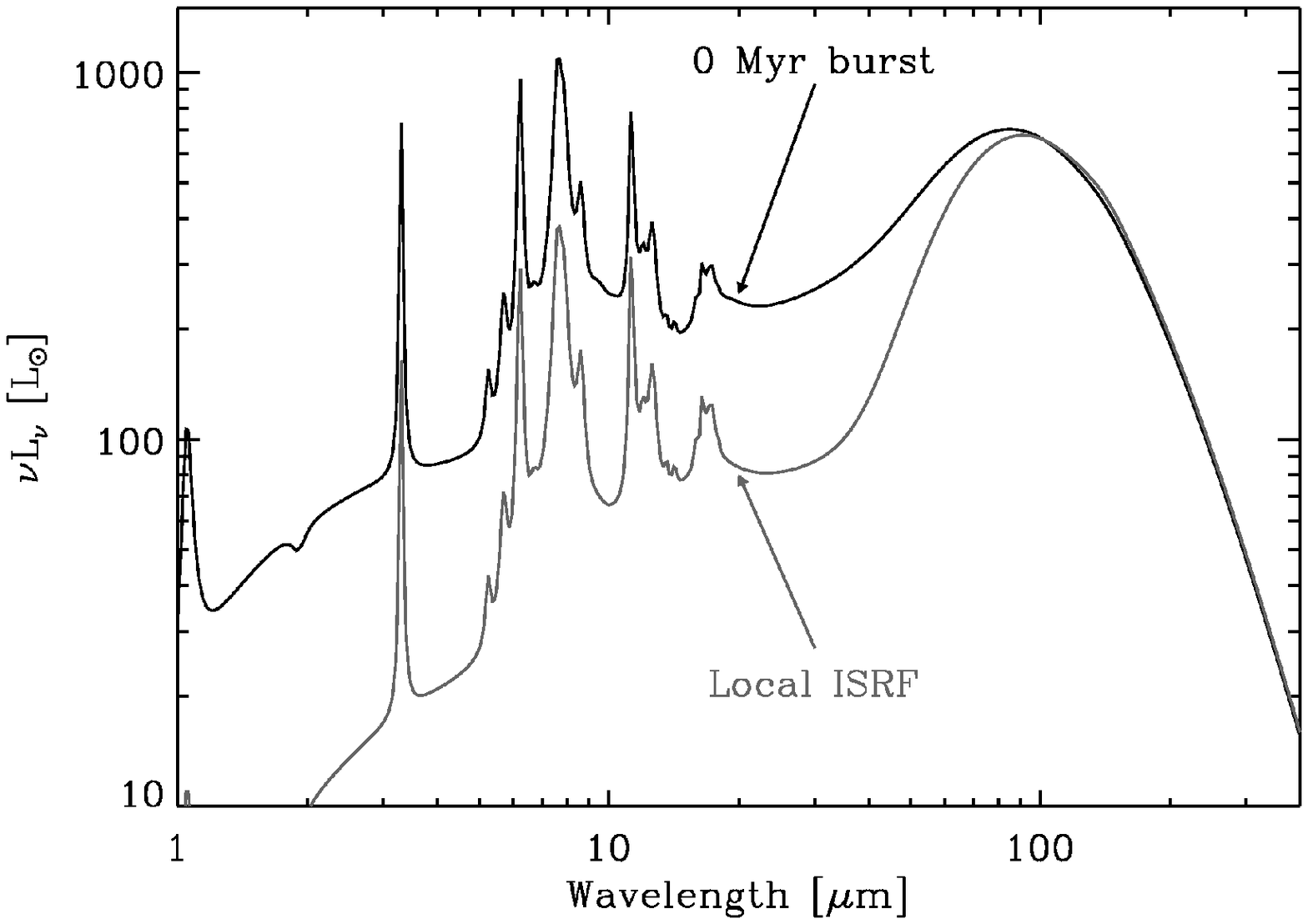}
  \caption{Sensitivity of the mid-IR SED to the age of the stellar populations.
           The two SEDs come from the same grain populations 
           \citep[abundances and size distribution;][]{zubko04};
           the only difference is the shape and intensity of the interstellar
           radiation field they are experiencing.
           The black SED is heated by a young UV-rich instantaneous burst,
           while the grey one is the subjected to the local ISRF.
           It shows that for a given far-IR spectrum, the PAH emission
           can vary by a factor of $\simeq3$ depending on the shape of the ISRF,
           therefore affecting the estimate of the PAH abundance by the same
           amount.}
  \label{fig:SEDvsage}
\end{figure}

Knowledge of the shape of the ISRF to determine the PAH abundance is important,
since the absorption properties of PAHs are significantly different from
the ones of solid-state grains (carbonaceous or silicates)
believed to be the carriers of the mid-IR continuum.
\reffig{fig:kappabs} shows the different wavelength-dependence of the mass absorption coefficient, $\kappa_\sms{abs}$, of various types of grains.
It shows that the value of $\kappa_\sms{abs}$ for PAHs drops by 4 orders of
magnitude between 0.1
and $1\mic$, where the stars emit most of their energy, while the 
$\kappa_\sms{abs}$ of graphite and silicate dust drop by less than 2 orders 
of magnitude.
However, most of the energy is absorbed at shorter wavelengths.
Thus, the PAHs are more sensitive to very young stellar populations than
the grains responsible for the $24\mic$ continuum (\reffig{fig:SEDvsage}), 
and a correct determination
of their excitation rate should take into account this property.

  \subsection{The Contribution of \hii\ Regions to the SED of a Galaxy}
  \label{sec:HII}

Models of the spatial distribution of the dust in our Galaxy indicate that 
\hii\ regions dominate the infrared SED up to wavelengths of $\sim 60\mic$
\citep{sodroski97,paladini07}. 
In principle, the contribution of \hii\ regions could be derived from any 
tracer of the recombination rate, like the radio free-free continuum.
However, the free-free-to-IR ratio depends significantly on the density and 
dust-to-gas mass ratio in the \hii\ region 
\citep[{e.g.}][]{petrosian72,mezger74}, requiring detailed modeling of \hii\
regions as a function of these parameters.

We consider \hii\ regions to be made of a central ionizing star, surrounded
by PAH-free dusty ionized gas.
The output of an individual \hii\ region therefore consists 
of:\textlist{\thetextlist~escaping non-ionizing stellar radiation; 
   \thetextlist~IR emission from the reprocessing of the stellar emission by 
     dust; and
   \thetextlist~radio free-free emission from the reprocessing of ionizing 
     photons by the gas.}We created a library of \hii\ region spectra for stars
of different spectral
types, embedded into a gas of different densities and dust-to-gas mass ratios.
The total contribution of \hii\ regions to a galaxy's SED is derived by 
integrating the spectra of indvidual \hii\ regions over the stellar initial 
mass function (IMF).

There are several simplifying assumptions in our procedure.
Individual \hii\ regions are considered to be spherical, homogeneous, consisting
of pure Hydrogen.
In calculating the infrared emission, we neglected scattering of stellar radiation by the dust.
We assumed that all the dust radiates at equilibrium dust temperature, and that
PAH are destroyed and therefore absent into the \hii\ region itself.
In calculating the emission from the ensemble of \hii\ regions, we assumed that
they are not clustered, so that the total radiation is an IMF weighted sum over
all spectral types, assumed to be surrounded by gas with the same density.

The theoretical evolution of the SED of \hii\ regions and their 
surrounding molecular material has been extensively discussed by
\citet{bressan02}, \citet{panuzzo03} and \citet{dopita05,dopita06c,dopita06b}.
In these studies, the spectrum of an \hii\ region depends on its evolutionary state, controlled by its age, density and the dissipation time scale of the surrounding molecular clouds.
In comparison, our approach is more flexible and empirical, in which the parameters can vary more freely instead of being constrained by evolutionary 
models.

In the following, we describe how we derive the SED of \hii\ regions.
The definition of each mathematical variable is summarised in \reftab{tab:var}.

\begin{deluxetable}{lll}
  \tabletypesize{\small}
  \rotate
  \tablecolumns{3}
  \tablewidth{\textheight}
  \tablecaption{Mathematical Variables Used Throughout the Paper.}
  \tablehead{
    \colhead{Name} & \colhead{Units} & \colhead{Definition}}
  \startdata
    \cutinhead{General Variables}
    $\nu$  & Hz     & Frequency of the radiation \\
    $\lambda$  
           & $\mu$m & Wavelength of the radiation \\
    $M_\sms{gas}$  
           & $M_\odot$       
                    & Gas mass (Hydrogen, Helium and metals) of the region 
                      or galaxy \\
    $M_\sms{dust}$  
           & $M_\odot$       
                    & Dust mass of the region or galaxy \\
    $Z_\sms{dust}=M_\sms{dust}/M_\sms{gas}$  
           &        & Dust-to-gas mass ratio \\
    $Z_\sms{gas}$
           &        & Metal mass fraction in the gas phase (or metallicity) \\
    $m_\sms{H}=1.67\E{-24}$ 
           & g      & Mass of a single H atom \\ 
    $m_\sms{dust}$  
           & g      & Average mass of a dust grain \\
    $a$    & $\rm\mu m$ 
                    & Grain radius \\
    $f(a)$ & $\rm grain\,\mu m^{-1}$
                    & Grain size distribution, normalized to unity \\
    $Q_\sms{abs}(a,\nu)$
           &        & Absorption efficiency of a grain of radius $a$ \\
    $\sigma_\sms{dust}(\nu)
     =\displaystyle\int_0^\infty \pi a^2 Q_\sms{abs}(a,\nu)\,f(a)\ddiff a$
           & $\rm cm^2\,grain^{-1}$
                    & Average dust cross-section per grain \\
    $m$    & $M_\odot$
                    & Mass of individual stars \\
    $\phi(m)$
           & $M_\odot^{-1}$
                    & Initial mass function \\
    \cutinhead{\hii\ Region Variables (\refS{sec:HII})}
    $\nu_1=3.29\E{15}$  
           & Hz     & Frequency of the Lyman break \\
    $R_s$  & cm     & Radius of the equivalent dust-free Str\"omgren sphere
                      of a single \hii\ region \\ 
    $y=r/R_s$   
           &        & Radial coordinate normalised to $R_s$ \\
    $n_\sms{H}$
           & $\rm H\,cm^{-3}$
                    & Number of H atoms (or protons) per unit volume \\
    $n_\sms{dust}$
           & $\rm grain\,cm^{-3}$
                    & Number of dust particles per unit volume \\
    $n_e(y)$  
           & $e\,\rm cm^{-3}$
                    & Number of electrons per unit volume \\
    $x(y)=n_e/n_\sms{H}$    
           &        & Ionization fraction of Hydrogen \\
    $\sigma_\sms{H}(\nu)$
           & $\rm cm^2\,H^{-1}$
                    & Ionization cross-section of Hydrogen 
                      \citep[Eq.~2.4 of][]{spitzer78} \\
    $T_e$  & K      & Electron temperature \\
    $T_\sms{dust}$  
           & K      & Equilibrium dust temperature \\
    $L_\nu(\nu,y)$
           & $L_\odot\,\rm Hz^{-1}$
                    & Total monochromatic luminosity, passing
                      through the shell of radius $y$ \\
    $L_\nu^\sms{OB}(\nu)$
           & $L_\odot\,\rm Hz^{-1}$
                    & Intrinsic monochromatic luminosity emitted by the central 
                      star, given by \citet{panagia73} \\
    $L_\sms{B}(y)$
           & $L_\odot$
                    & Luminosity of the case B recombination lines at a  
                      radius $y$ \\
    $L_\nu^\sms{dust}(\nu,y)$
           & $L_\odot\,\rm Hz^{-1}$
                    & Monchromatic luminosity reprocessed by the dust at a
                      radius $y$ \\
    $L_\nu^{ff}(\nu,y)$
           & $L_\odot\,\rm Hz^{-1}$
                    & Monchromatic luminosity of the free-free cooling at 
                      a radius $y$ \citep[Eq.~3.54 of ][]{spitzer78} \\
    $\alpha_\sms{A}(T_e)$
           & $\rm cm^3\,s^{-1}$
                    & Case A recombination coefficient 
                      \citep[Table~2.1 of ][]{osterbrock89} \\
    $\alpha_\sms{B}(T_e)$
           & $\rm cm^3\,s^{-1}$
                    & Case B recombination coefficient 
                      \citep[Table~2.1 of ][]{osterbrock89} \\
    $L_\nu^\sms{\hii}(\nu,m,Z_\sms{dust},n_\sms{H})$
           & $L_\odot\,\rm Hz^{-1}$
                    & Monochromatic luminosity of an individual \hii\ region \\
    $t_\sms{burt}$
           & Myr    & Age of the burst of star formation \\
    $L_\nu^\sms{burst}(\nu,t_\sms{burst},Z_\sms{dust},n_\sms{H})$
           & $L_\odot\,\rm Hz^{-1}$
                    & Monocromatic luminosity of a distribution 
                      of \hii\ regions \\
    \cutinhead{Global SED Variables (\refS{sec:SED})}
    $t_\sms{SF}$
           & Myr    & Age of the galaxy \\
    $M_\star$
           & $M_\odot$
                    & Mass of the non-ionizing stars \\
    $f_\sms{sync}$
           &        & Ratio between the synchrotron and the free-free 
                      at $\lambda=1\;\rm cm$ \\
    $M_\sms{dust}$
           & $M_\odot$
                    & Dust mass in the PDRs \\
    $f_\sms{PAH}=M_\sms{PAH}/M_\sms{dust}$
           &        & PAH mass fraction \\
    $f_+=M_\sms{PAH$^+$}/M_\sms{PAH}$
           &        &  Fraction of ionized PAHs \\
    $L_i$
           & $L_\odot$       
                    & Luminosity of a mid-IR ionic line \\
    $U_\nu(\nu)$
           & $\rm erg\,s^{-1}\,cm^{-3}\,Hz^{-1}$
                    & Monochromatic radiation density in the PDRs \\
    $U=\displaystyle\int_0^\infty U_\nu(\nu)\ddiff\nu$
           & $\rm erg\,s^{-1}\,cm^{-3}$
                    & Integrated radiation density in the PDRs \\
    $L_\nu^\star(\nu,t_\sms{SF},M_\star)$
           & $L_\odot\,\rm Hz^{-1}$    
                    & Monochromatic luminosity emitted by the non-ionizing
                      stars \\
    $L_\nu^\sms{sync}(\nu,f_\sms{sync})$
           & $L_\odot\,\rm Hz^{-1}$    
                    & Monochromatic luminosity of the synchrotron radiation \\    
    $L_\nu^\sms{Zub}(\nu,U,M_\sms{dust},f_\sms{PAH},f_+)$
           & $L_\odot\,\rm Hz^{-1}$    
                    & Monochromatic luminosity emitted by the dust in 
                      a given $U$ \\  
    $L_\nu^\sms{PDR}(\nu,\alpha,U_-,U_+,M_\sms{dust},f_\sms{PAH},f_+)$
           & $L_\odot\,\rm Hz^{-1}$    
                    & Monochromatic luminosity emitted by the dust in PDRs \\  
    $P_e^\sms{\hii}(\nu,A_\sms{V}^\sms{\hii})$
           &        & Escaping fraction of the \hii\ region photons \\  
    $P_e^\sms{PDR}(\nu,A_\sms{V}^\sms{PDR})$
           &        & Escaping fraction of the PDR photons \\  
    $L_\sms{burst}=\displaystyle\int_0^\infty L_\nu^\sms{burst}\ddiff\nu$
           & $L_\odot$ 
                    & Intrinsic luminosity radiated by the \hii\ regions \\
    $\langle P_e L_\sms{burst}\rangle
     =\displaystyle\int_0^\infty P_e^\sms{\hii} L_\nu^\sms{burst}\ddiff\nu$
           & $L_\odot$ 
                    & Escaping luminosity from the \hii\ regions \\
    $L_\sms{PDR}=\displaystyle\int_0^\infty L_\nu^\sms{PDR}\ddiff\nu$
           & $L_\odot$ 
                    & Luminosity radiated by the dust in the PDRs \\
    $\langle P_e L_\sms{PDR}\rangle
     =\displaystyle\int_0^\infty P_e^\sms{PDR} L_\nu^\sms{PDR}\ddiff\nu$
           & $L_\odot$ 
                    & Escaping luminosity from the PDRs \\
    $L_\star=\displaystyle\int_0^\infty L_\nu^\star\ddiff\nu$
           & $L_\odot$ 
                    & Intrinsic luminosity radiated by the non-ionizing stars \\
    $\langle P_e L_\star\rangle
     =\displaystyle\int_0^\infty P_e^\sms{PDR}L_\nu^\star\ddiff\nu$
           & $L_\odot$ 
                    & Escaping luminosity from the non-ionizing stars \\
    \cutinhead{Elemental and Dust Evolution Variables (\refS{sec:dustvol})}
    $\Sigma_\sms{gas}(t)$
           & $M_\odot\,\rm pc^{-2}$ 
                    & Gas mass surface density \\
    $\Sigma_\sms{gas}^{0}$
           & $M_\odot\,\rm pc^{-2}$ 
                    & Initial gas mass surface density \\
    $\Sigma_\sms{SFR}(t)$
           & $M_\odot\,\rm yr^{-1}\,pc^{-2}$ 
                    & Star formation rate surface density \\
    $Z_\sms{ISM}$
           &        & Total metal mass fraction (gas and dust) in the ISM \\
    $\tau(m)$
           & Myr    & Lifetime of a star of mass $m$ \\
    $m_\sms{ej}(t)$
           & $M_\odot$
                    & Mass of gas ejected by a star of mass $m$, after a time
                      $\tau(m)$ \\
    $Y_\sms{Z}(m)$
           & $M_\odot$
                    & Mass of metals ejected by a star of mass $m$, after a time
                      $\tau(m)$ \\
    $\mu_\sms{gas}(t)=\Sigma_\sms{gas}(t)/\Sigma_\sms{gas}^0$
           &        & Reduced gas mass of the system \\
    $\tau_\sms{dust}(t)$
           & Myr    & Dust lifetime \\
    $\langle m_\sms{ISM}\rangle$
           & $M_\odot$
                    & Average gas mass swept-up by a single \snii\ \\ 
  \enddata
  \label{tab:var}
\end{deluxetable}

  \subsubsection{The Modeling of Individual \hii\ Regions}
  
We assume that PAHs are totally depleted inside the ionized gas phase (the actual \hii\ region), and that their mass fraction is constant outside of these 
regions, inside the photodissociation regions.
Furthermore, we assume that the \hii\ regions are homogeneous, spherical, 
and that they contain only Hydrogen and dust.
The dust-free Str\"omgren radius $R_s$ \citep[{e.g.}][]{spitzer78,osterbrock89} 
of such a region is defined by the balance between the rate of ionizing photons
emitted by the central star and by the electrons recombining to the ground level, and the rate of recombinations to any level higher than the fundamental
state:
\begin{equation}
  \frac{4\pi}{3}\,R_s^3\, n_\sms{H}^2\, \alpha_\sms{B}(T_e)
     = \int_{\nu_1}^\infty \frac{L_\nu^\sms{OB}(\nu)}{h\nu}\ddiff \nu.
\end{equation}
We use this quantity to normalize the radial coordinate $r$ to the dimensionless
radius $y=r/R_s$.
In each shell of radius $y$ and thickness $\dd y$, the Hydrogen optical depth is given by:
\begin{equation}
  \dd\tau_\sms{H}(\nu,y) = \left[1-x(y)\right]\,n_\sms{H}\,R_s\times
                           \sigma_\sms{H}(\nu)\ddiff y,
  \label{eq:tauH}
\end{equation}
the dust optical depth is:
\begin{equation}
  \dd\tau_\sms{dust}(\nu,y) = n_\sms{dust}(y)\,R_s \times 
                              \sigma_\sms{dust}(\nu)\ddiff y,
  \label{eq:taud}
\end{equation}
and the volume of the shell is $\dd V(y) = 4\pi\,R_s^3\,y^2\ddiff y$.
We solve the radiative transfer equation, from the central
star to the photoionization front:
\begin{equation}
  \frac{\dd L_\nu(\nu,y)}{\dd y} = 
    - L_\nu(\nu,y)\times\left(\frac{\dd\tau_\sms{H}(\nu,y)}{\dd y} 
                              +\frac{\dd\tau_\sms{dust}(\nu,y)}{\dd y}\right)
    +\frac{\dd L_\nu^\sms{dust}(\nu,y)}{\dd y}
    +\frac{\dd L_\nu^{ff}(\nu,y)}{\dd y},
   \label{eq:ionirate}
\end{equation}
together with the photoionization equilibrium in each shell:
\begin{equation}
  \alpha_\sms{A}(T_e)\, x(y)^2 n_\sms{H}^2\ddiff V(y) \times h\nu_1
    = \int_{\nu_1}^\infty L_\nu(\nu,y)\,\dd\tau_\sms{H}(\nu,y)
                          \ddiff\nu.
  \label{eq:photoioneq}
\end{equation}
\refeq{eq:photoioneq} gives the value of $x(y)$.

We assume that all the case B recombination lines are resonantly scattered 
by the gas and are finally absorbed locally by the dust.
Therefore the dust is heated both by the stellar continuum and these recombination lines:
\begin{equation}
  \int_0^\infty \frac{\dd \tau_\sms{dust}(\nu,y)}{\dd y} L_\nu(\nu,y)\ddiff\nu
    +L_\sms{B}(y)
    = n_\sms{dust}\frac{\dd V}{\dd y}(y)
      \int_0^\infty \sigma_\sms{dust}(\nu)\,4\pi B_\nu(\nu,T_\sms{dust})
            \ddiff\nu.
  \label{eq:Tdust}
\end{equation}
In \refeq{eq:Tdust}, we assume that the dust is at thermal equilibrium with the
radiation field.
Very small grains may still undergo temperature fluctuations, but since their size distribution is poorly known, we neglect this effect in calculating the IR emission. 
This will lead to an underestimate of the $\sim 1 - 20\mic$ continuum. 
We will discuss in \refS{sec:SED} an empirical way to compensate this 
underestimation.

Finally, we take into account the fact that dust sublimates in the center of
\hii\ regions \citep[{e.g.}][]{inoue02}, by considering that each shell where
the equilibrium temperature of the dust exceeds its sublimation temperature
is free of this dust specie.
We adopt sublimation temperatures of $2500\;\rm K$ and $1800\;\rm K$ 
for graphite and silicate, respectively \citep{kruegel03}.

We solved \refeqs{eq:ionirate} and (\ref{eq:photoioneq}) for stars of 
masses $8\msun<m<100\msun$, with a grid of densities ranging from 
$n_\sms{H}=1\;\rm cm^{-3}$ to $n_\sms{H}=3\E{4}\;\rm cm^{-3}$, and a grid 
of dust-to-gas mass ratios ranging from $Z_\sms{dust}=1/120$ 
\citep[Galactic value;][]{zubko04} to $Z_\sms{dust}=1/12000$. 

  \subsubsection{The Total Contribution of \hii\ Regions}
  
\begin{figure*}[htbp]
  \centering
  \includegraphics[width=\textwidth]{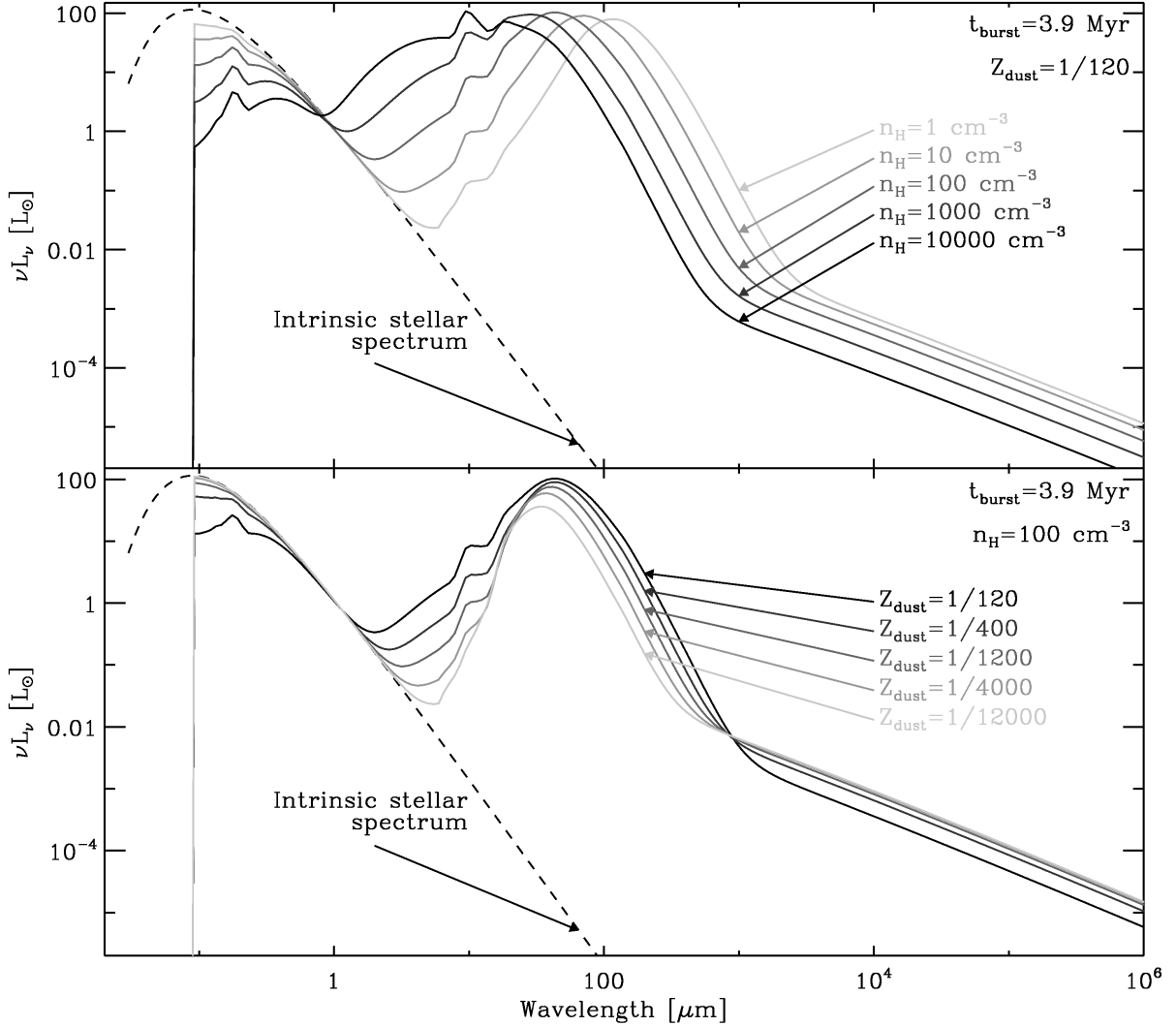}
  \caption{SED of an ensemble of \hii\ regions \refeqp{eq:HII}:
           for a given age and dust-to-gas mass ratio, varying the density
           (top panel);
           and for a given age and density, varying the dust-to-gas mass ratio
           (bottom panel).
           The density tends to increase the dust temperature, as well as
           the free-free-to-IR ratio (top panel).
           Indeed, when the density is high, the dust absorbs directly a
           significant fraction of the ionizing photons.
           Inversely, when $Z_\sms{dust}$ drops, more ionizing photons
           are absorbed by the gas.
           The free-free-to-IR ratio is then higher, and the dust luminosity
           lower (bottom panel).}
  \label{fig:HII}
\end{figure*}

At the scale of a galaxy, the SED of the ionized gas phase
is the combination of several \hii\ regions.
We assume that all the \hii\ regions of a given galaxy have the same density,
and the same dust-to-gas mass ratio.
We adopt a Salpeter initial mass function:
\begin{eqnarray}
  \phi(m) \propto m^{-2.35} & \mbox{for} & m_- < m < m_+ \label{eq:imf} \\
  & \mbox{normalised to} & \int_{m_-}^{m_+} \phi(m) \ddiff m = 1 \nonumber\\
  & \mbox{with} & 
  \left\{
  \begin{array}{rcl}
    m_- & = & 0.1\msun \\ 
    m_+ & = & 100\msun \\ 
  \end{array}
  \right. \nonumber
\end{eqnarray}
The total SED of an ensemble of \hii\ regions is then given by:
\begin{equation}
  L^\sms{burst}_\nu(\nu,t_\sms{burst},Z_\sms{dust},n_\sms{H})
    = \int_{m_-}^{m_+(t_\sms{burst})}
      L_\nu^\sms{\hii}(\nu,m,Z_\sms{dust},n_\sms{H})\times \phi(m)\ddiff m.
  \label{eq:HII}
\end{equation}
The upper mass decreases with the age of the burst of star formation.
\reffig{fig:HII} shows several of these SEDs.

In our case, the age of the \hii\ region does not have an important effect
on the shape of the \hii\ SED.
Indeed, the age affects essentially the ionizing-to-non-ionizing photon rate.
Since our \hii\ SEDs are integrated only out to the photoionization front, the 
ionizing photons dominate the total power input. 
Therefore, in what follows, we will consider \hii\ SEDs with an age of 4~Myr.

  \subsection{The Inclusion of Emission from PDRs and the Resulting 
              Galactic SEDs}
  \label{sec:SED}

  \subsubsection{The Building Blocks}
  
\begin{figure*}[htbp]
  \centering
  \includegraphics[width=\textwidth]{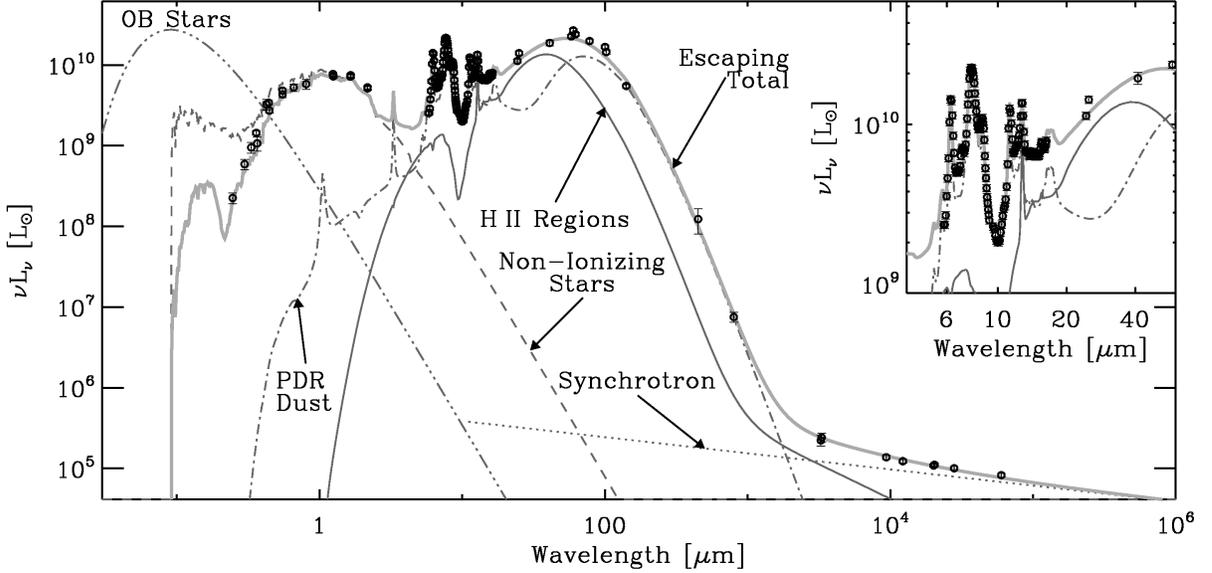}
  \caption{Demonstration of the panchromatic SED model applied on a galaxy.}
  \label{fig:methSED}
\end{figure*}
\begin{figure*}[htbp]
  \centering
  \includegraphics[width=0.9\textwidth]{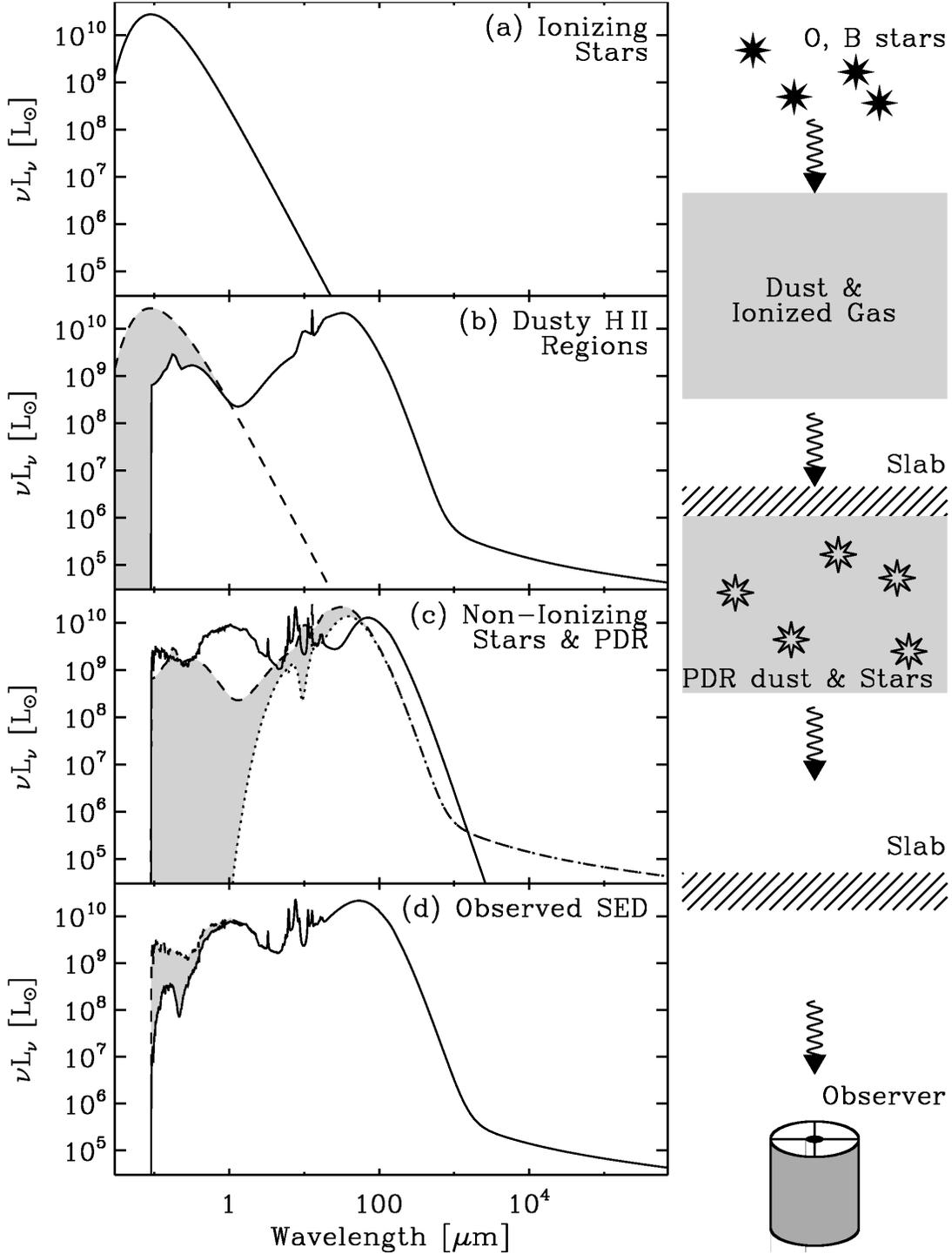}
  \caption{Illustration of the geometry of the model \refeqp{eq:SED}.
           The left panels, from the top to the bottom shows the combination
           of the various SED building blocks from the massive star clusters
           to the observer.
           The solid lines are the total SED at each step; 
           the dashed lines are the SED of the previous step;
           the part of the SED that has been absorbed is shown in grey.
           The right panel illustrate the path of the photons from the star 
           clusters to the observer.}
  \label{fig:bigSED}
\end{figure*}

  
A significant fraction of the radiation escaping from \hii\ regions is 
absorbed by the dust in the surrounding molecular clouds, as well as in
the diffuse ISM.
These dense and diffuse PDR components are also heated by the ambient
non-ionizing radiation field.
Technically, all the parameters of our model are adjusted simultaneously.
However, for clarity, we can decompose the procedure in the following steps:
\begin{enumerate}
  \item Observations of at least two data points sampling the radio 
        continuum constrain the synchrotron-to-free-free ratio.
        The synchrotron emission is $L_\nu^\sms{sync}\propto\nu^{-0.8}$.
  \item The resulting free-free continuum, together with observations of
        the mid-IR continuum between $\sim 5$ and $\sim 60\mic$ are used to 
        constrain the parameters of the \hii\ regions.
        We use the \hii\ templates presented in \refS{sec:HII}, attenuated
        with a slab extinction 
        $P_e^\sms{\hii}(\nu)=\exp\left[-\tau_\sms{dust}(\nu)\right]$,
        to account for absorption by intervening dust to the \hii\ region.
        The power absorbed, $L_\sms{burst}-\langle P_e L_\sms{burst}\rangle$, 
        contributes to the heating of the dust in PDRs.
        As mentioned in \refS{sec:HII}, we have not taken into account 
        the emission from stochastically heated grains in this phase.
        To correct empirically for this effect, we redistribute a part of 
        the \hii\ dust power into the sum of two modified black bodies
        of temperatures 130~K and 300~K (peaking around 15 and $5\mic$).
        The luminosity of each individual black body is free to vary.
        We enforce that this redistribution does not exceed $40\%$ of the total
        \hii\ dust luminosity.
        The dust-to-gas mass ratio in the \hii\ phase is assumed to be equal to
        the one in the PDRs.
  \item Optical/near-IR broadbands are used to constrain the escaping diffuse 
        stellar radiation, $L_\nu^\star$, using the 
        stellar population synthesis code \peg\ \citep{fioc97}, with a 
        Salpeter IMF, and a star formation rate as a function of time 
        proportional to a power-law of the gas surface density 
        with an index of 1.4 \citep{kennicutt98}.
        We vary the initial gas mass surface density $\Sigma_\sms{gas}^{0}$
        from $3\msun\rm\, pc^{-2}$ to $300\msun\rm\, pc^{-2}$.
        Here again, we assume a slab extinction, for the sake of simplicity.
        The power absorbed, $L_\star-\langle P_e L_\star\rangle$, 
        contributes to the heating of the dust in PDRs.
  \item The far-IR/submm observed SED constrains the dust emission from the
        PDRs.
        In order to account for variations of the radiation density, $U$, 
        in these regions, we assume a power-law distribution, following the 
        empirical prescription introduced by \citet{dale01}:
        \begin{equation}
          L_\nu^\sms{PDR}(\nu) \propto \int_{U_-}^{U_+} L_\nu^\sms{Zub}(\nu,U)
                               \times U^{-\alpha}\ddiff U,
          \label{eq:dale}
        \end{equation}
        where $L_\nu^\sms{Zub}(\nu,U)$ is the dust SED corresponding to a single 
        radiation density $U$.
        We adopt the dust properties of the Galactic diffuse ISM modeled by 
        \citet{zubko04}, for bare grains with solar abundance constraints.
        We fix the silicate-to-graphite mass ratio, but let free to vary 
        the PAH-to-dust mass ratio, $f_\sms{PAH}$, as well as the fraction of 
        ionized PAHs, $f_+$.
        These PAH properties are constrained by the detailed fit of the 
        features seen on the mid-IR spectrum.
        The shape of the radiation field exciting the dust is:
        \begin{equation}
          U_\nu(\nu) \propto L_\nu^\sms{burst}(\nu)+L_\nu^\star(\nu).
        \end{equation}
        This dust component is attenuated by the same factor than the diffuse 
        ISRF, $P_e^\sms{PDR}(\nu)$, in order to reproduce the silicate 
        extinction feature at $9.7\mic$.
  \item Our mid-IR spectra exhibit several fine structure ionic lines: 
        \ariiline, \ariiiline, \sivline, \neiiline, \neiiiline, 
        \siiiline, \siiilineb\ and \siIIline.
        We fit these lines in order to get a better $\chi^2$, but
        we do not use them for our physical interpretation.
        We adopt a Gauss profile whose width is determined by the resolution 
        of the spectrograph.
        The luminosity, $L_i$, of each line is free to vary.
\end{enumerate}
The total SED is then:
\begin{equation}
  L_\nu(\nu) = P_e^\sms{\hii}(\nu)\times L_\nu^\sms{burst}(\nu)
             + P_e^\sms{PDR}(\nu)\times
               \left[L_\nu^\sms{PDR}(\nu)+L_\nu^\star(\nu)\right]
             + L_\nu^\sms{sync}(\nu) 
             + \sum_i L_\nu^{(i)}(\nu).
  \label{eq:SED}
\end{equation}
The energy conservation implies that:
        \begin{equation}
          \langle P_e L_\sms{PDR}\rangle 
          = L_\sms{burst}-\langle P_e L_\sms{burst}\rangle
          + L_\star-\langle P_e L_\star\rangle.
        \end{equation}
\reffig{fig:methSED} demonstrates this model, and \reffig{fig:bigSED} 
illustrates its geometry.
  
  \subsubsection{The Model Results}

We fit the observed UV-to-radio SED, 
$L_\nu^\sms{obs}(\lambda_i)\pm\Delta L_\nu^\sms{obs}(\lambda_i)/2$ 
($i$ denoting an individual wavelength), 
of each one of the sources in \reftab{tab:source}, with \refeq{eq:SED}, 
minimizing the $\chi^2$.
We weight each data point depending on the density of adjacent points:
\begin{equation}
  \chi^2 = \sum_i \frac{{\rm min}_j(\lambda_j-\lambda_i)}{\lambda_i}
           \left(\frac{L_\nu^\sms{obs}(\lambda_i)-L_\nu(\lambda_i)}
                      {\Delta L_\nu^\sms{obs}(\lambda_i)}\right)^2.
  \label{eq:chi2}
\end{equation}
\refeq{eq:chi2} prevents the $\chi^2$ to be dominated by the parts of the
electromagnetic spectrum where its sampling by the observations is dense.
From this fit, we derive the value of the various free parameters.
In particular, we are interested in the total PAH-to-gas 
(and PDR-dust-to-gas) mass ratio, 
$Z_\sms{PAH(dust)}$:
\begin{equation}
  Z_\sms{PAH(dust)} 
    = \frac{M_\sms{PAH(dust)}}{\mu\,\left(M_\sms{\hi}+M_\sms{\hmol}\right)} 
  \qquad\mbox{ where }\qquad \mu=\frac{1}{1-Y_\odot-Z_\sms{gas}}.
\end{equation}


\clearpage
\section{THE PAH AND DUST ABUNDANCES IN GALAXIES}
\label{sec:results}

\begin{figure*}[htbp]
  \plotone{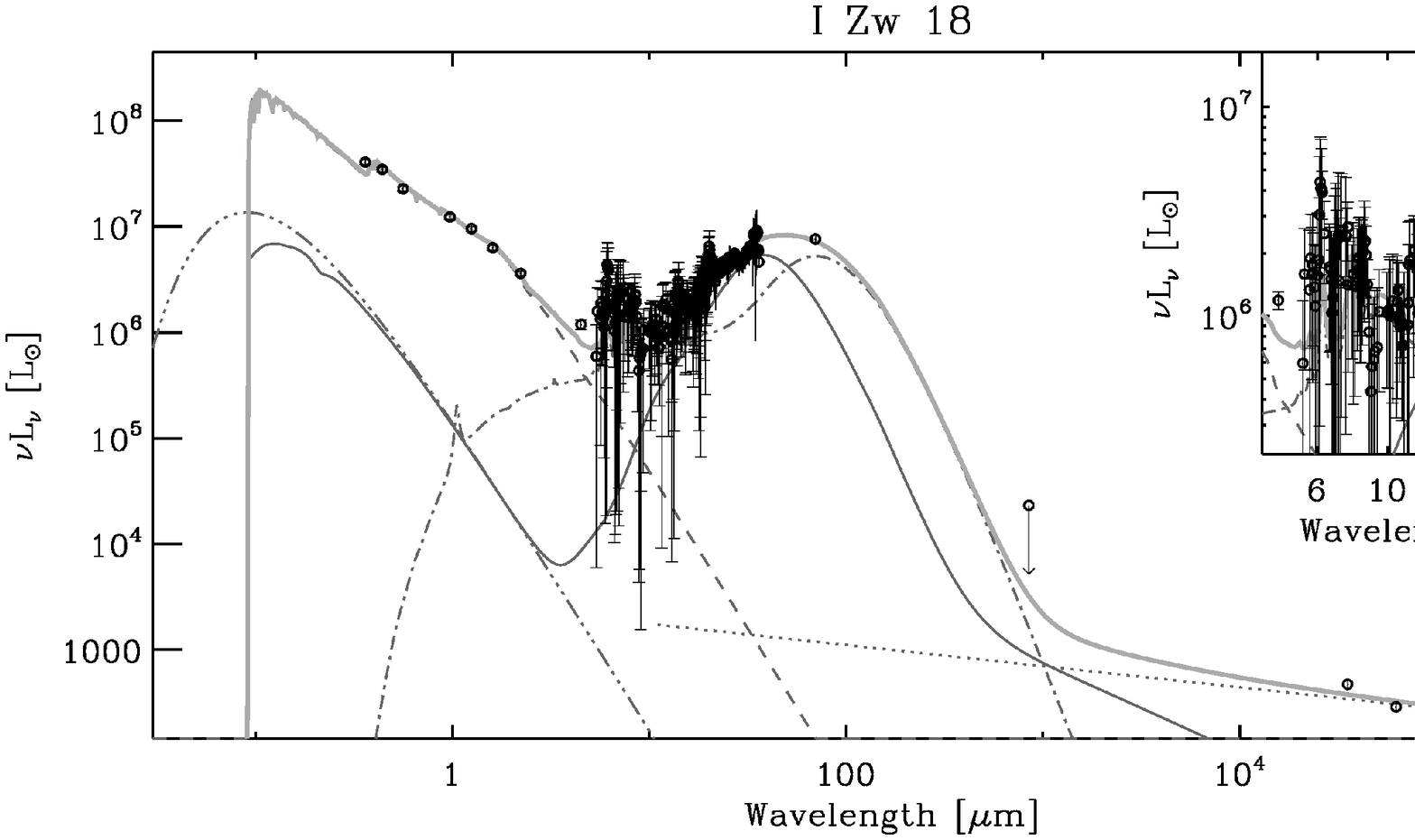}
  \plotone{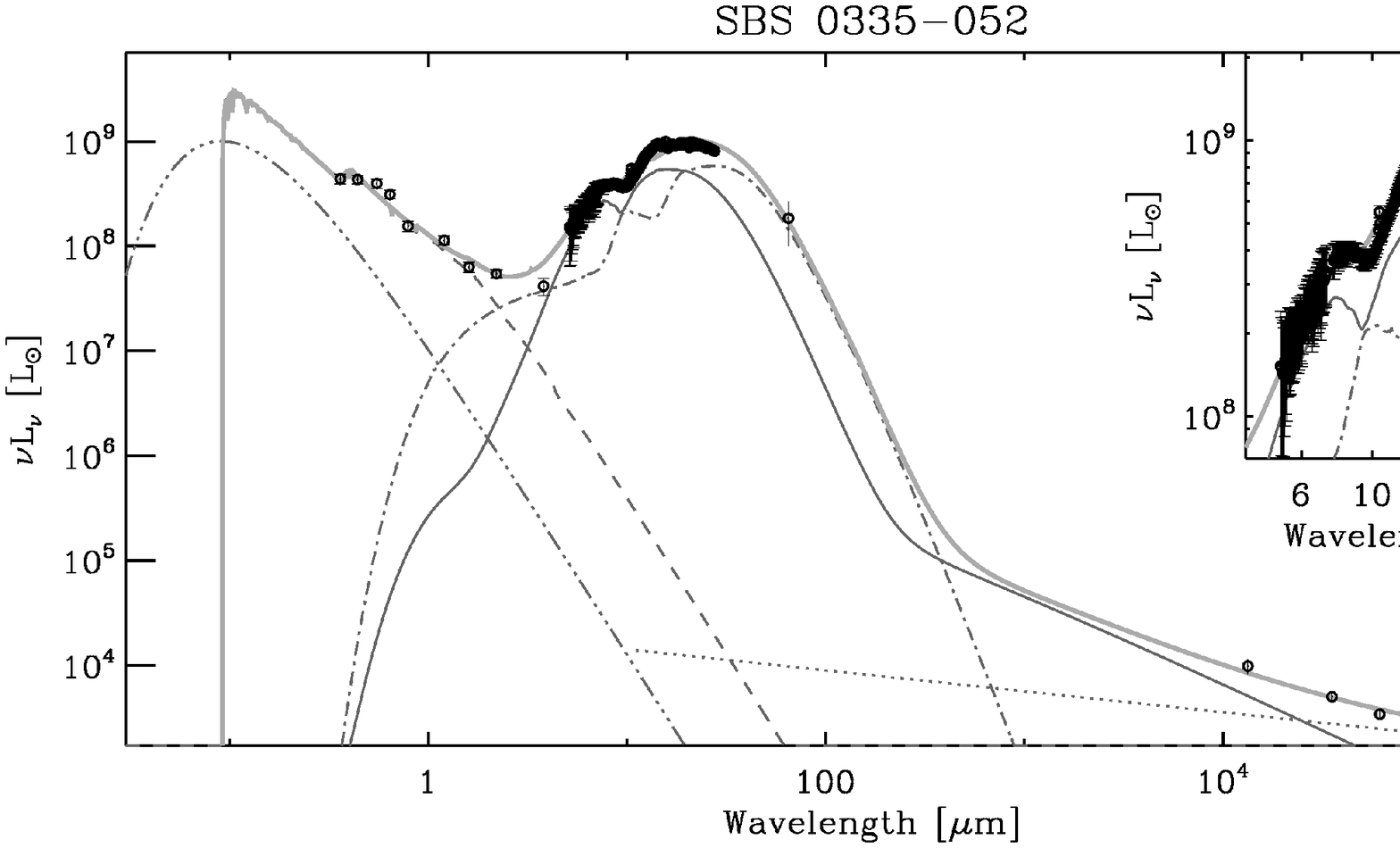}
  \caption{Fit of the galaxies' SEDs.
           The black lines are the components of the model 
           (\reffig{fig:methSED}) and the thick grey line is the total
           observed SED.
           The circles with error bars are the broad-band and spectral
           observations. 
           The {\it top-right panel} shows the detailed fit of the mid-IR 
           spectrum.}
  \label{fig:fits1}
\end{figure*}

\begin{figure*}[htbp]
  \plotone{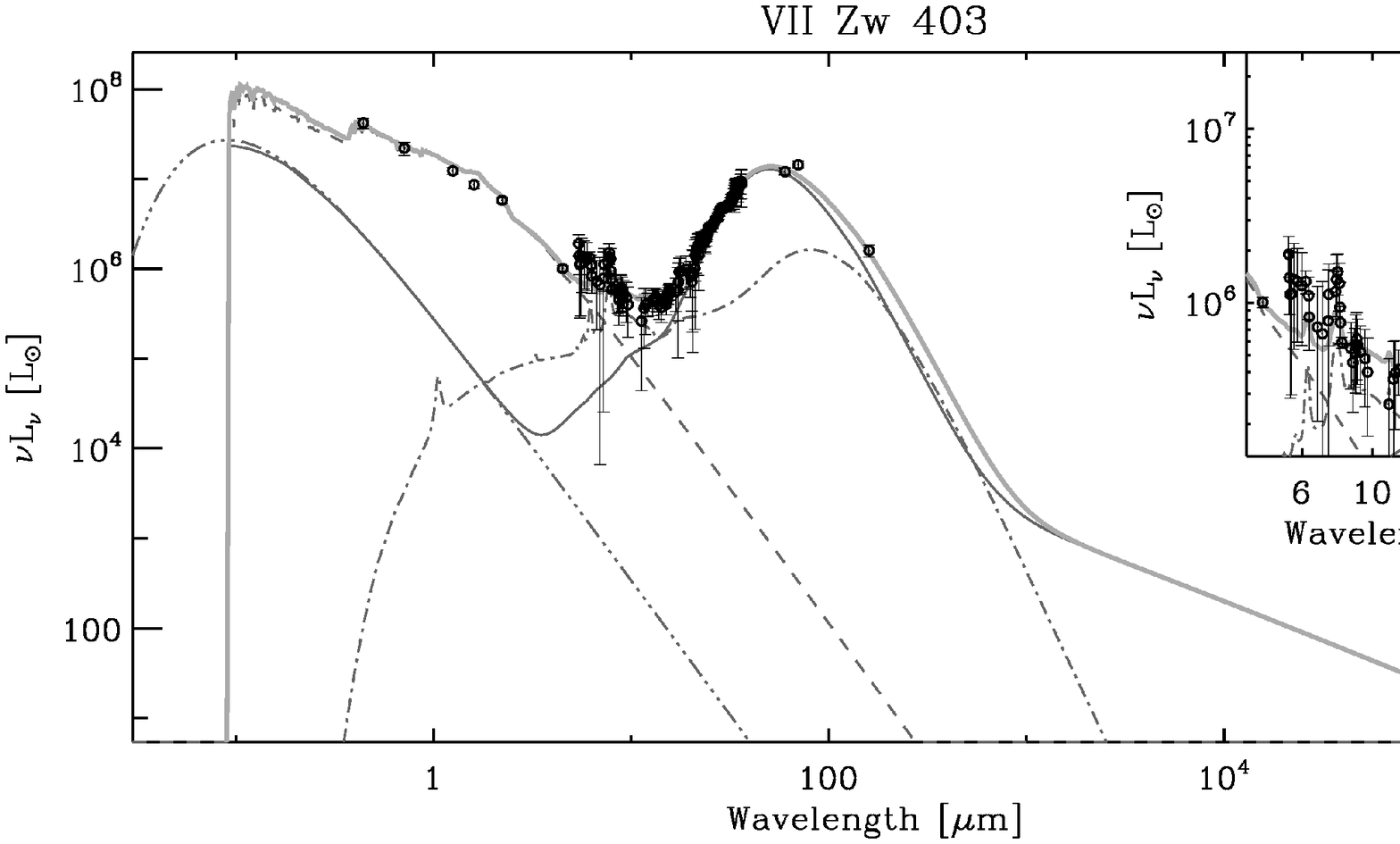}
  \plotone{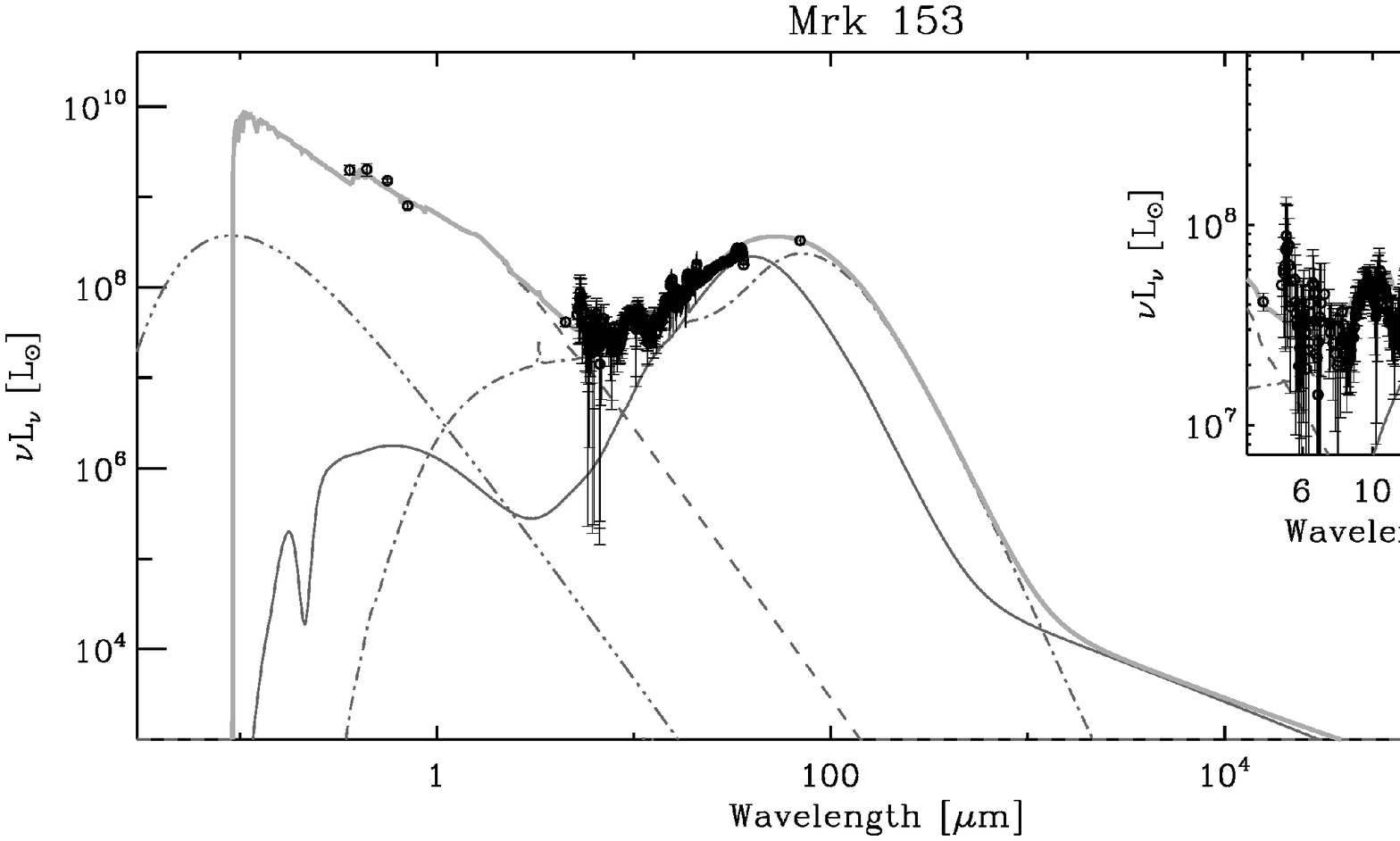}
  \caption{Fit of the galaxies' SEDs.
           See \reffig{fig:fits1} for details.}
  \label{fig:fits2}
\end{figure*}

\begin{figure*}[htbp]
  \plotone{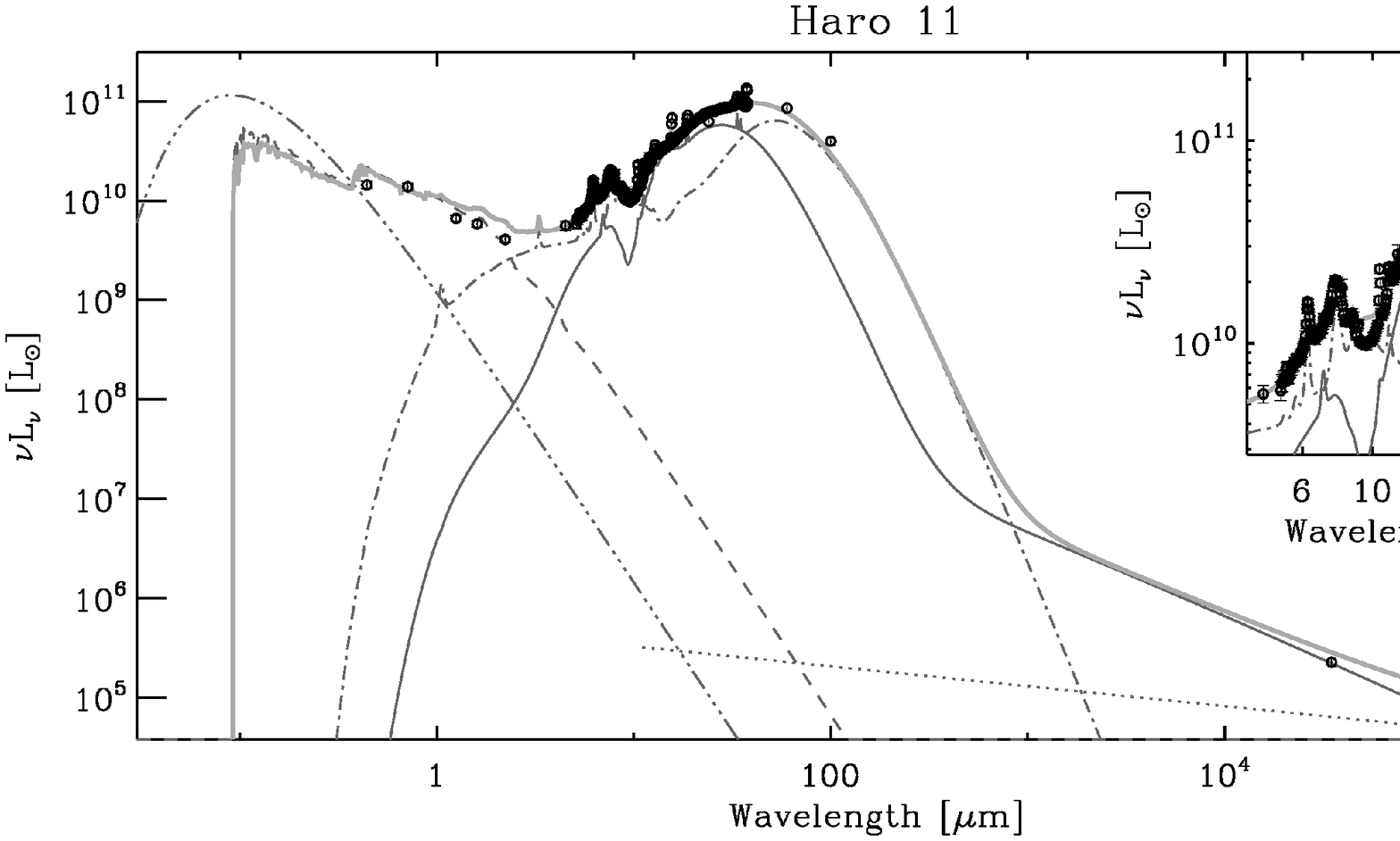}
  \plotone{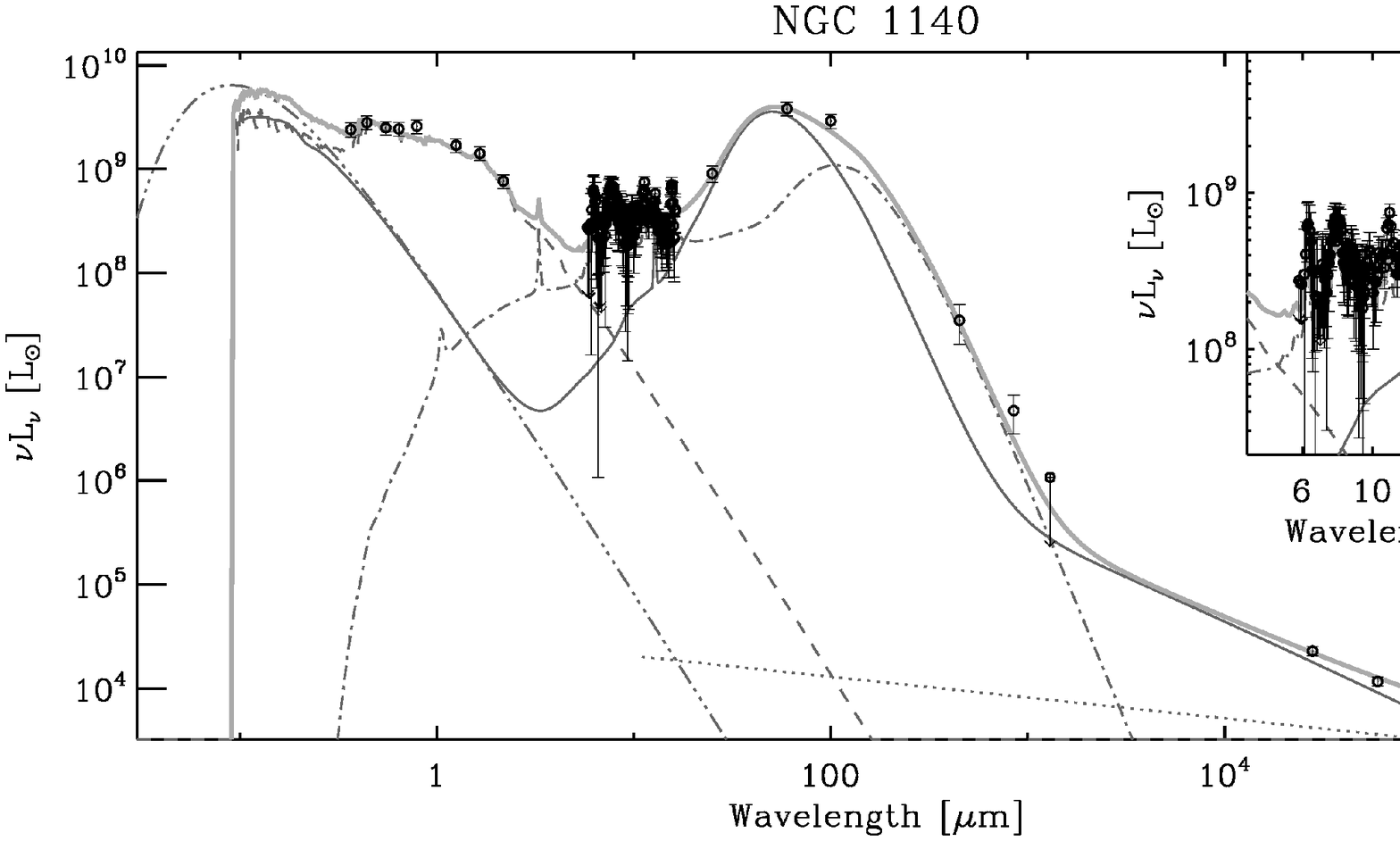}
  \caption{Fit of the galaxies' SEDs.
           See \reffig{fig:fits1} for details.}
  \label{fig:fits3}
\end{figure*}

\begin{figure*}[htbp]
  \plotone{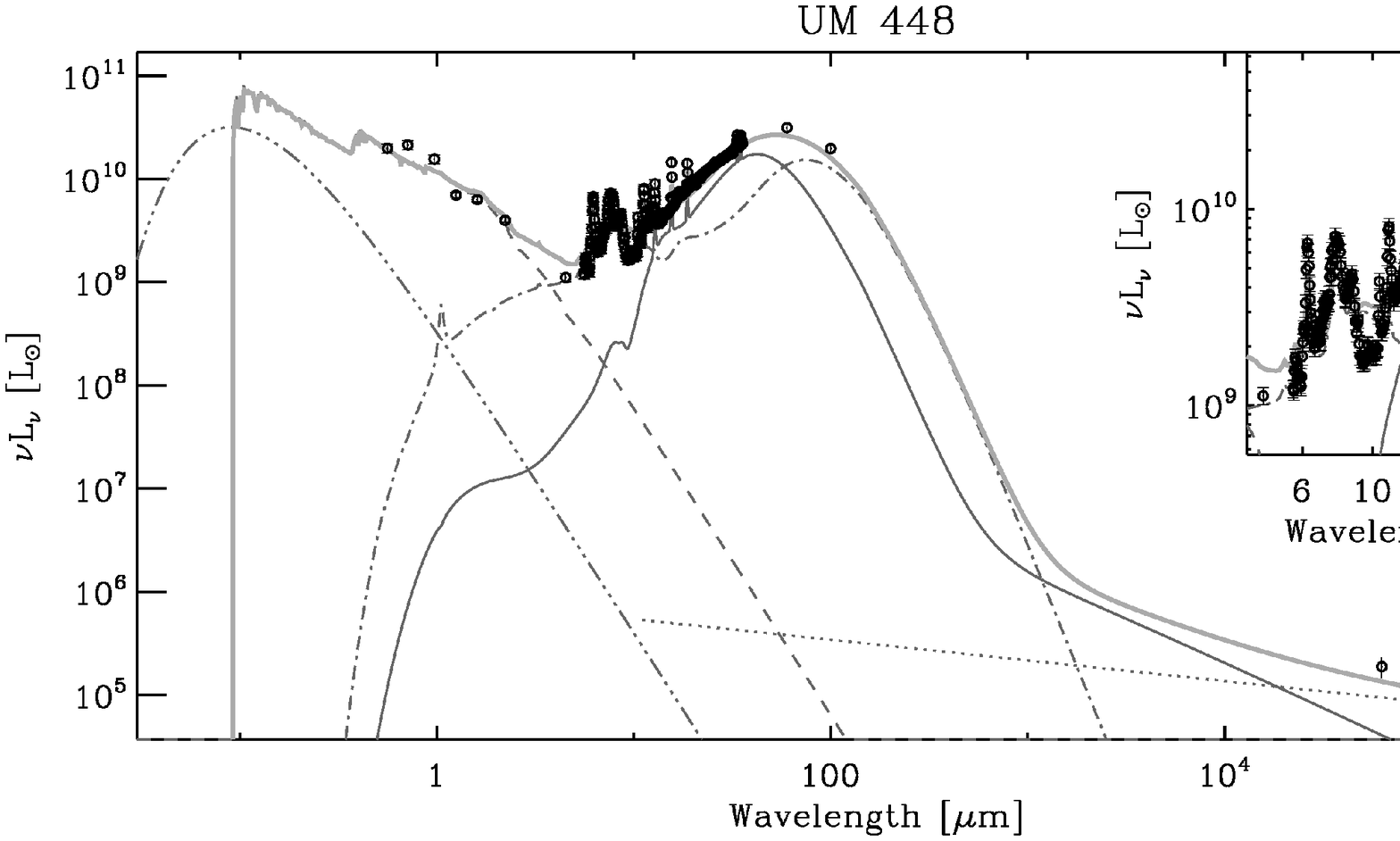}
  \plotone{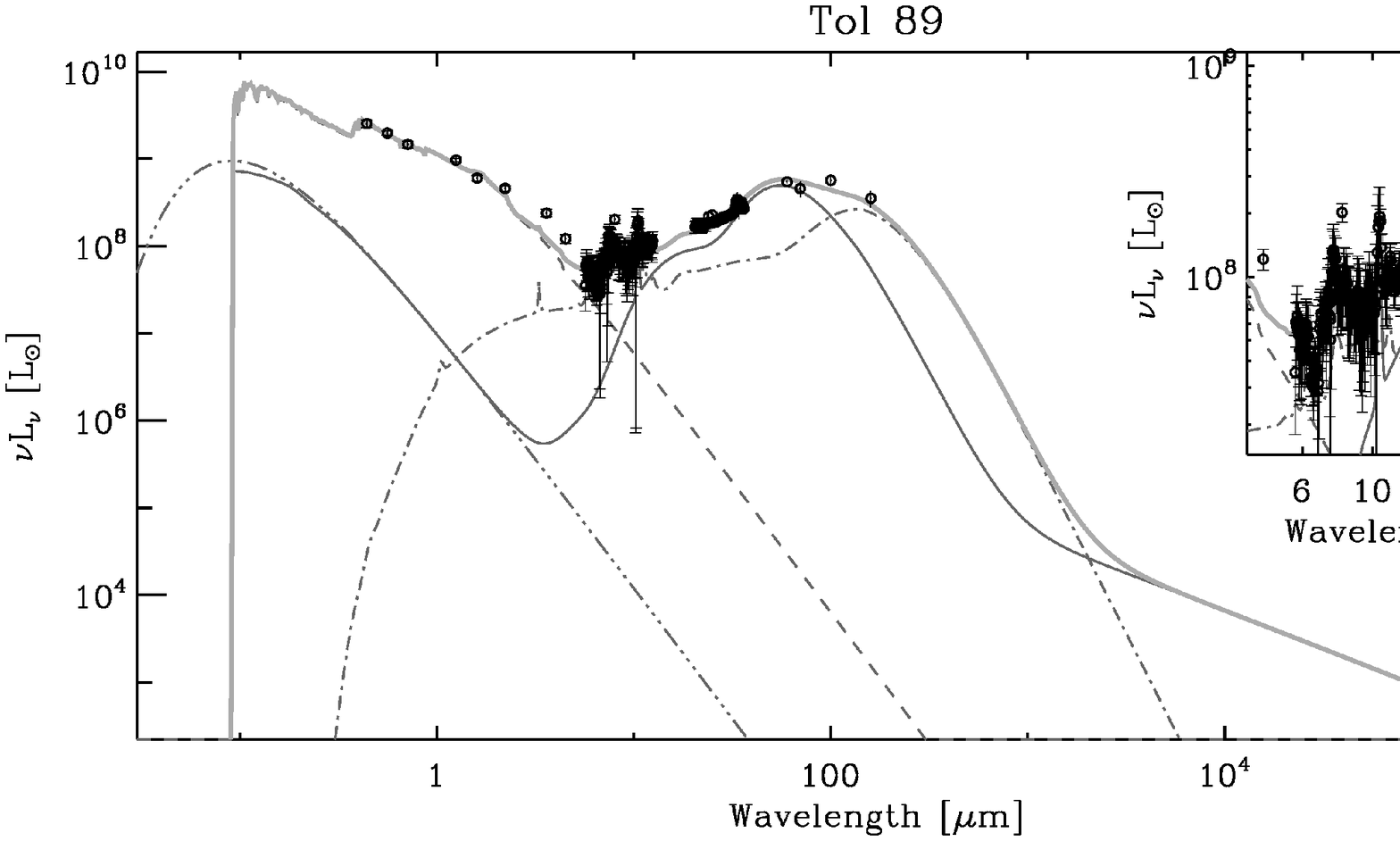}
  \caption{Fit of the galaxies' SEDs.
           See \reffig{fig:fits1} for details.}
  \label{fig:fits4}
\end{figure*}

\begin{figure*}[htbp]
  \plotone{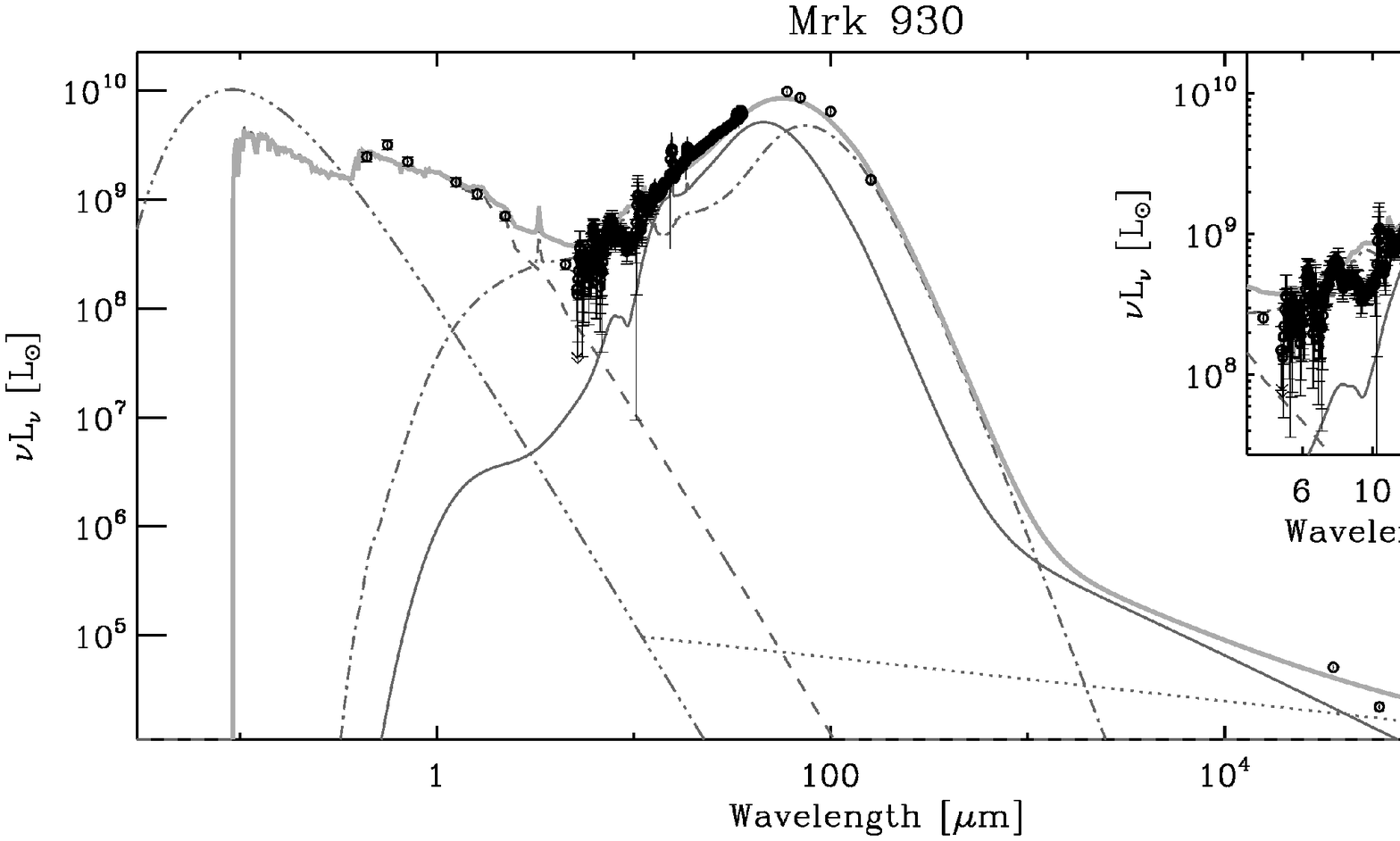}
  \plotone{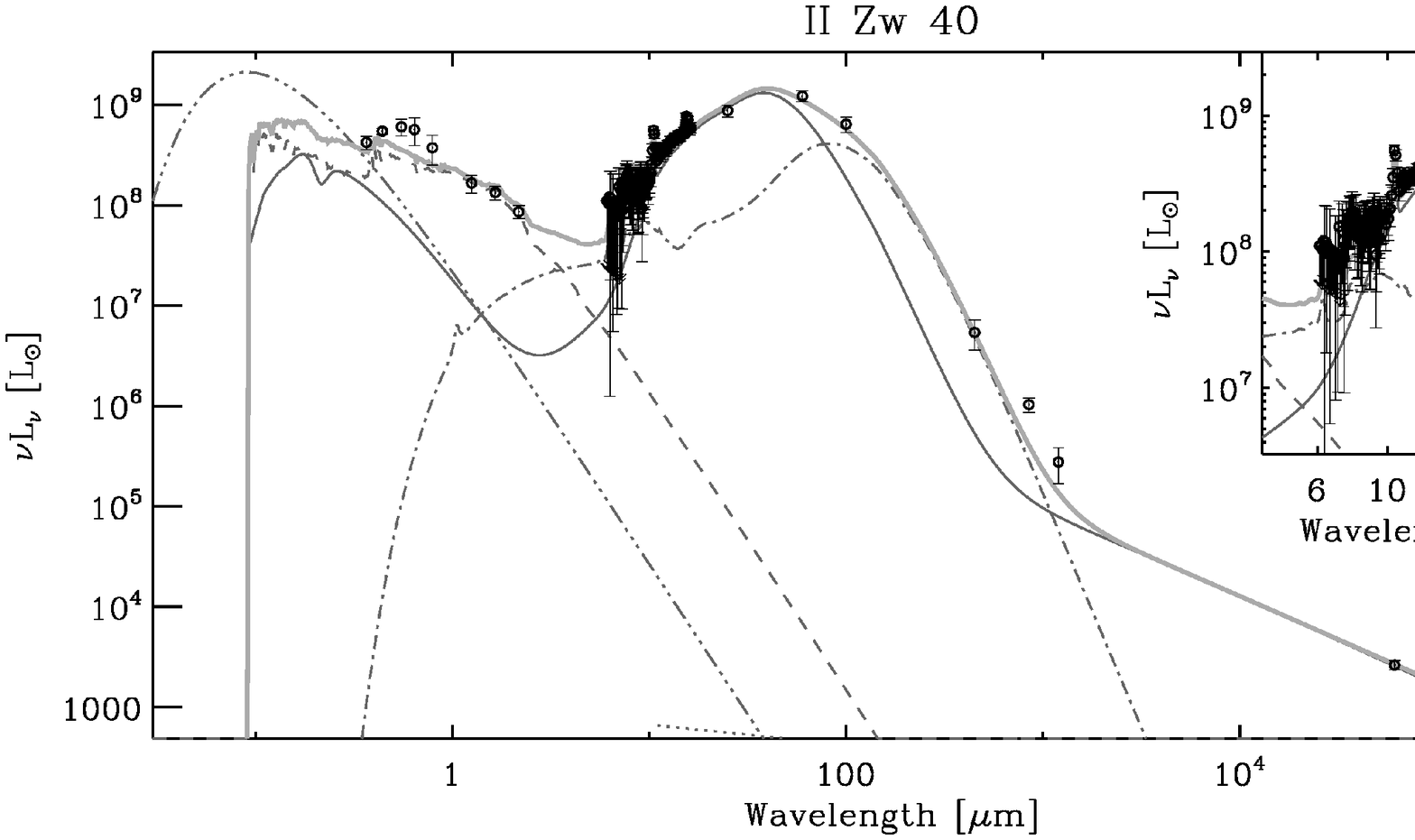}
  \caption{Fit of the galaxies' SEDs.
           See \reffig{fig:fits1} for details.}
  \label{fig:fits5}
\end{figure*}

\begin{figure*}[htbp]
  \plotone{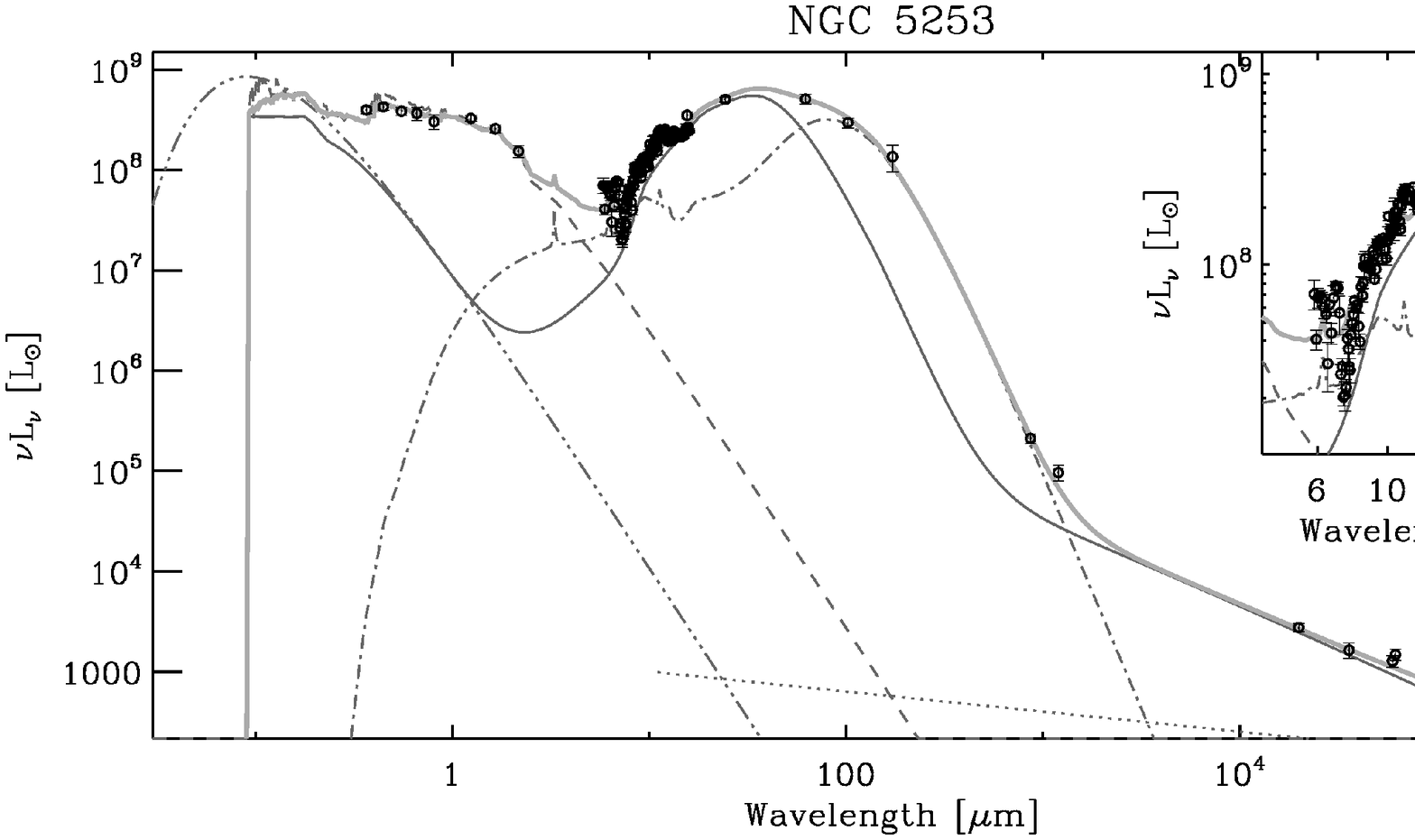}
  \plotone{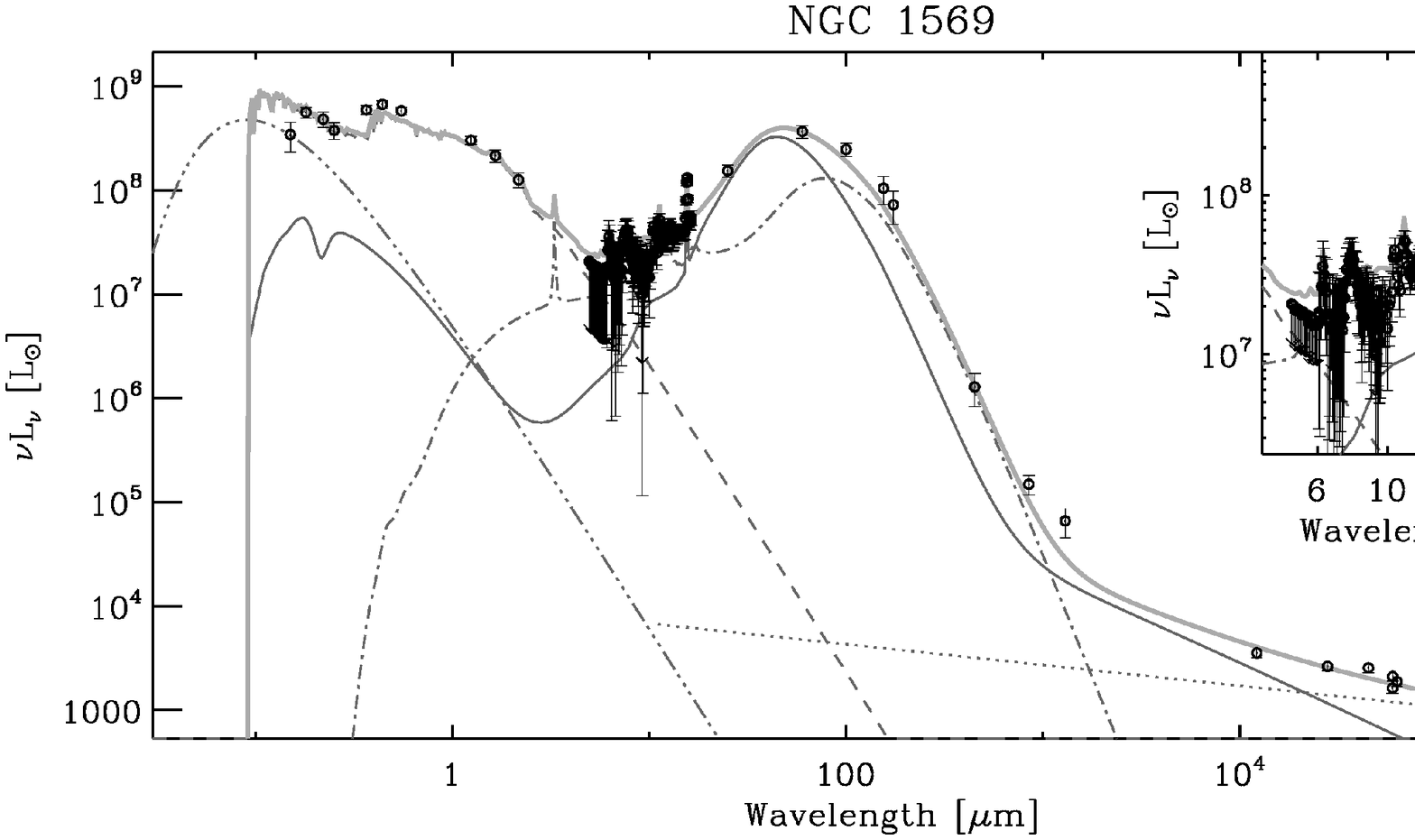}
  \caption{Fit of the galaxies' SEDs.
           See \reffig{fig:fits1} for details.}
  \label{fig:fits6}
\end{figure*}

\begin{figure*}[htbp]
  \plotone{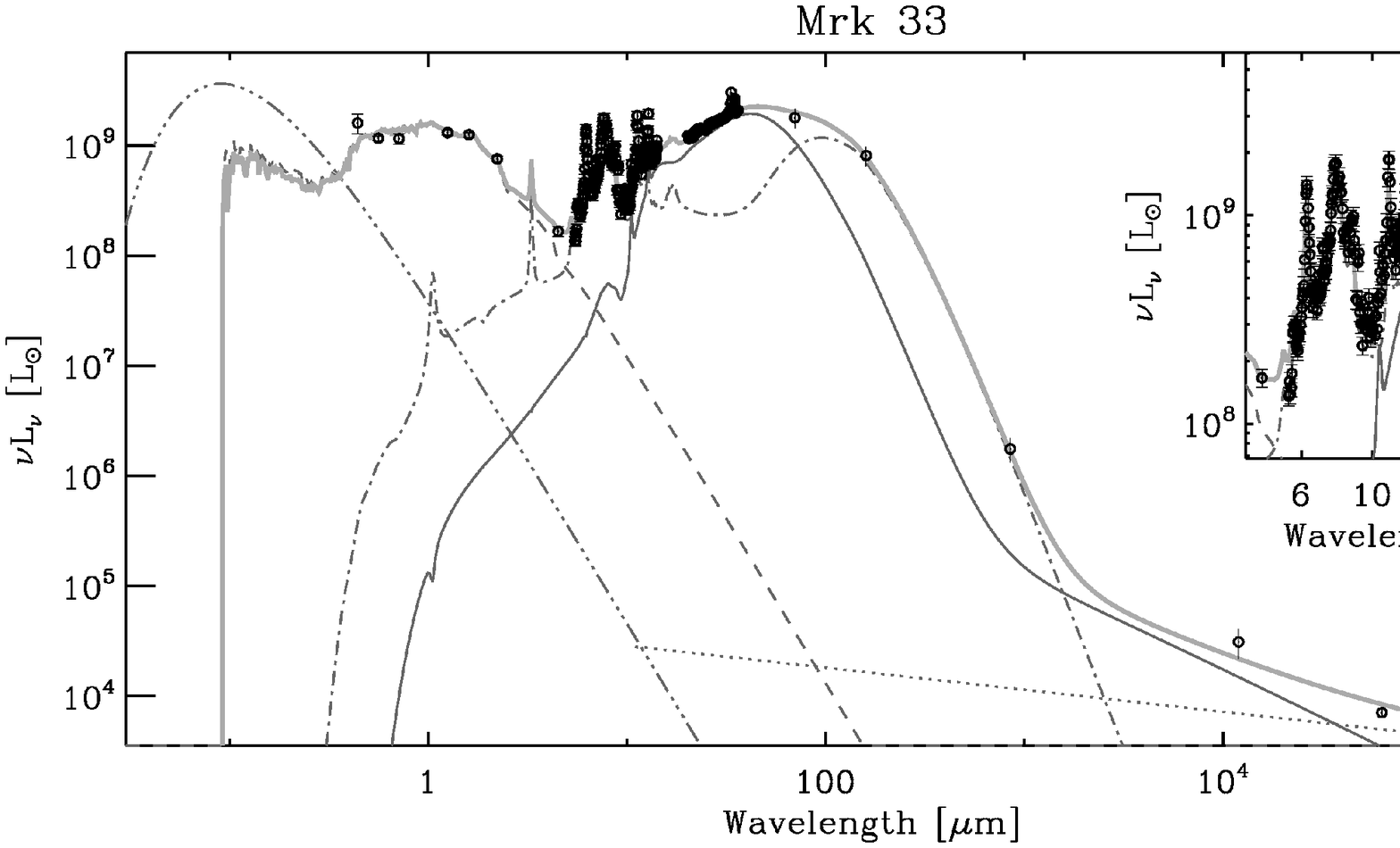}
  \plotone{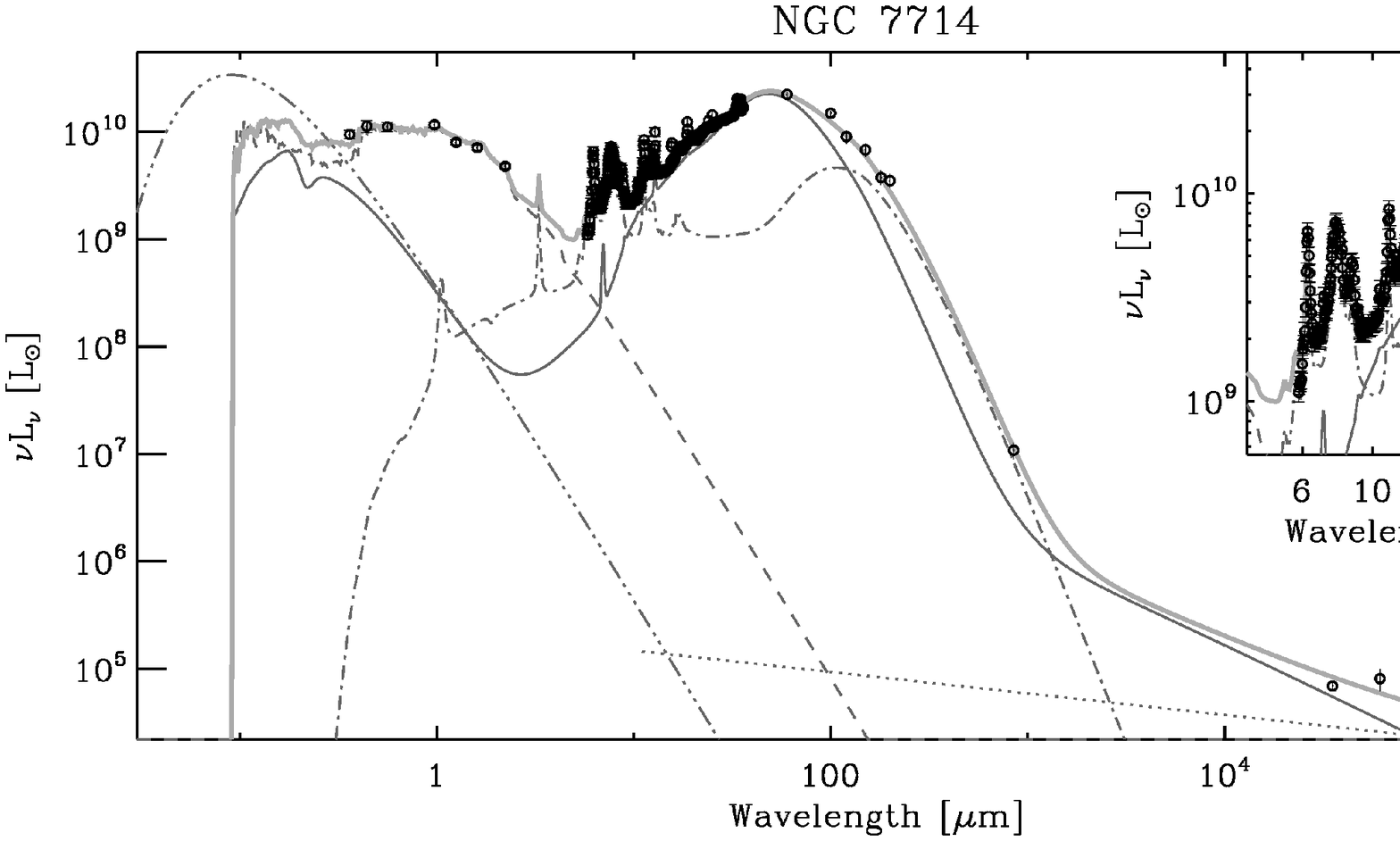}
  \caption{Fit of the galaxies' SEDs.
           See \reffig{fig:fits1} for details.}
  \label{fig:fits7}
\end{figure*}

\begin{figure*}[htbp]
  \plotone{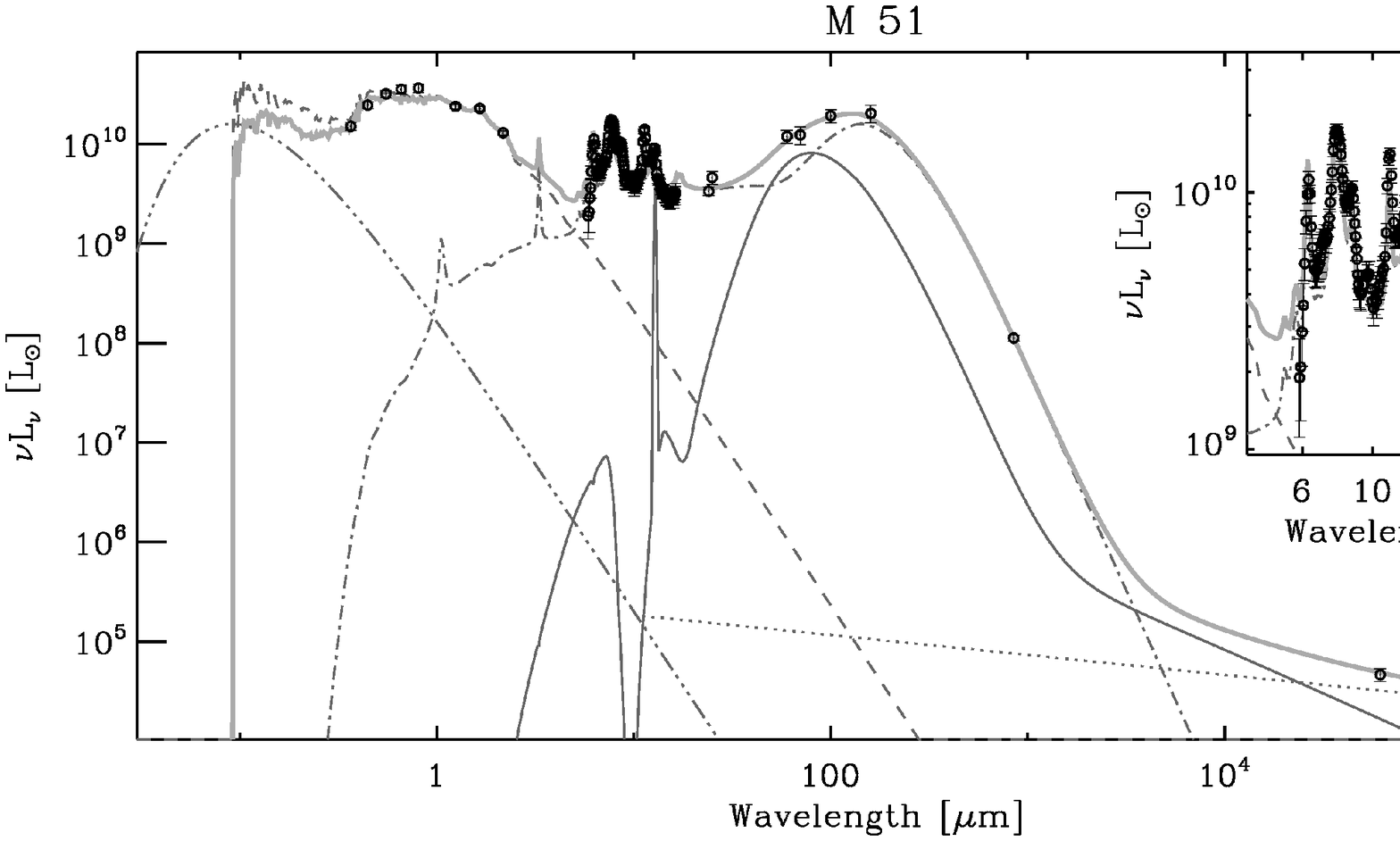}
  \plotone{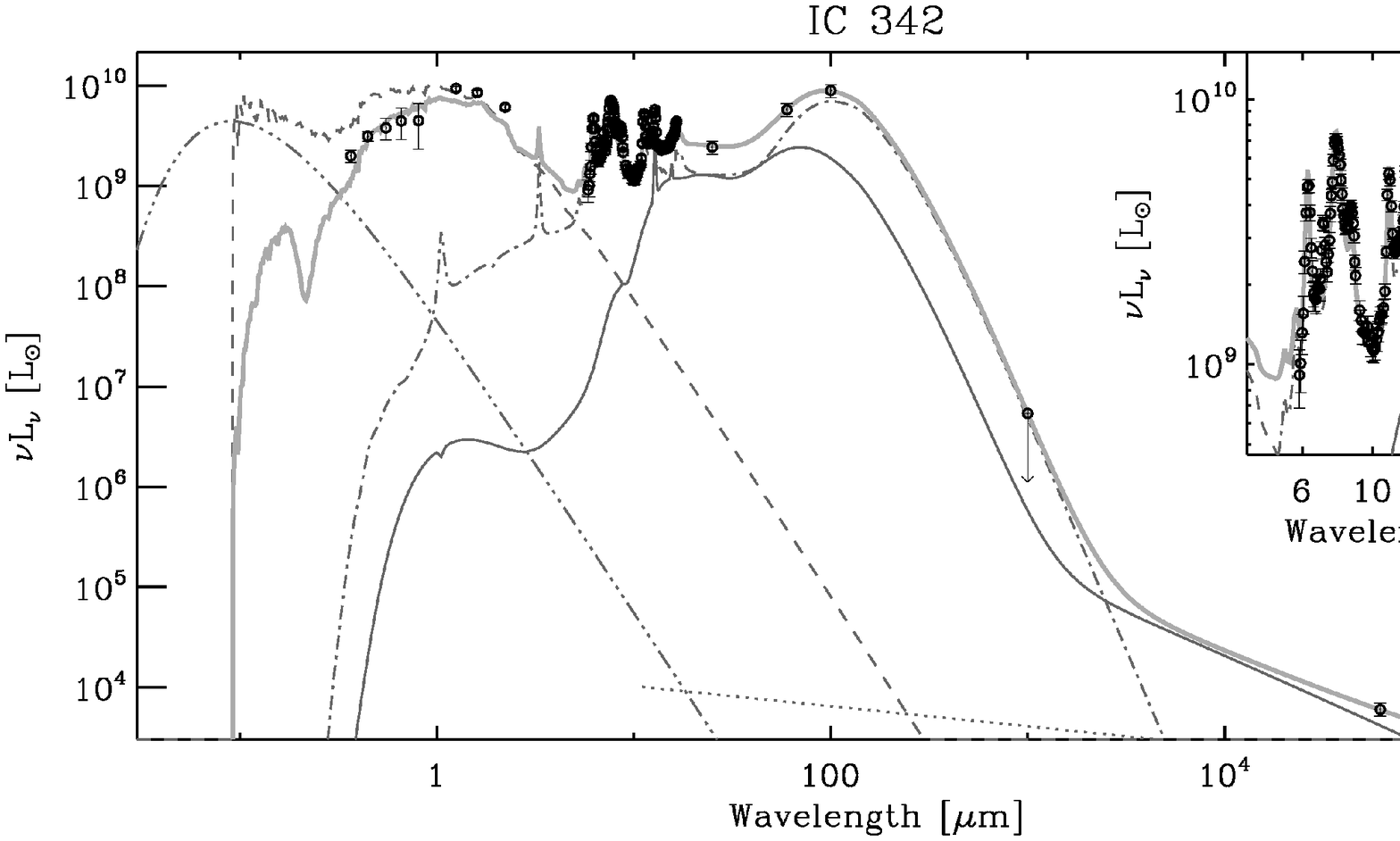}
  \caption{Fit of the galaxies' SEDs.
           See \reffig{fig:fits1} for details.}
  \label{fig:fits8}
\end{figure*}

\begin{figure*}[htbp]
  \plotone{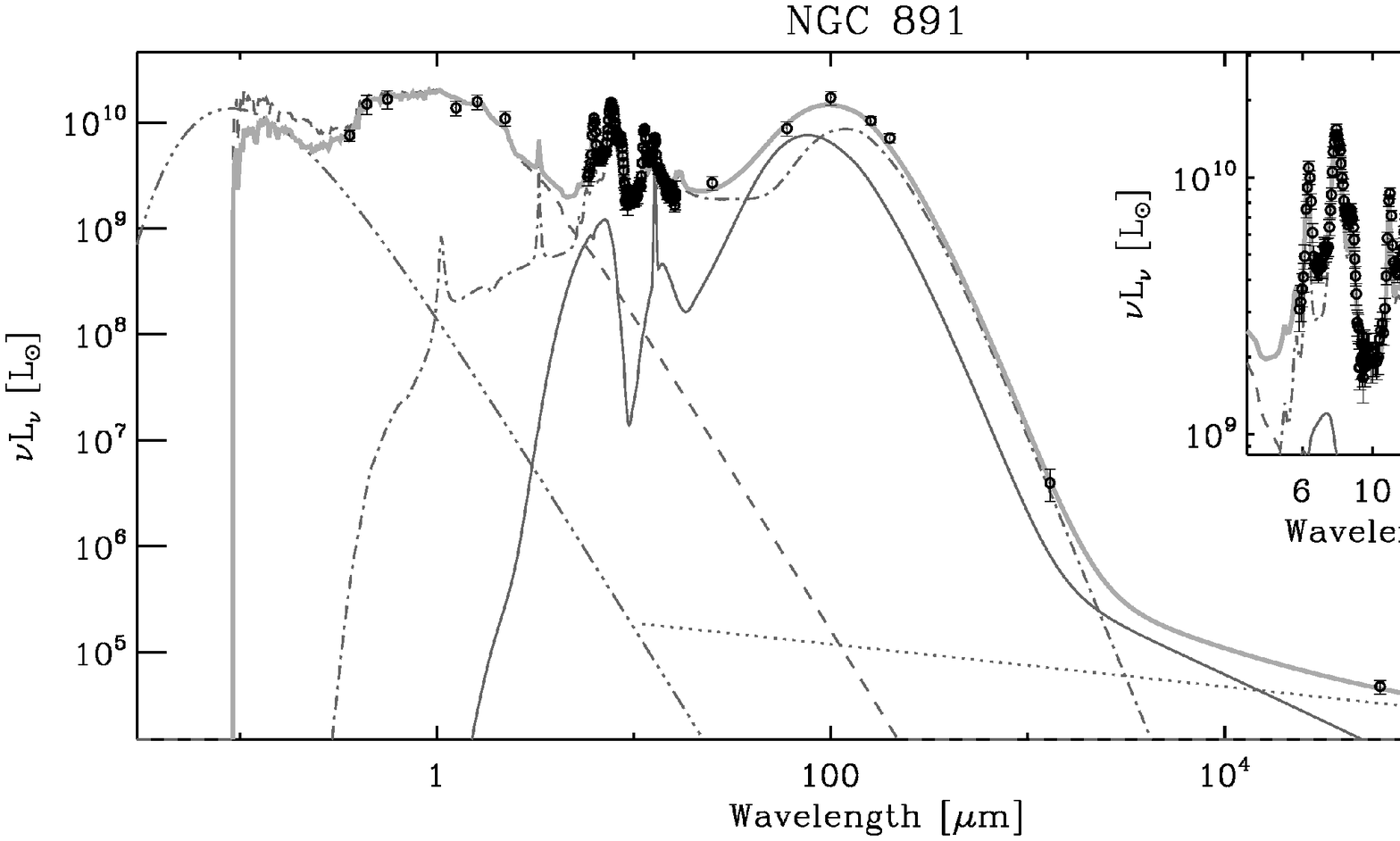}
  \plotone{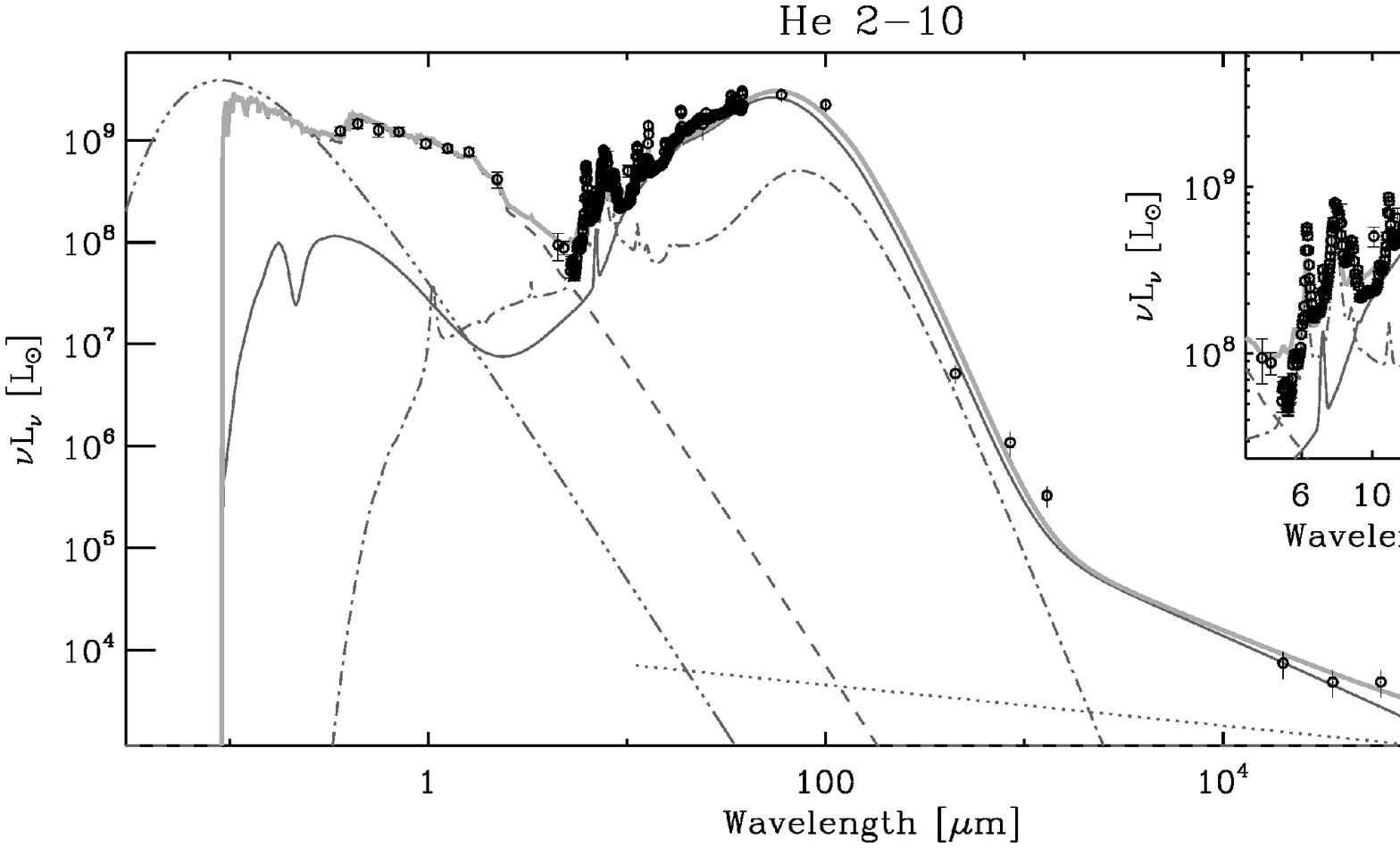}
  \caption{Fit of the galaxies' SEDs.
           See \reffig{fig:fits1} for details.}
  \label{fig:fits9}
\end{figure*}

\begin{figure*}[htbp]
  \plotone{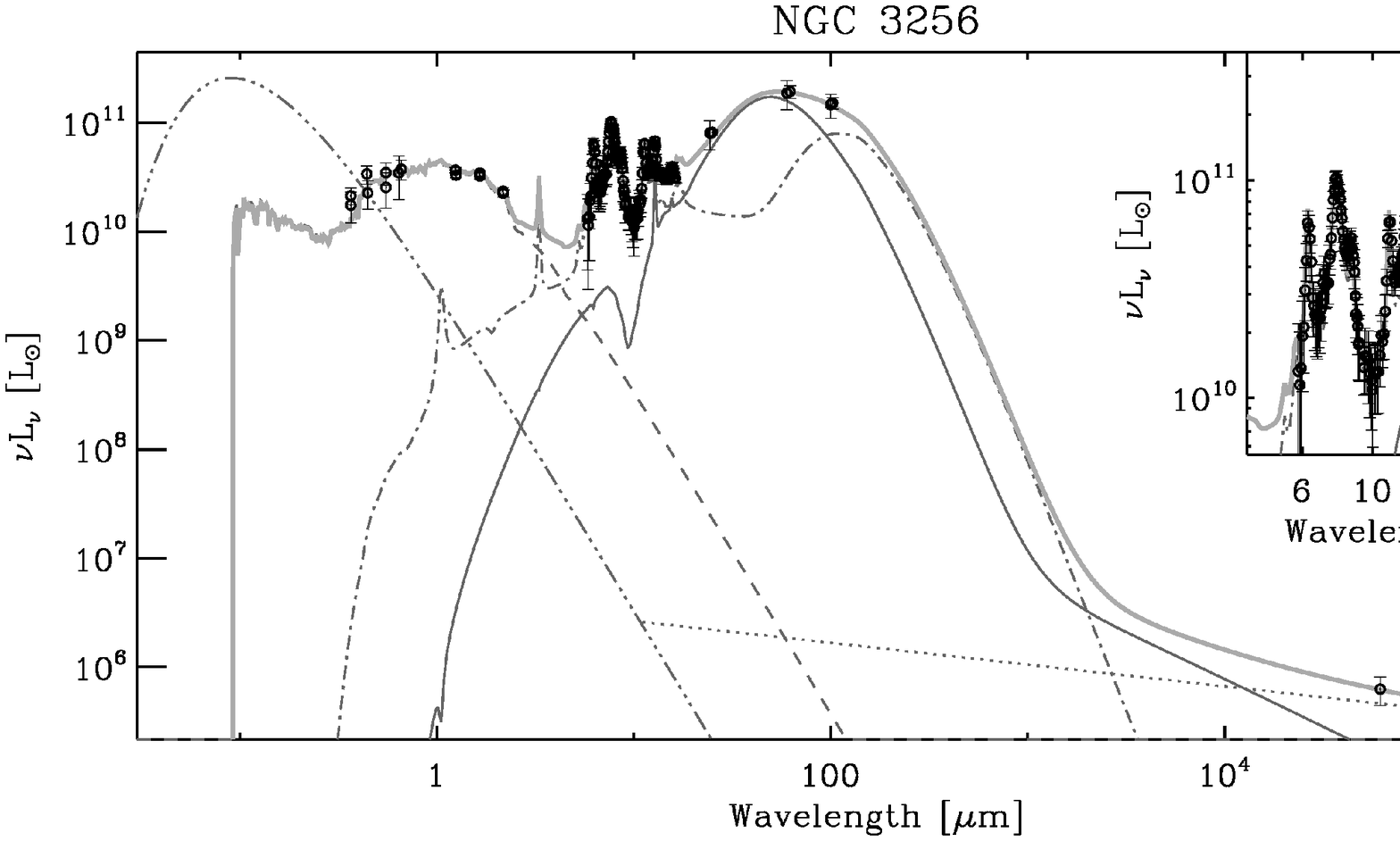}
  \plotone{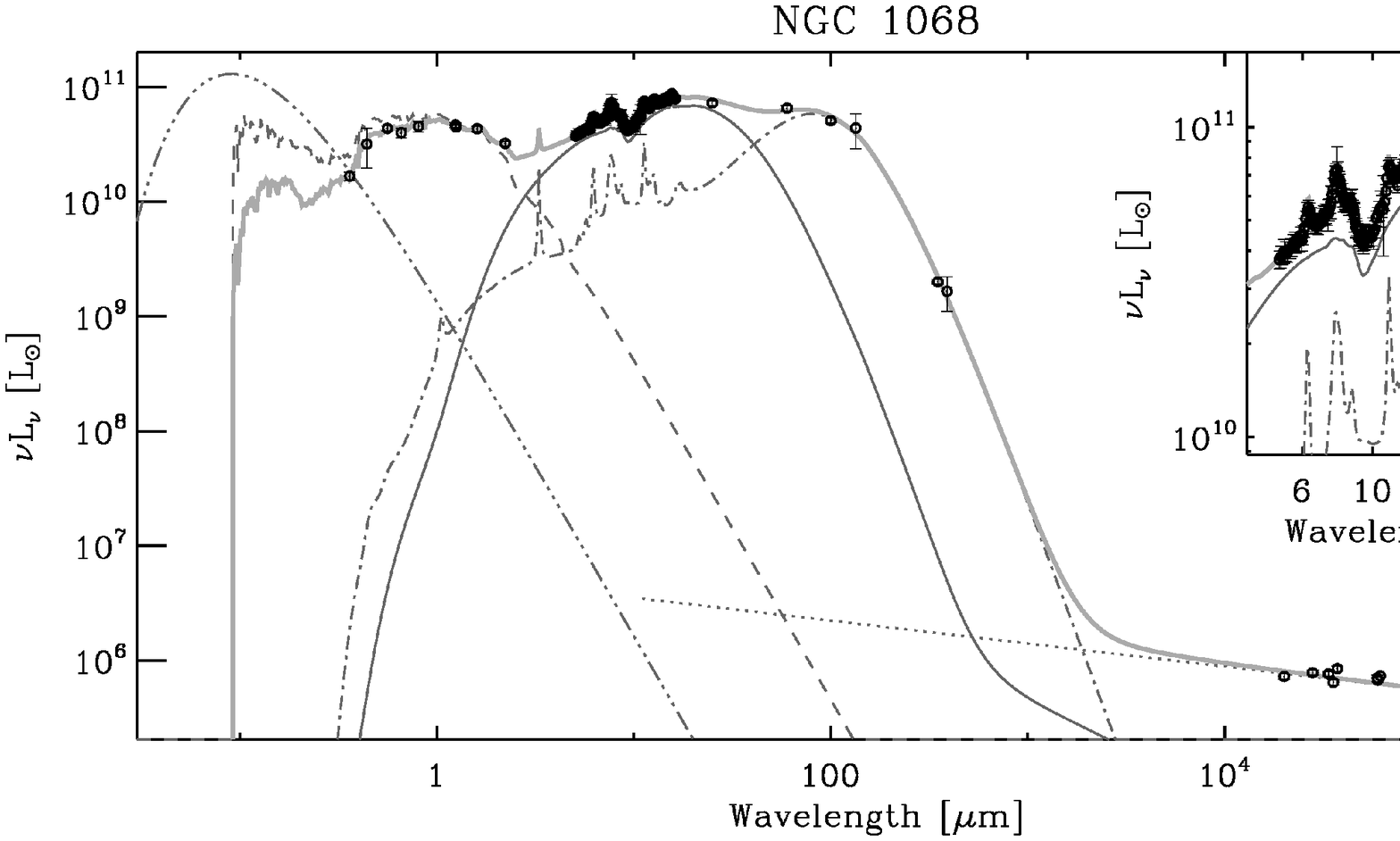}
  \caption{Fit of the galaxies' SEDs.
           See \reffig{fig:fits1} for details.}
  \label{fig:fits10}
\end{figure*}

\begin{figure*}[htbp]
  \plotone{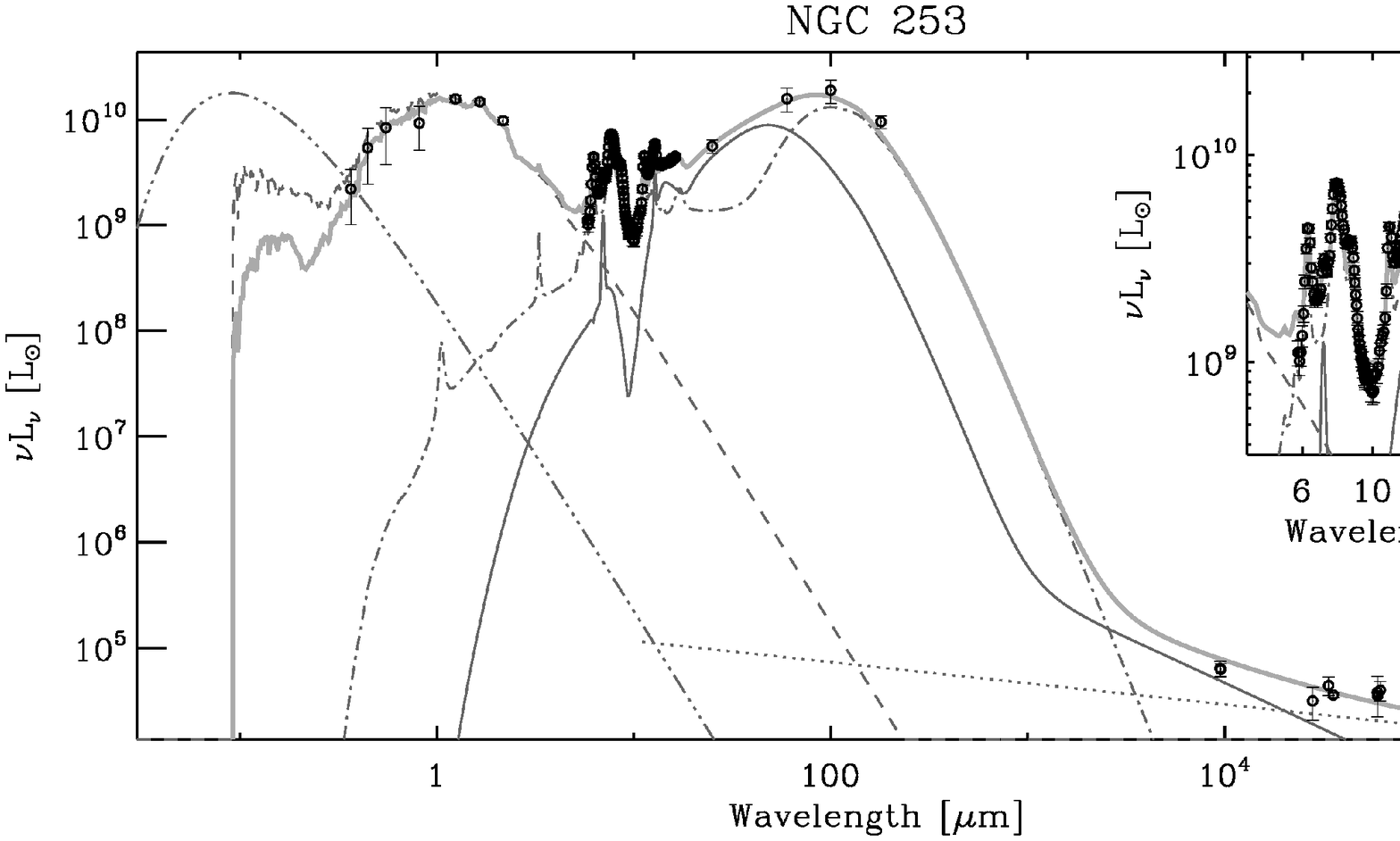}
  \plotone{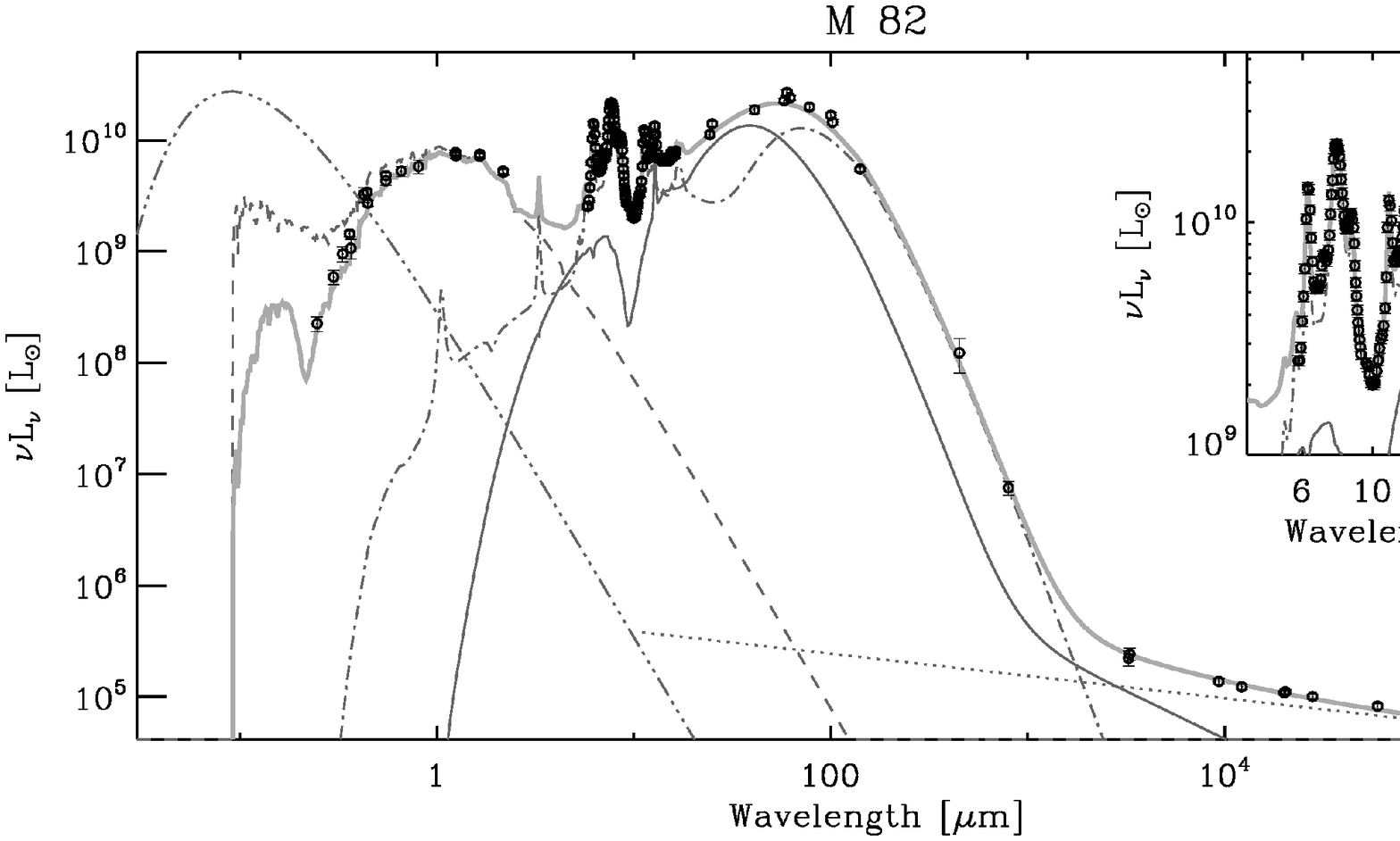}
  \caption{Fit of the galaxies' SEDs.
           See \reffig{fig:fits1} for details.}
  \label{fig:fits11}
\end{figure*}

\begin{figure*}[htbp]
  \plotone{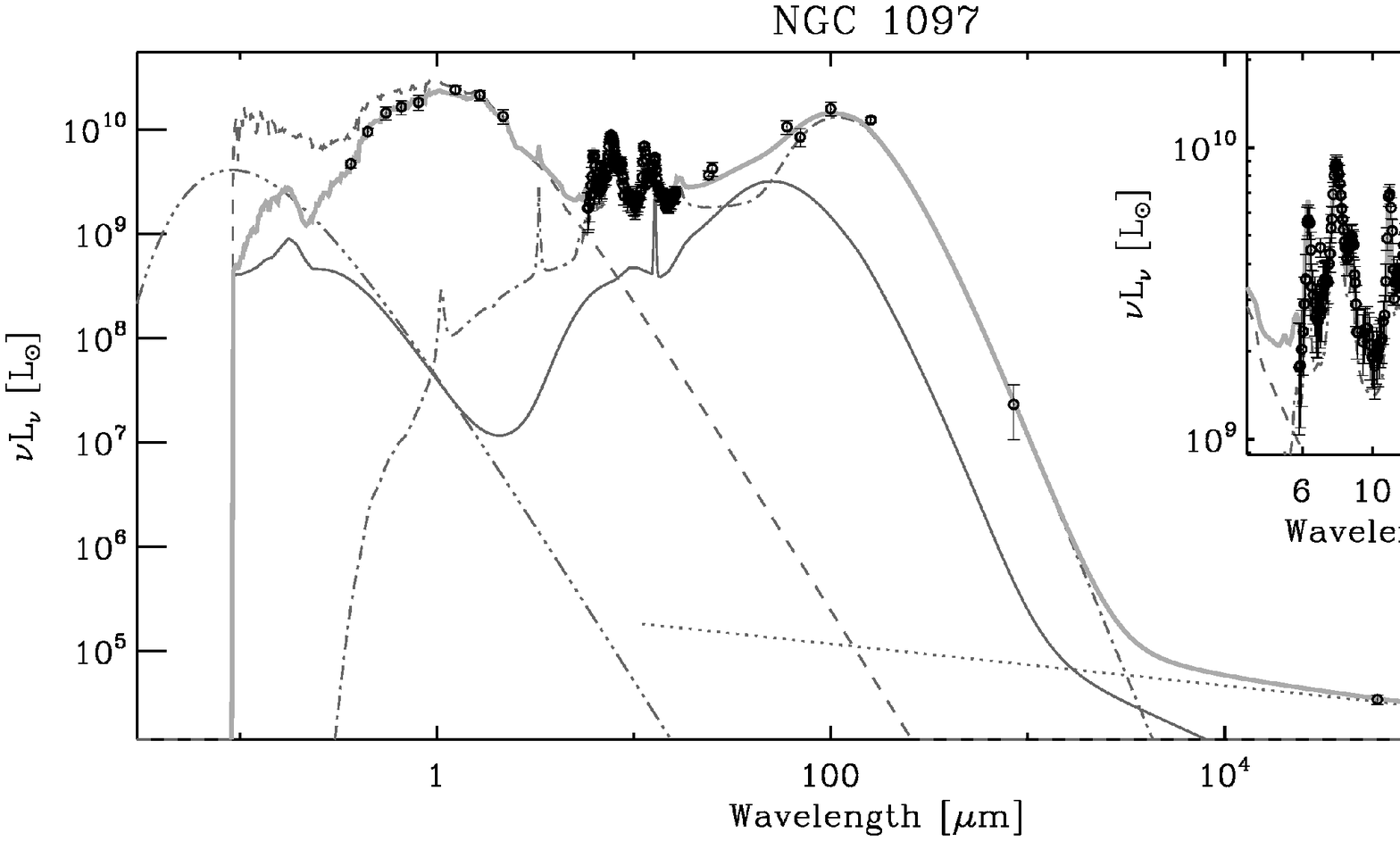}
  \plotone{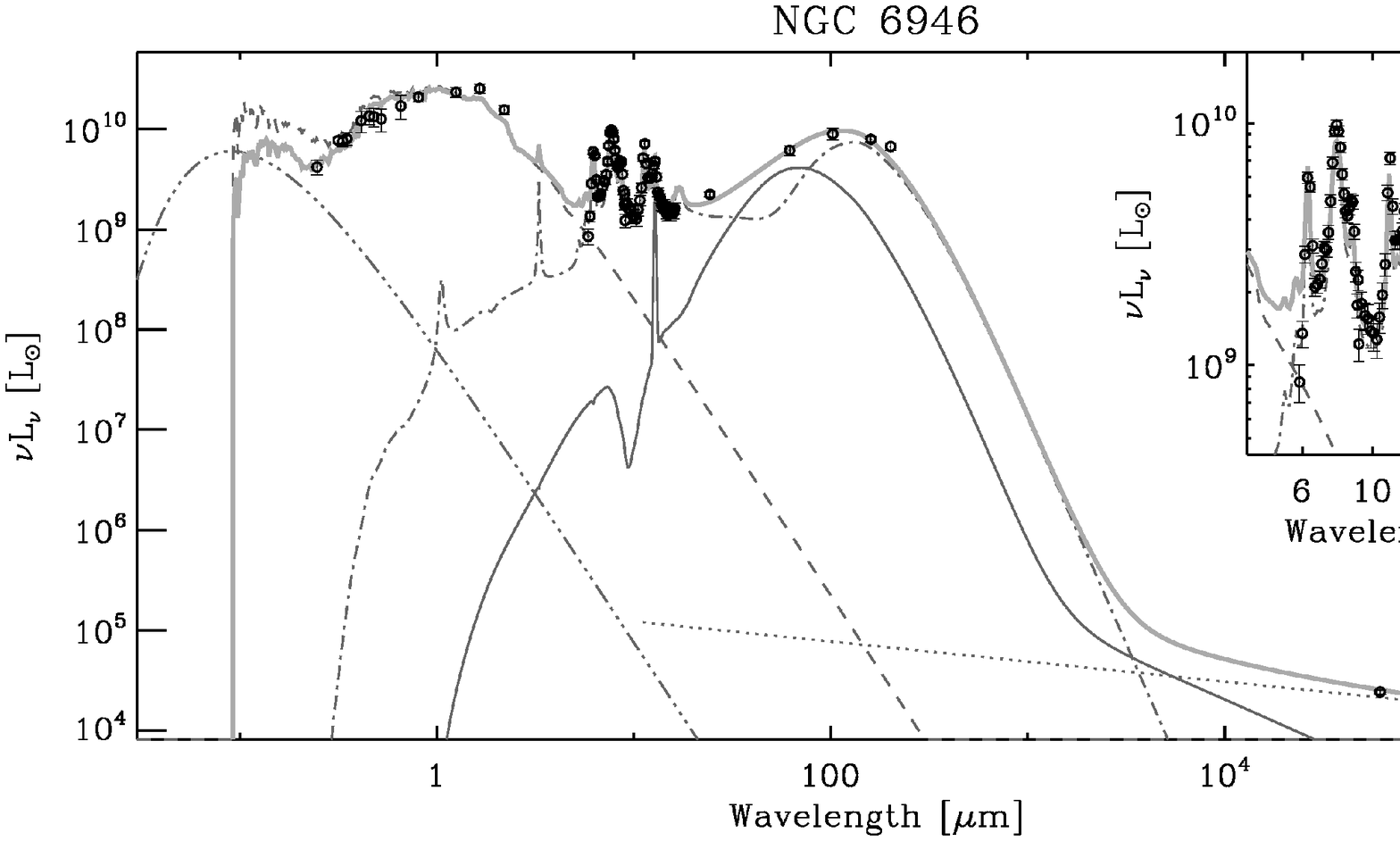}
  \caption{Fit of the galaxies' SEDs.
           See \reffig{fig:fits1} for details.}
  \label{fig:fits12}
\end{figure*}

\begin{figure*}[htbp]
  \plotone{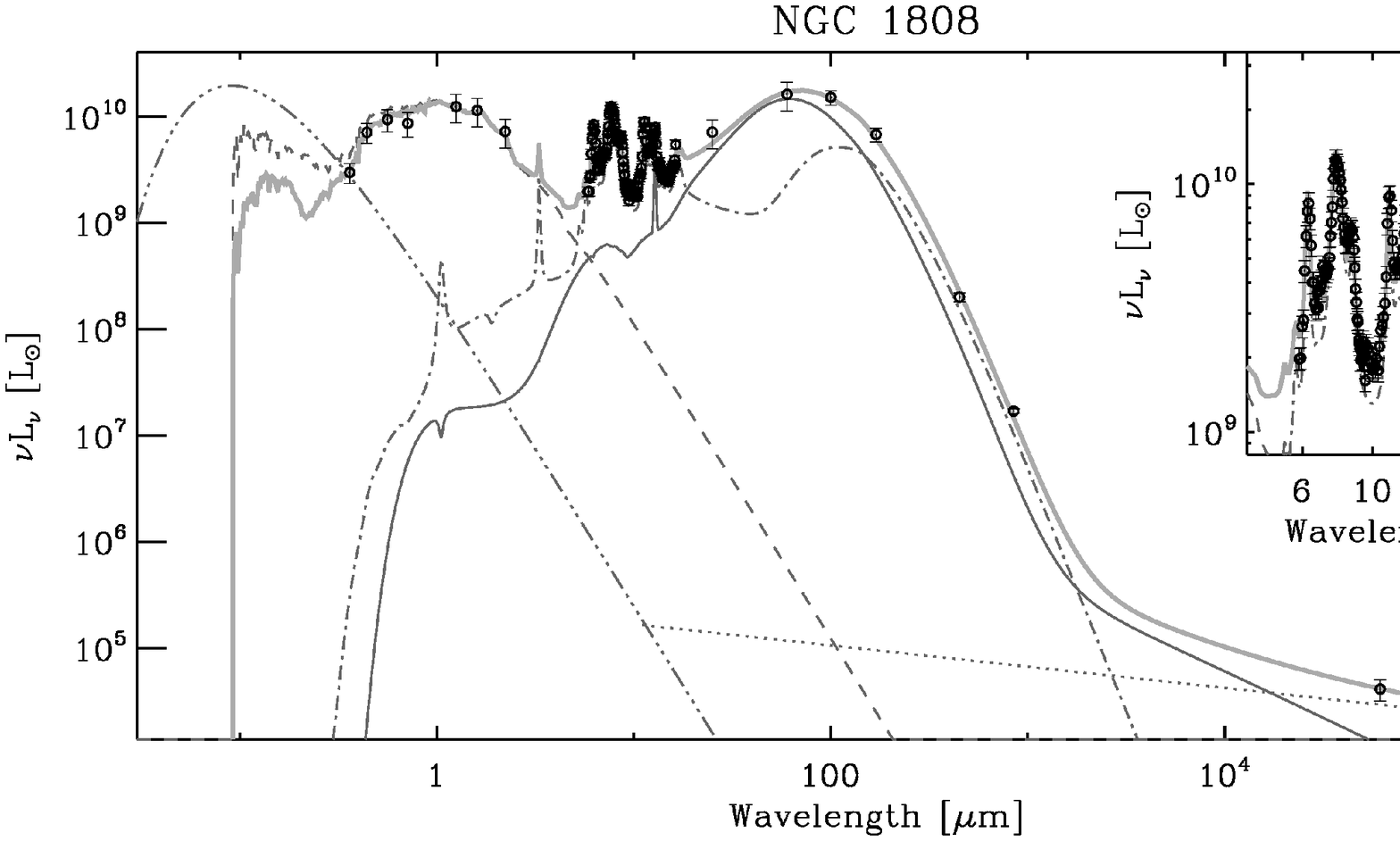}
  \plotone{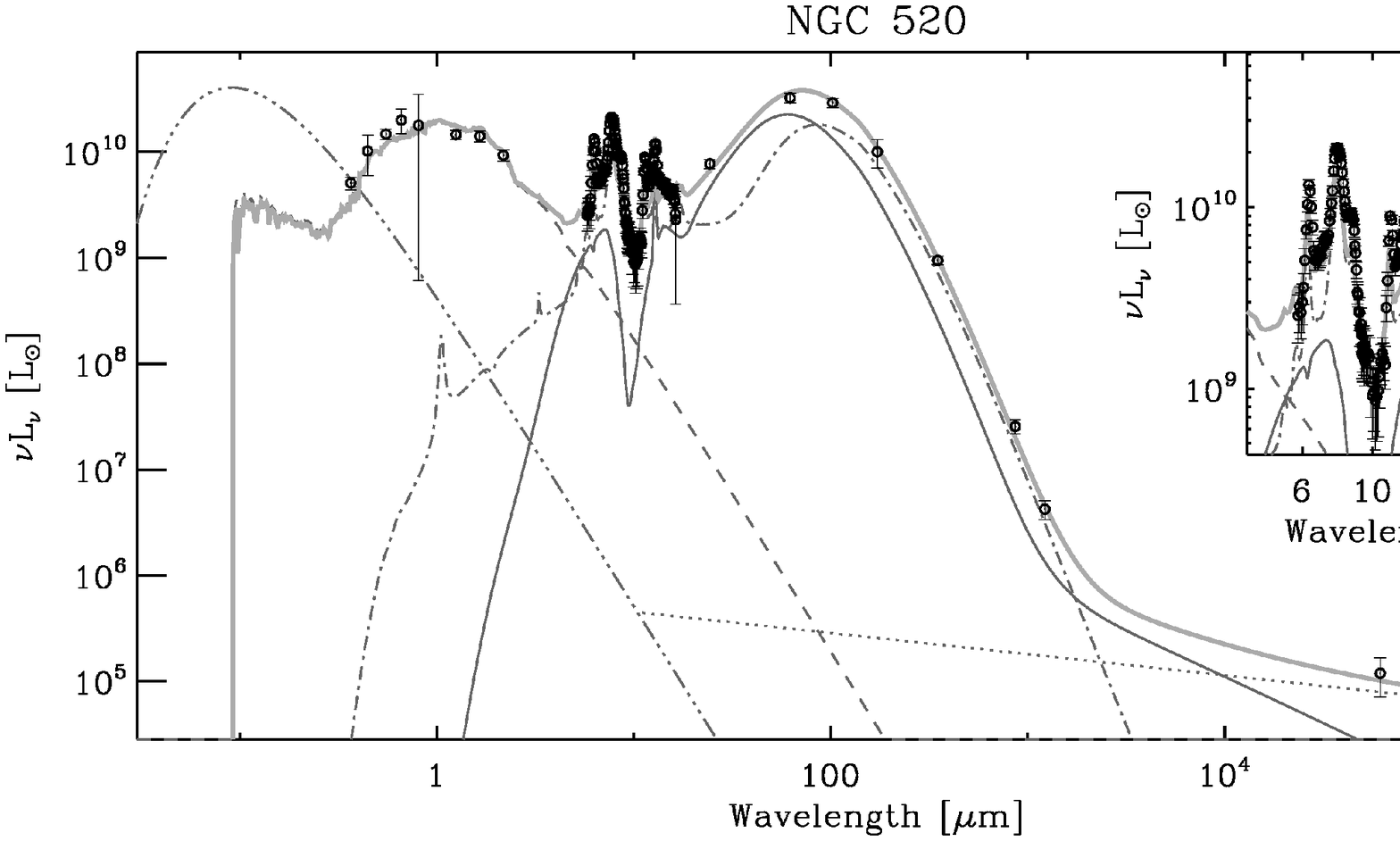}
  \caption{Fit of the galaxies' SEDs.
           See \reffig{fig:fits1} for details.}
  \label{fig:fits13}
\end{figure*}

\begin{figure*}[htbp]
  \plotone{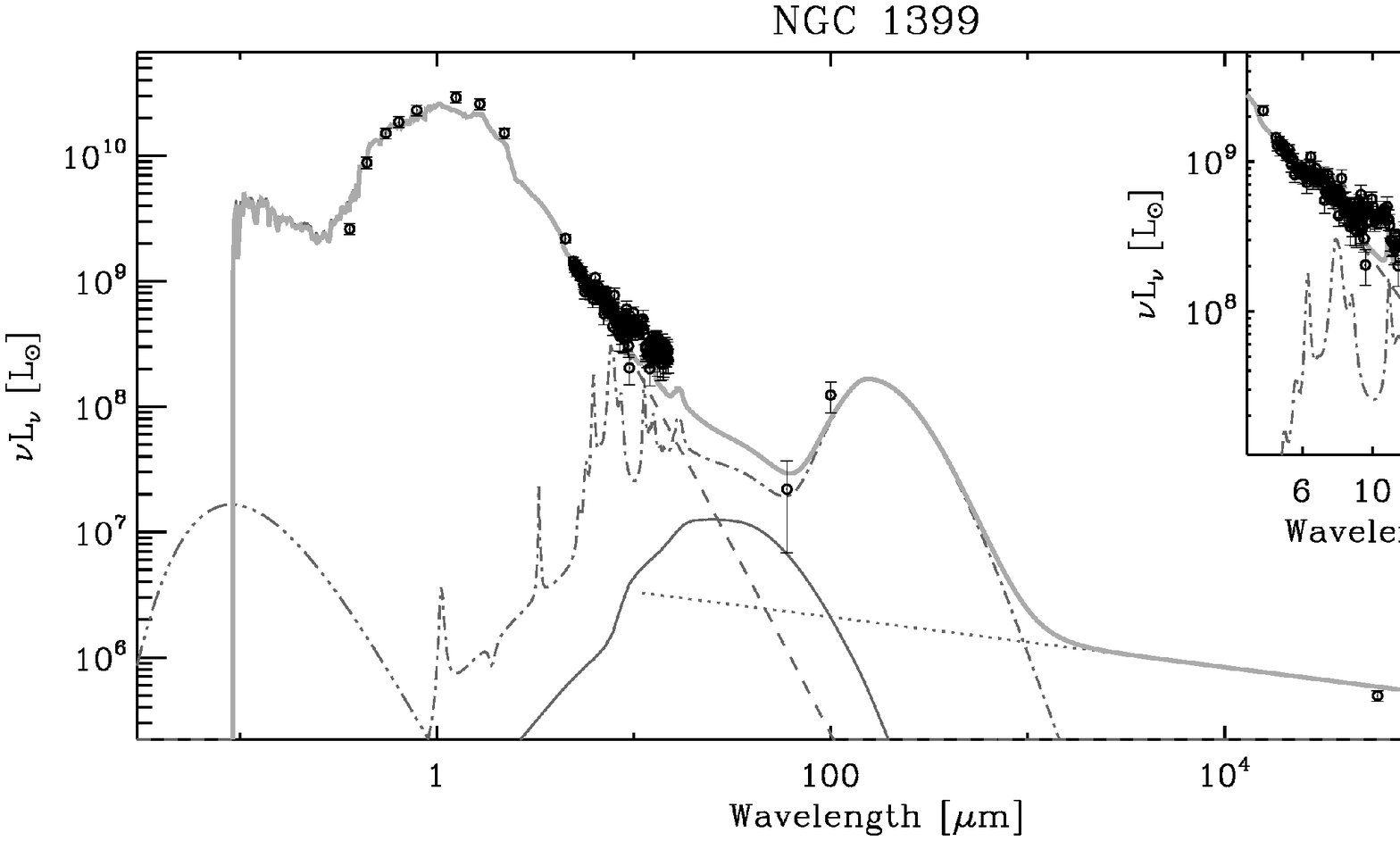}
  \plotone{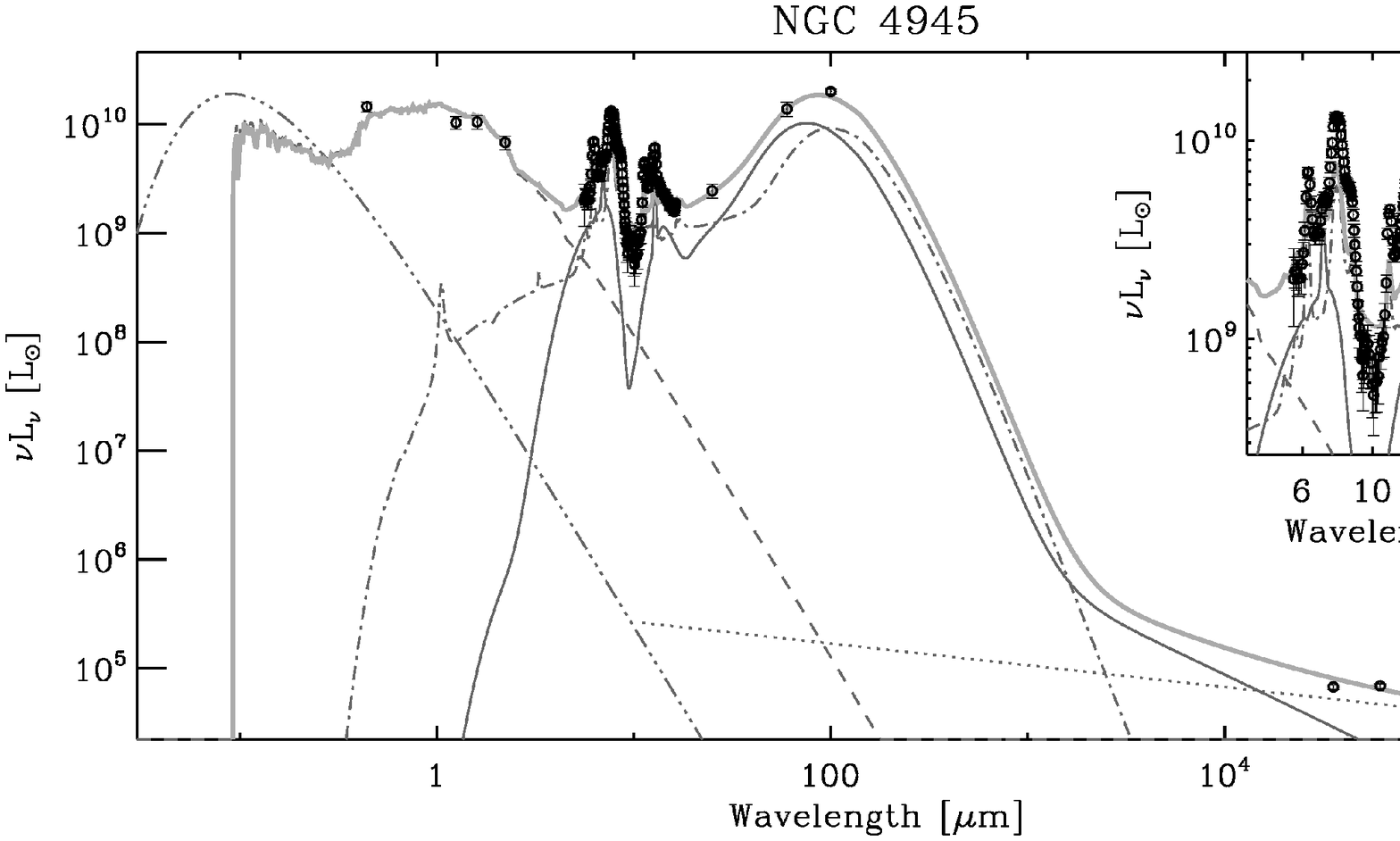}
  \caption{Fit of the galaxies' SEDs.
           See \reffig{fig:fits1} for details.}
  \label{fig:fits14}
\end{figure*}

\begin{figure*}[htbp]
  \plotone{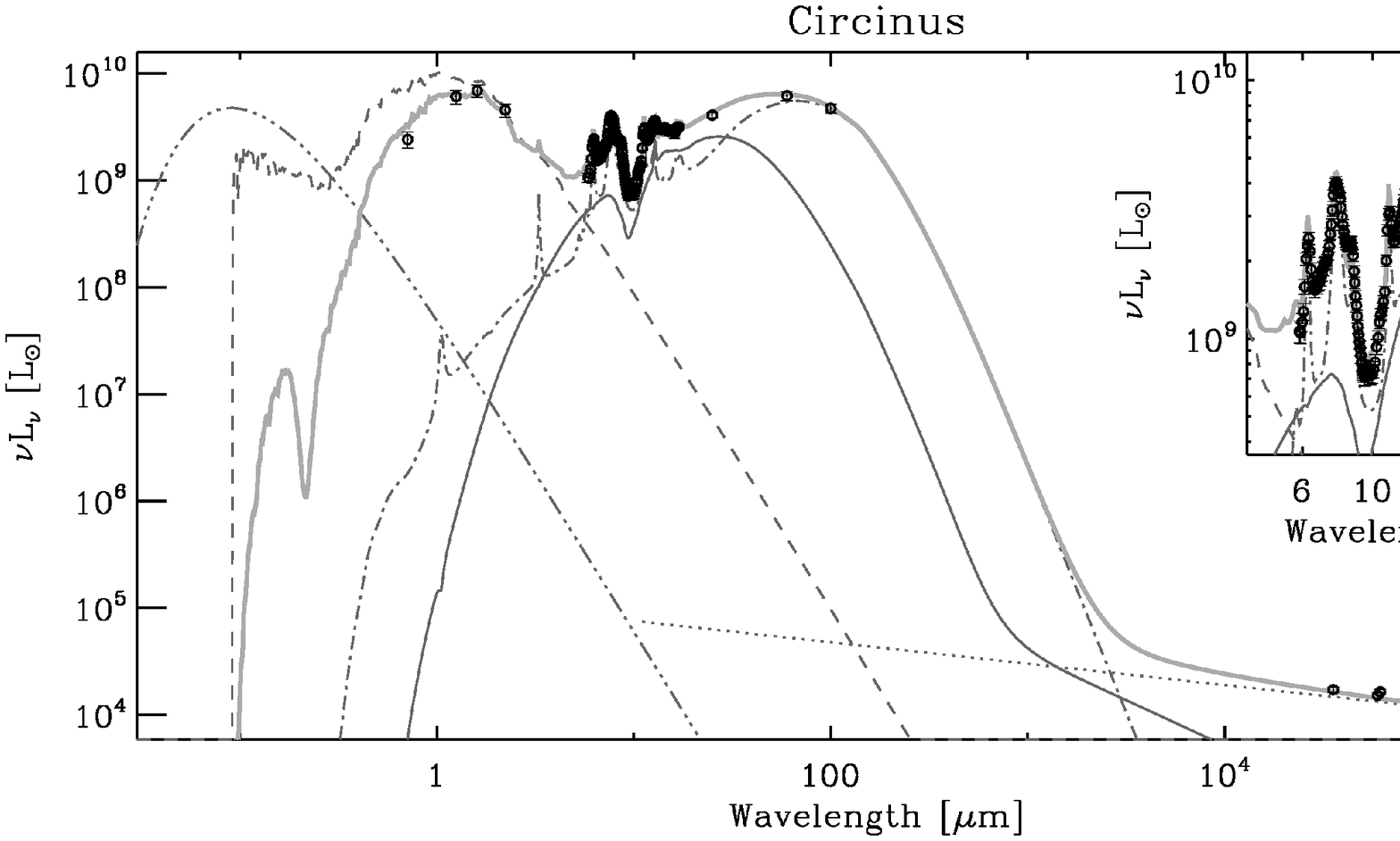}
  \plotone{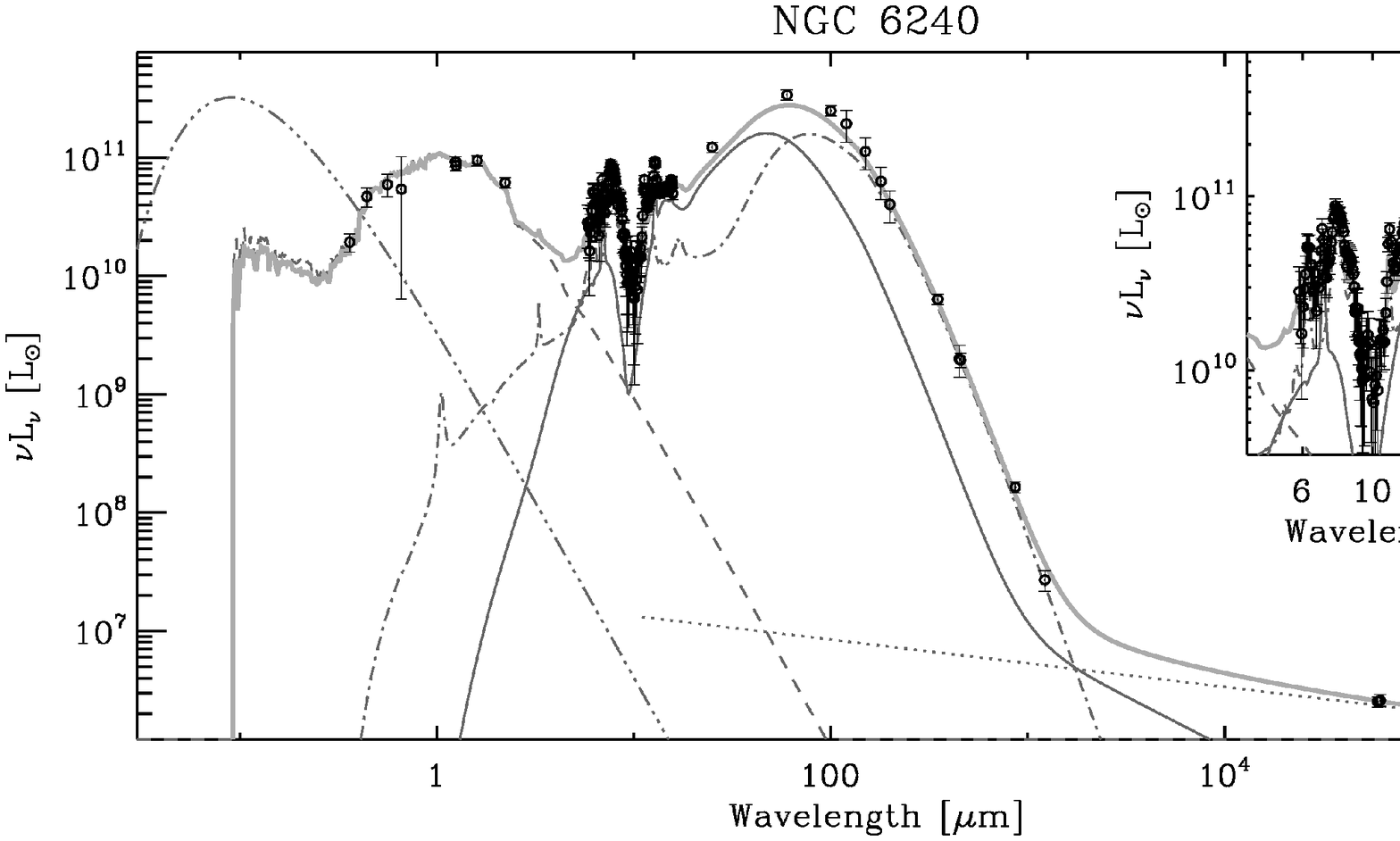}
  \caption{Fit of the galaxies' SEDs.
           See \reffig{fig:fits1} for details.}
  \label{fig:fits15}
\end{figure*}

\begin{figure*}[htbp]
  \plotone{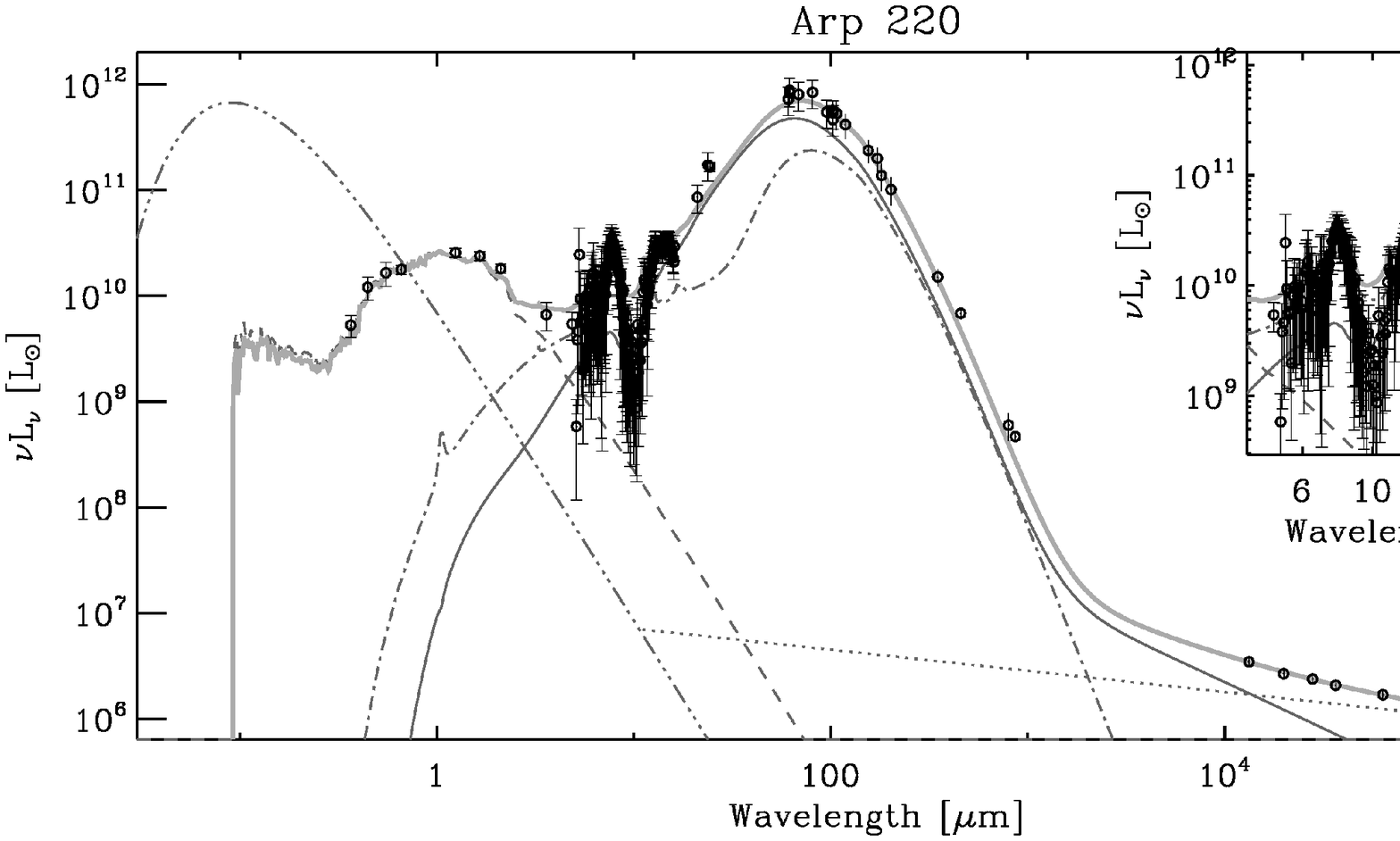}
  \plotone{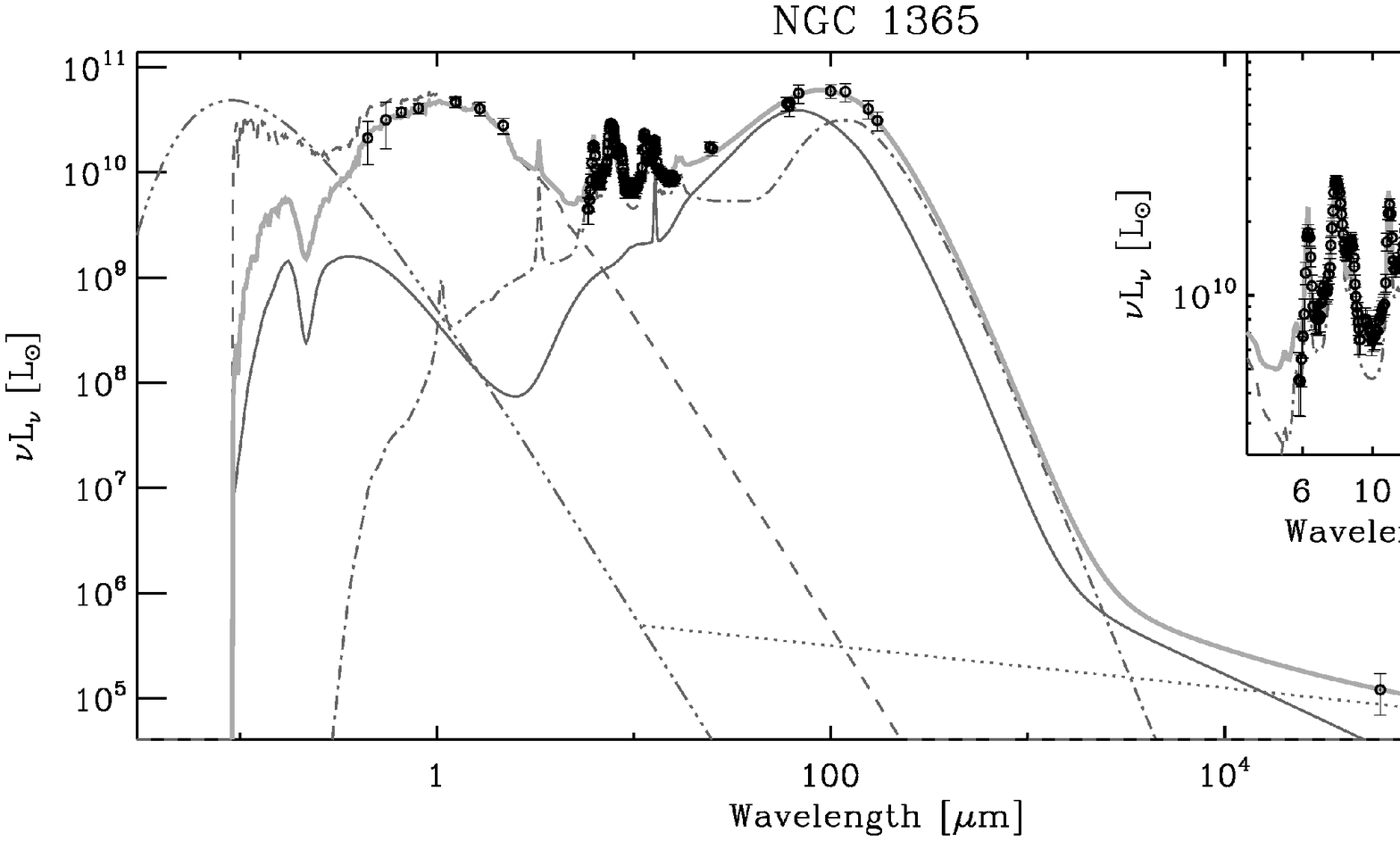}
  \caption{Fit of the galaxies' SEDs.
           See \reffig{fig:fits1} for details.}
  \label{fig:fits16}
\end{figure*}

\begin{figure*}[htbp]
  \plotone{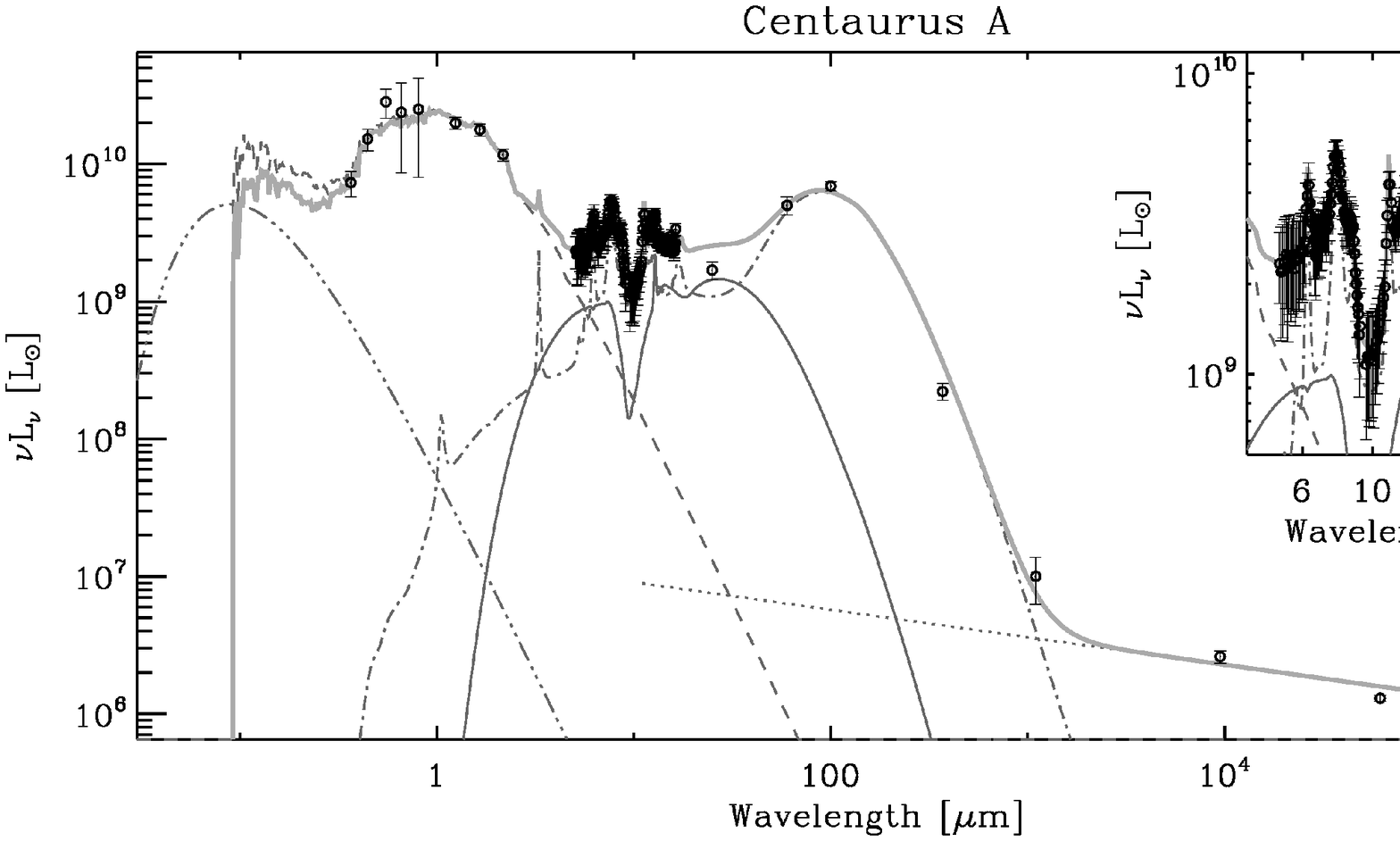}
  \plotone{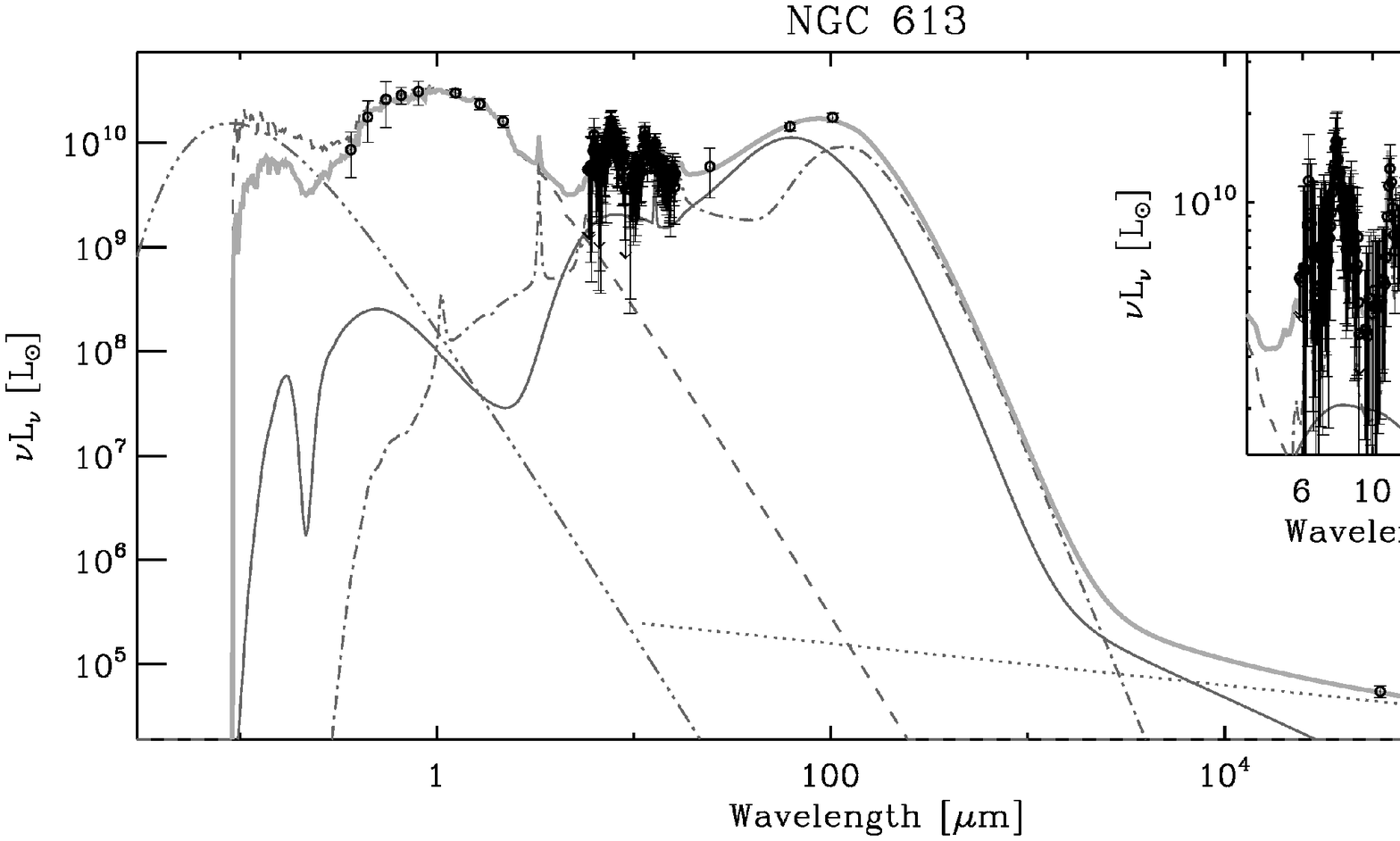}
  \caption{Fit of the galaxies' SEDs.
           See \reffig{fig:fits1} for details.}
  \label{fig:fits17}
\end{figure*}

\begin{figure*}[htbp]
  \plotone{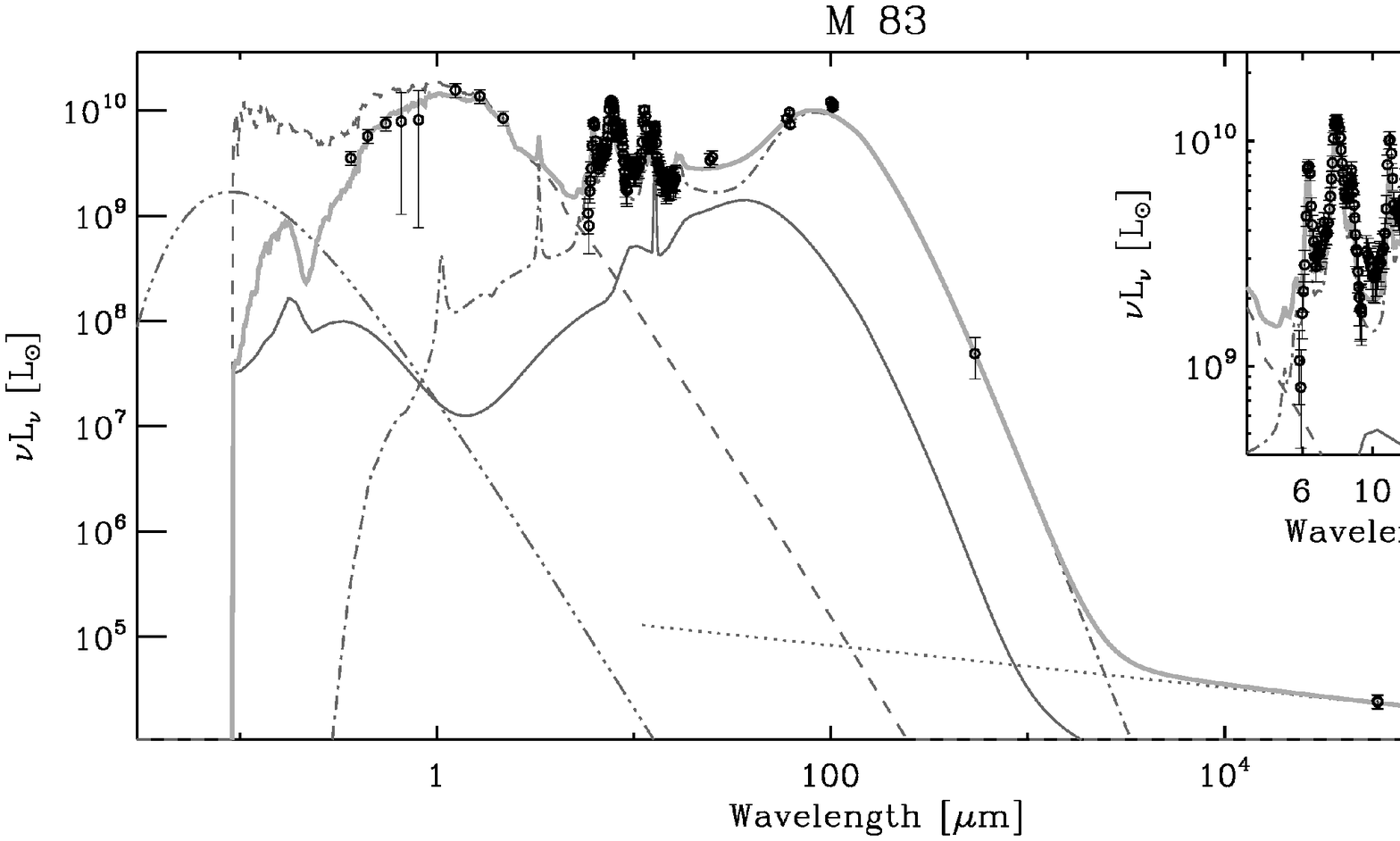}
  \caption{Fit of the galaxies' SEDs.
           See \reffig{fig:fits1} for details.}
  \label{fig:fits18}
\end{figure*}

\begin{figure*}[htbp]
  \centering
  \includegraphics[width=0.8\textwidth]{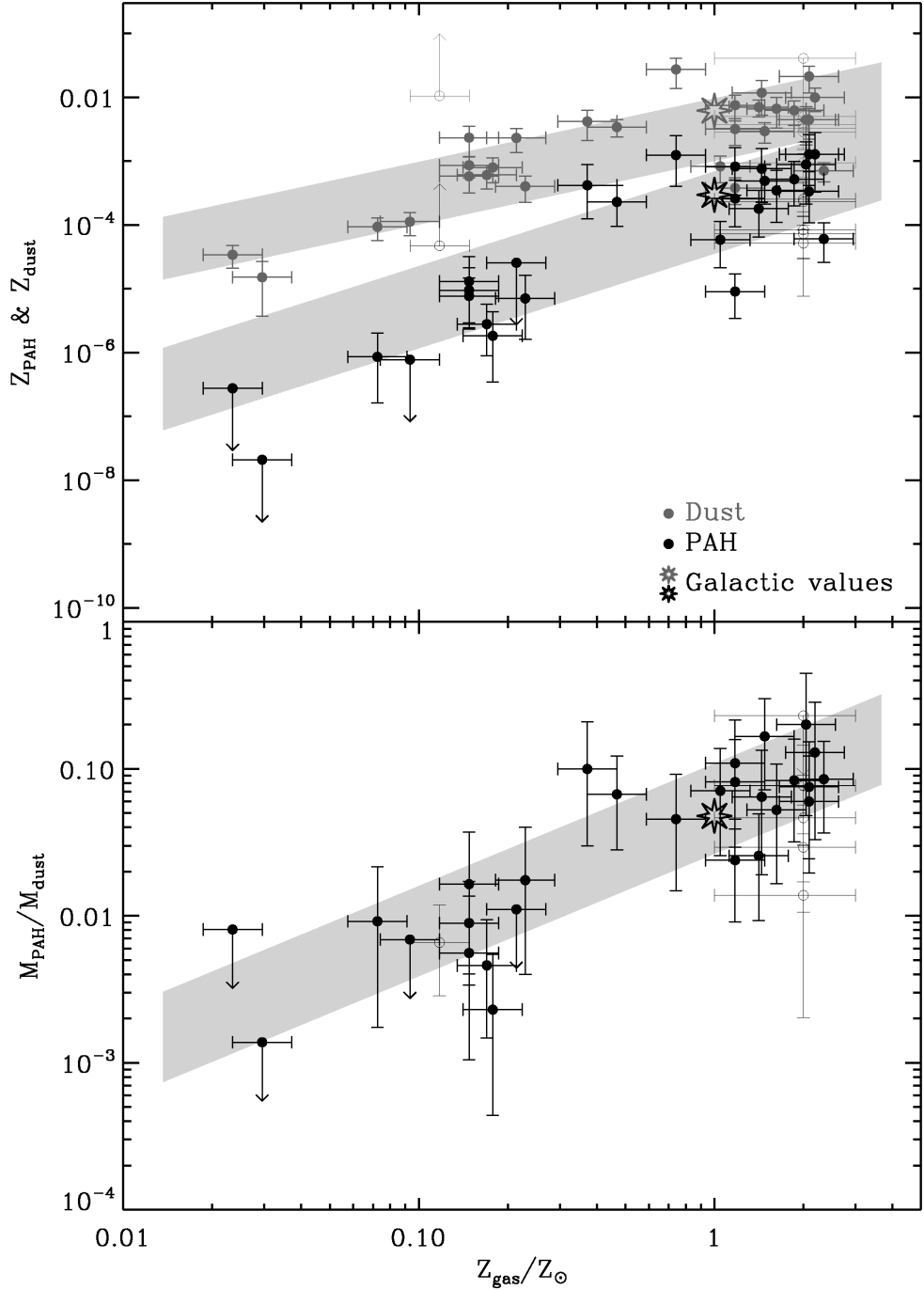}
  \caption{\uline{Top panel}: dust-to-gas mass ratio for PAHs, $Z_\sms{PAH}$,
           and the dust that gives rise to the far-IR emission, 
           $Z_\sms{dust}$, as a function of metallicity.
           \uline{Bottom panel}: mass ratio of PAH-to-dust, as a function
           of metallicity. 
           The circles correspond to galaxies and the open stars to the 
           the diffuse Galactic ISM \citep{zubko04}.
           The filled circles are the reliable measurements, and
           the open circles are the ones which are considered uncertain.
           The grey stripes are the $\pm1\sigma$ linear correlation, in 
           logarithmic scale.
           These figures show the two distinct evolutionary trends of 
           $Z_\sms{PAH}$ and $Z_\sms{dust}$ with metallicity.}
  \label{fig:pahvsOovH}
\end{figure*}

We applied the method described in \refS{sec:method}
to the sample in \reftab{tab:source} (\reffigs{fig:fits1} to \ref{fig:fits18}).
The results are presented in \reffig{fig:pahvsOovH} and 
\reftab{tab:results}.

PAH features are not detected in \izw, \sbs, \mrk{153}, \ngc{5253} and 
\ngc{1399}.
For these galaxies, we get an upper limit by fitting the short-wavelength part
of the mid-IR spectrum with the maximum PAH amount allowed by the observational 
errors.
\begin{description}
  \item[\izw:] The \hi\ extends far out of the star forming region.
    Instead of normalizing the dust masses by the value of $M_\sms{\hi}$
    in \reftab{tab:source}, we consider that the \hi\ mass associated to the
    star forming region is the one of the object \hi-A \citep{van-zee98},
    where $M_\sms{\hi} = 4.4\E{7}\msun$.
  \item[\sbs:] The $65\mic$ flux is the one reported by \citet{hunt05}.
    The fit of this flux gives a far-IR dust temperature slightly colder
    than what we would obtain by fitting only the slope of the \irs\ spectrum.
    However, this gives a conservative solution, since we derive only an upper
    limit on the mass of PAHs, in this galaxy.
    The \hi\ halo extends also far out of the star forming region.
    We assume that the star forming region has the same size than the optical
    galaxy, i.e.\ a radius of 1.6~kpc ($6\arcsec$).
    We derive the corresponding mass of \hi, using the average column density of
    $N_\sms{\hi}=7.4\E{20}\;\rm cm^{-2}$ reported by \citet{pustilnik01}.
    We find $M_\sms{\hi}=4.6\E{7}\msun$. 
  \item[\viizw:] This \irs\ spectrum of this galaxy has been studied by 
    \citet{wu06}.
    However, they did not report any PAH detection.
    The degradation of the spectral resolution that we performed on this spectra
    (see \refS{sec:irs}) increases the signal-to-noise ratio significantly.
    We report a $4\sigma$ detection of the $7.7 \mic$ feature, and marginal
    detections of the $6.2$ and $8.6\mic$ features.
    This is the lowest metallicty PAH detection to date.
    Similarly to \izw\ and \sbs\ the \hi\ halo extends far out of the star
    forming region.
    To correct for this effect, we consider that the star forming region has 
    a size $1\arcmin\times0.5\arcmin$, with a column density 
    $N_\sms{\hi}=1.6\E{21}\;\rm cm^{-2}$ \citep{thuan04}.
    This leads to an effective $M_\sms{\hi}=1.1\E{7}\msun$.
    This galaxy has only one radio point.
    To remain conservative, we assume that it is free-free dominated.
  \item[\mrk{153}:]
    The mid-infrared spectrum of this galaxy show prominent silicate emission.
    We do not have any radio observation of this object, therefore the 
    contribution of \hii\ regions to the total SED is rather uncertain.
  \item[\haro{11}:] As quoted by \citet{bergvall00}, the ratio 
    $M_\sms{\hi}/L_\sms{B}<0.01$ is abnormally low in this galaxy.
    The low value of the \hi\ mass explains the high value of its dust-to-gas
    mass ratio (\reffig{fig:pahvsOovH}).
    On the contrary, its PAH-to-dust mass ratio is not peculiar.
    This source, at 92~Mpc, is one of the most distant object in our sample.
  \item[\IC{342}:] This galaxy is located at $10\degr$ of Galactic 
    latitude. 
    Hence, it is highly extincted by the foreground, which explains the 
    peculiar shape of its optical/near-IR observed SED.
  \item[\tol{89}:] The mid-IR spectrum of this galaxy samples only its nucleus.
    That is the reason why it underestimates the \iraciv\ flux.
    To be conservative, we derive the lower limit on the PAH mass by fitting
    the spectrum, and the upper limit, by fitting the broad band.
    This galaxy has only one radio point.
    Therefore, we assume that it is free-free dominated.
  \item[\ngc{1068}:]
    This galaxy contains a powerful AGN.
    In principle our model can not be applied to this object, since it does
    not take into account the contribution of the accretion disc.
    However, we fit this galaxy as if it was a starburst, in order to test
    the robustness of our approach.
  \item[\ngc{1399}:] This object is a cD galaxy.
    Its interstellar medium is likely very tenuous and
    the mid-IR emission is dominated by the contribution of evolved stars.
    It has not been represented on the upper panel of \reffig{fig:pahvsOovH},
    since we only know the upper limits on the PAH and \hi\ gas masses.
    However, it is shown on the lower panel; the higher open circle is the 
    ratio between the upper limit on the PAH mass and the lower limit on the
    dust mass.
  \item[Circinus:] This galaxy is located at Galactic latitude below $5\degr$.
    Like \IC{342}, its optical fluxes are very uncertain.
\end{description}

The top panel of \reffig{fig:pahvsOovH} shows the variation of the PAH and 
dust to gas mass ratios with the metallicity of the interstellar medium.
Each individual galaxy can be seen as a snapshot of galaxy evolution, at a 
given time.
First, we note that the Galactic values of the dust-to-gas mass ratios
are in agreement with the one of the other galaxies, around the
same metallicity.
These Galactic values were derived by \citet{zubko04}, from the fit of the emission and extinction of the diffuse interstellar medium, with further constraints from the elemental depletion pattern.
Thus, it is a very reliable estimation.
This comparison confirms that our method does not overlook a significant amount
of dust, at least around the solar metallicity.
Second, the trends of $Z_\sms{PAH}$ and $Z_\sms{dust}$, with the 
metallicity are not identical.
Our sample spreads two orders of magnitude in metallicity.
We can see that the PAH-to-gas mass ratio rises by five orders of magnitudes,
while the dust-to-gas mass ratio, by only three.
This differential evolution is illustrated in the lower panel of 
\reffig{fig:pahvsOovH}, showing the PAH-to-dust mass ratio.
This figure is the analog of \reffig{fig:band_ratios}, but instead of considering integrated fluxes, it deals with abundances.
The PAH-to-dust mass ratio rises by two orders of magnitude, over our sample, 
while the \IRACiv/\MIPSi\ band ratio varies only by one order of magnitude.

\clearpage
\clearpage
\clearpage
\clearpage
\clearpage
\clearpage
\clearpage
\clearpage
\clearpage
\clearpage
\clearpage
\clearpage
\clearpage
\clearpage
\clearpage
\clearpage
\clearpage
\clearpage

\begin{deluxetable}{lrrrrrrrr}
  \tabletypesize{\scriptsize}
  \rotate
  \tablecolumns{9}
  \tablewidth{0pc}
  \tablecaption{SED Modeling Results}
  \tablehead{
    \colhead{Name} & \colhead{$Z_\sms{gas}$} 
    & \colhead{$Z_\sms{PAH}$}
    & \colhead{$Z_\sms{dust}$} 
    & \colhead{$t_\sms{SF}$}
    & \colhead{$n_\sms{H}$}
    & \colhead{$f_+$}
    & \colhead{$L_\sms{burst}$} 
    & \colhead{$L_\sms{PDR}$} 
    \\
    \colhead{}     & \colhead{$[Z_\odot]$} 
    & \colhead{}
    & \colhead{}
    & \colhead{[Myr]}
    & \colhead{$[\rm cm^{-3}]$}
    & \colhead{}
    & \colhead{$[L_\odot]$}
    & \colhead{$[L_\odot]$} }
\startdata
I Zw 18 & $0.023\petm{0.005}{0.006}$ & $\lesssim2.7\E{-7}$ & $3.4\petm{1.3}{1.3}\E{-5}$ & 160 & 70 & \ldots & $1.6\E{7}$ & $8.3\E{6}$\unc \\
SBS 0335-052 & $0.030\petm{0.006}{0.008}$ & $\lesssim2.1\E{-8}$ & $1.5\petm{1.1}{1.1}\E{-5}$ & 110 & $2.0\E{4}$ & \ldots & $1.6\E{9}$ & $8.6\E{8}$ \\
VII Zw 403 & $0.072\petm{0.015}{0.019}$ & $8.6\petm{6.9}{11.5}\E{-7}$ & $9.3\petm{3.6}{3.6}\E{-5}$ & 650 & 30\unc & 1.0\unc & $3.4\E{7}$\unc & $2.6\E{6}$ \\
Mrk 153 & $0.093\petm{0.019}{0.024}$ & $\lesssim7.7\E{-7}$ & $1.1\petm{0.4}{0.4}\E{-4}$ & 190 & 110\unc & 0.0 & $5.1\E{8}$\unc & $3.6\E{8}$\unc \\
Haro 11 & $0.12\petm{0.02}{0.03}$ & $1.1\petm{0.6}{0.9}\E{-4}$\unc & $1.6\petm{0.6}{0.6}\E{-2}$\unc & 680 & $1.0\E{3}$ & 0.7 & $1.6\E{11}$ & $9.2\E{10}$ \\
NGC 1140 & $0.15\petm{0.03}{0.04}$ & $9.5\petm{7.2}{11.9}\E{-6}$ & $5.8\petm{2.6}{2.6}\E{-4}$ & 1610 & 30 & 0.4 & $8.2\E{9}$ & $1.8\E{9}$ \\
UM 448 & $0.15\petm{0.03}{0.04}$ & $7.7\petm{4.8}{7.0}\E{-6}$ & $8.6\petm{3.2}{3.2}\E{-4}$ & 450 & 100 & 1.0 & $4.3\E{10}$ & $2.4\E{10}$ \\
Tol 89 & $0.15\petm{0.03}{0.04}$ & $1.3\petm{1.1}{1.9}\E{-5}$ & $2.3\petm{1.2}{1.2}\E{-3}$ & 470 & 20\unc & 0.3\unc & $1.2\E{9}$\unc & $4.6\E{8}$ \\
Mrk 930 & $0.17\petm{0.03}{0.04}$ & $2.8\petm{1.9}{2.9}\E{-6}$ & $6.1\petm{2.4}{2.4}\E{-4}$ & 1360 & 70 & \ldots & $1.3\E{10}$ & $7.2\E{9}$ \\
II Zw 40 & $0.18\petm{0.04}{0.05}$ & $1.8\petm{1.5}{2.5}\E{-6}$ & $7.9\petm{3.4}{3.4}\E{-4}$ & 1330 & 150 & 1.0\unc & $2.8\E{9}$ & $6.4\E{8}$ \\
NGC 5253 & $0.21\petm{0.04}{0.06}$ & $\lesssim2.6\E{-5}$ & $2.3\petm{0.9}{0.9}\E{-3}$ & 1520 & 320 & \ldots & $1.1\E{9}$ & $5.0\E{8}$ \\
NGC 1569 & $0.23\petm{0.05}{0.06}$ & $7.0\petm{5.4}{9.1}\E{-6}$ & $4.0\petm{1.8}{1.8}\E{-4}$ & 1280 & 80 & 0.0 & $6.3\E{8}$ & $2.2\E{8}$ \\
Mrk 33 & $0.37\petm{0.08}{0.10}$ & $4.2\petm{2.9}{4.6}\E{-4}$ & $4.2\petm{2.1}{2.1}\E{-3}$ & 5110 & 130 & 0.6 & $4.8\E{9}$ & $2.2\E{9}$ \\
NGC 7714 & $0.47\petm{0.10}{0.12}$ & $2.3\petm{1.3}{1.9}\E{-4}$ & $3.4\petm{1.0}{1.0}\E{-3}$ & 3100 & 60 & 0.6 & $4.3\E{10}$ & $9.4\E{9}$ \\
M 51 & $0.74\petm{0.15}{0.19}$ & $1.2\petm{0.8}{1.3}\E{-3}$ & $2.7\petm{1.4}{1.4}\E{-2}$ & 2570 & 10 & 0.5 & $2.2\E{10}$ & $3.1\E{10}$ \\
IC 342 & $1.0\petm{0.2}{0.3}$ & $5.8\petm{3.7}{5.5}\E{-5}$ & $8.3\petm{3.7}{3.7}\E{-4}$ & 4610 & 10\unc & 0.5 & $6.2\E{9}$\unc & $1.2\E{10}$ \\
NGC 891 & $1.2\petm{0.2}{0.3}$ & $2.6\petm{1.7}{2.5}\E{-4}$ & $3.2\petm{1.4}{1.4}\E{-3}$ & 3690 & 10 & 0.6 & $1.9\E{10}$ & $1.7\E{10}$ \\
He 2-10 & $1.2\petm{0.2}{0.3}$ & $9.0\petm{5.6}{8.1}\E{-6}$ & $3.8\petm{1.7}{1.7}\E{-4}$ & 1240 & 40 & 1.0 & $5.1\E{9}$ & $8.4\E{8}$ \\
NGC 3256 & $1.2\petm{0.2}{0.3}$ & $8.2\petm{5.3}{7.9}\E{-4}$ & $7.5\petm{3.2}{3.2}\E{-3}$ & 6740 & 90 & 0.6 & $3.7\E{11}$ & $1.5\E{11}$ \\
NGC 1068 & $1.4\petm{0.3}{0.4}$ & $1.8\petm{1.1}{1.7}\E{-4}$ & $7.0\petm{2.1}{2.1}\E{-3}$ & 3710 & $1.0\E{4}$ & 0.2 & $2.0\E{11}$ & $1.1\E{11}$ \\
NGC 253 & $1.4\petm{0.3}{0.4}$ & $7.6\petm{5.3}{8.2}\E{-4}$ & $1.2\petm{0.6}{0.6}\E{-2}$ & 14000 & 100 & 0.7 & $2.7\E{10}$ & $1.9\E{10}$ \\
M 82 & $1.5\petm{0.3}{0.4}$ & $4.9\petm{2.8}{3.9}\E{-4}$ & $2.9\petm{1.0}{1.0}\E{-3}$ & 9040 & 530 & 0.7 & $3.8\E{10}$ & $2.4\E{10}$ \\
NGC 1097 & $1.6\petm{0.3}{0.4}$ & $3.5\petm{2.4}{3.6}\E{-4}$ & $6.6\petm{3.3}{3.3}\E{-3}$ & 6060 & 50 & 0.5 & $5.7\E{9}$ & $2.1\E{10}$ \\
NGC 6946 & $1.9\petm{0.4}{0.5}$ & $5.2\petm{3.2}{4.7}\E{-4}$ & $6.2\petm{2.6}{2.6}\E{-3}$ & 5210 & 20 & 0.5 & $8.4\E{9}$ & $1.3\E{10}$ \\
NGC 1808 & $2.0\petm{0.4}{0.5}$ & $8.9\petm{6.7}{11.0}\E{-4}$ & $4.4\petm{2.7}{2.7}\E{-3}$ & 6040 & 30 & 0.6 & $2.8\E{10}$ & $1.2\E{10}$ \\
NGC 520 & $2.0\petm{1.0}{1.0}$\unc & $2.5\petm{1.5}{2.2}\E{-4}$ & $3.3\petm{1.3}{1.3}\E{-3}$ & 13900 & 50 & 1.0 & $5.8\E{10}$ & $2.8\E{10}$ \\
NGC 1399 & $2.0\petm{1.0}{1.0}$\unc & $\lesssim9.3\E{-3}$\unc & $\lesssim4.1\E{-2}$ & 14000 & 300\unc & \ldots & $2.4\E{7}$\unc & $3.2\E{8}$\unc \\
NGC 4945 & $2.0\petm{1.0}{1.0}$\unc & $8.3\petm{5.3}{7.9}\E{-5}$ & $2.8\petm{1.3}{1.3}\E{-3}$ & 4660 & 10 & 1.0 & $2.7\E{10}$ & $1.4\E{10}$ \\
Circinus & $2.0\petm{1.0}{1.0}$\unc & $7.3\petm{4.4}{6.4}\E{-5}$ & $9.4\petm{3.8}{3.8}\E{-4}$ & 14000 & $1.1\E{3}$ & 0.5 & $7.3\E{9}$ & $1.1\E{10}$ \\
NGC 6240 & $2.0\petm{1.0}{1.0}$\unc & $2.3\petm{1.5}{2.2}\E{-4}$ & $5.0\petm{2.0}{2.0}\E{-3}$ & 12940 & 100 & 0.8 & $4.5\E{11}$ & $2.3\E{11}$ \\
Arp 220 & $2.0\petm{1.0}{1.0}$\unc & $5.2\petm{4.4}{8.3}\E{-5}$ & $3.7\petm{2.2}{2.2}\E{-3}$ & 14000 & 20 & 1.0\unc & $9.4\E{11}$ & $3.0\E{11}$ \\
NGC 1365 & $2.1\petm{0.4}{0.5}$ & $3.3\petm{2.3}{3.4}\E{-4}$ & $4.4\petm{2.2}{2.2}\E{-3}$ & 5490 & 20 & 0.4 & $7.1\E{10}$ & $5.5\E{10}$ \\
Centaurus A & $2.1\petm{0.4}{0.5}$ & $1.3\petm{0.9}{1.3}\E{-3}$ & $2.1\petm{1.0}{1.0}\E{-2}$ & 5480 & $8.6\E{3}$ & 0.3 & $8.0\E{9}$ & $1.1\E{10}$ \\
NGC 613 & $2.2\petm{0.4}{0.6}$ & $1.3\petm{1.0}{1.5}\E{-3}$ & $9.9\petm{3.9}{3.9}\E{-3}$ & 5590 & 20 & 0.3 & $2.2\E{10}$ & $1.8\E{10}$ \\
M 83 & $2.3\petm{0.5}{0.6}$ & $6.0\petm{3.4}{4.8}\E{-5}$ & $7.1\petm{2.3}{2.3}\E{-4}$ & 5440 & 290 & 0.6 & $2.5\E{9}$ & $1.7\E{10}$ \\
\enddata

  \tablecomments{Most of these quantities are defined in 
                 \refS{sec:method} and \reftab{tab:var}.
                 $Z_\sms{PAH}$ and $Z_\sms{dust}$ are the PAH-to-gas and 
                 dust-to-gas mass ratios;
                 $t_\sms{SF}$ is the age of the
                 galaxy constrained by the optical and near-IR broad bands;
                 $n_\sms{H}$ is the average gas density in \hii\ regions;
                 $f_+$ is the mass fraction of cationic PAHs;
                 $L_\sms{burst}$ by the recent burst of star formation;
                 $L^\sms{PDR}$ is the intrinsic bolometric luminosity emitted
                 by the dust inside the neutral phase.
                 The sources are ordered according to their metallicity.
                 The symbol \unc\ identifies uncertain values.}
  \label{tab:results}
\end{deluxetable}


\section{A MODEL FOR THE GAS AND DUST EVOLUTION}
\label{sec:dustvol}

We have developed a one-zone single-phase chemical evolution model, 
to follow the abundances and composition of the dust and the metallicty as a
function of time, in order to interpret the results of our SED modeling.
In this section, we will give a brief description of the physical processes which are incorporated.
A more detailed discussion can be found in \citet{dwek98} and \citet{dwek07}.
Assuming a common star formation history, we will use this model to interpret
the observed evolutionary trend of galaxies' SED with metallicity,
on a global scale.

  \subsection{Metal Enrichment and Gas Evolution}
  \label{sec:evZ}

In the present paper, we consider a closed box model.
We consider the delayed injection of material by different stellar progenitors,
but we assume that the mixing of the elements in the ISM is instantaneous.
To be consistent with our stellar population synthesis, we adopt a 
Salpeter initial mass function, $\phi(m)$ \refeqp{eq:imf}, 
where $m$ is the mass of individual stars.
We define the average stellar mass:
\begin{equation}
  \langle m\rangle \equiv \int_{m_l}^{m_u} m\,\phi(m)\ddiff m.
\end{equation}
The evolution of the gas mass surface density, $\Sigma_\sms{gas}(t)$, with 
the time $t$, is:
\begin{equation}
  \frac{\dd\Sigma_\sms{gas}(t)}{\dd t} =
    - \Sigma_\sms{SFR}(t) +
    \int_{m_l}^{m_u} \Sigma_\sms{SFR}(t-\tau(m))\,
      \frac{m_\sms{ej}(m)}{\langle m\rangle}\,\phi(m)\ddiff m,
  \label{eq:diffg}
\end{equation}
where $\Sigma_\sms{SFR}(t)$ is the mass of star formed per unit time and
per unit surface area, 
$\tau(m)$, the lifetime of a star of mass $m$, and $m_\sms{ej}(m)$, 
its mass of gas returned to the ISM.
The first term of the right hand side of \refeq{eq:diffg} is the amount of gas
removed by the star formation, and the second term, is the delayed injection
of the gas (H, He and metals), by the various progenitors.
The ISM metallicity, is defined as:
\begin{equation}
  Z_\sms{ISM}(t) \equiv Z_\sms{gas}(t)+Z_\sms{dust}(t),
\end{equation}
where $Z_\sms{gas}(t)$ is the metal-to-gas mass ratio, and 
$Z_\sms{dust}(t)$, the dust-to-gas mass ratio that will be discussed at 
\refS{sec:evd}.
Its evolution is analog to \refeq{eq:diffg}:
\begin{equation}
  \frac{\dd \left[\Sigma_\sms{gas}(t)\,Z_\sms{ISM}(t)\right]}{\dd t} =
    - Z_\sms{ISM}(t)\,\Sigma_\sms{SFR}(t)
    + \int_{m_l}^{m_u} \Sigma_\sms{SFR}(t-\tau(m))\,
      \frac{Y_\sms{Z}(m)}{\langle m\rangle}\,\phi(m)\ddiff m,
  \label{eq:diffZ}
\end{equation}
where $Y_Z(m)$ is the yield of elements heavier than He, by the stars of mass
$[m,m+\dd m]$.

\begin{figure}[htbp]
  \centering
  \plotone{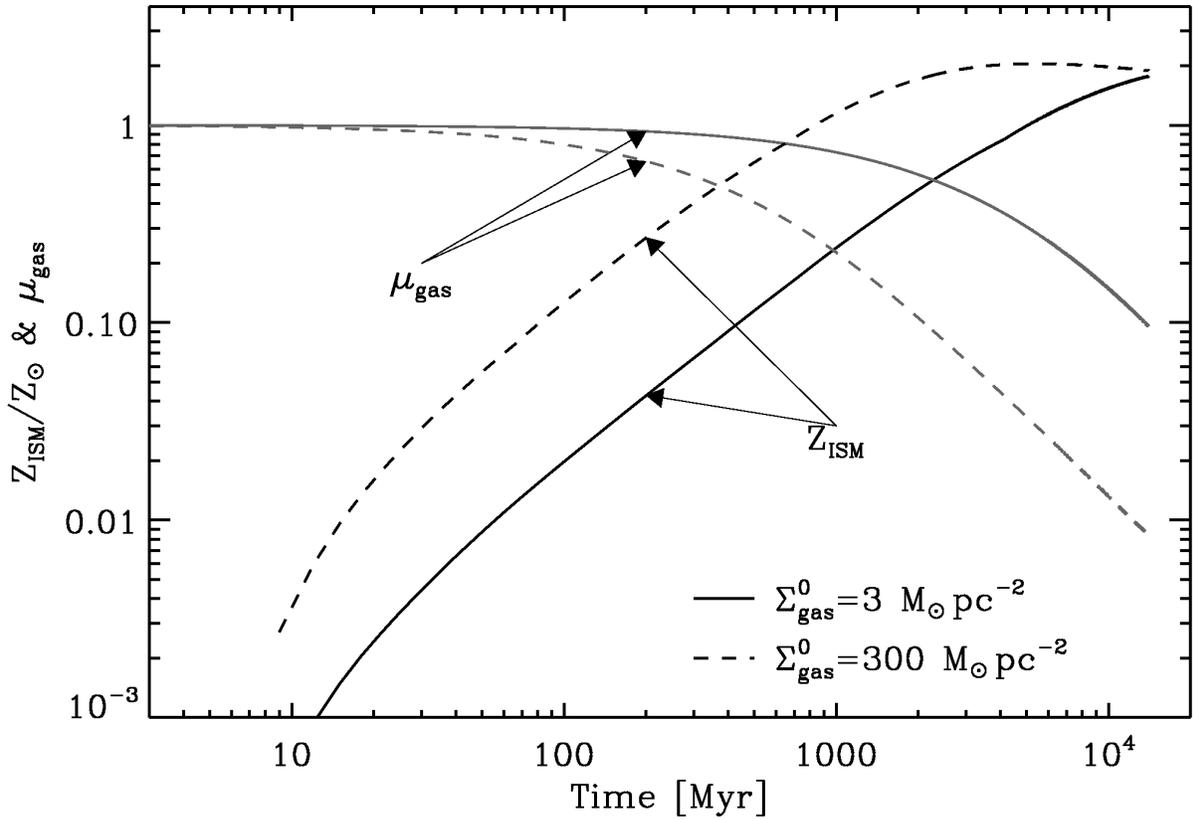}
  \caption{Evolution of the total metallicity (in units of $Z_\odot$), 
           and of the reduced gas mass ($\mu_\sms{gas}$),
           for two different initial surface gas mass densities, 
           $\Sigma_\sms{gas}^0$.}
  \label{fig:gasvol}
\end{figure}

The elemental yields, $Y_Z(m)$, for the low mass stars ($m\leq8\msun$)
are taken from \citet{karakas03b,karakas03}, and from \citet{woosley95} for 
the high mass stars ($m>8\msun$).
Furthermore, a prescription for the star formation rate is required, in order
to solve these equations.
This prescription is given by the \citet{schmidt59} law, with the coefficients
derived by \citet{kennicutt98}:
\begin{equation}
  \frac{\Sigma_\sms{SFR}(t)}{1\msun\,{\rm yr^{-1}\,pc^{-2}}}
   = (2.5\pm0.7)\E{-10}\,\left(\frac{\Sigma_\sms{gas}(t)}
                                    {1\msun\,{\rm pc^{-2}}}\right)^{1.40\pm0.15}
\end{equation}
and is used to replace $\Sigma_\sms{SFR}(t)$ in \refeq{eq:diffg}.
\reffig{fig:gasvol} shows the evolution of the total metallicity, and of the 
reduced gas mass 
$\mu_\sms{gas}(t) \equiv \Sigma_\sms{gas}(t)/\Sigma_\sms{gas}^0$, where
$\Sigma_\sms{gas}^0 = \Sigma_\sms{gas}(0)$ is the initial gas mass surface
density.

  \subsection{Dust Formation and Destruction}
  \label{sec:evd}
  
The evolution of the mass surface density, $\Sigma_i(t)$, of a given
dust specie, is the sum of three 
contributions:\textlist{\thetextlist~the rate of dust destruction by 
    star formation,
  \thetextlist~the rate of dust condensation in stellar progenitors,
  \thetextlist~the rate of dust destruction in the ISM, by SN blast waves:}
\begin{equation}
  \frac{\dd\Sigma_\sms{i}(t)}{\dd t} 
  = - Z_\sms{i}(t)\,\Sigma_\sms{SFR}(t) 
    + \int_{m_l}^{m_u} 
      \frac{Y_i(m)}{\langle m\rangle}\,\Sigma_\sms{SFR}(t-\tau(m))
      \,\phi(m)\ddiff m
    - \frac{\Sigma_\sms{i}(t)}{\tau_\sms{dust}(t)},
\end{equation}
where $Y_\sms{i}(m)$ is the yield of the considered dust specie by stars
of mass $[m,m+\dd m]$, and $\tau_\sms{dust}(t)$, the dust lifetime in the ISM.

The dust yields are derived from the elemental stellar yields described at 
\refS{sec:evZ}, following \citet{dwek98}.
For low mass stars ($m\leq 8\msun$), the dust yields depend on the value of 
the C/O ratio.
We assume that the ejecta is microscopically mixed, so that all the excess
carbon is locked-up in dust, if C$>$O.
If C$<$O, then we combine all the available Fe, Si, Mg, Ca and Ti, with one 
O atom to produce silicate dust, and titanium oxydes.
For high mass stars ($m>8\msun$), we assume that the ejecta is only macroscopically mixed, so that both carbon and oxygen rich dust can condense.
We assume a condensation efficiency of unity for all dust species.

The lifetime of dust, $\tau_\sms{dust}(t)$ is directly related to the 
\snii\ rate
\citep{dwek80,mckee86}:
\begin{equation}
  \tau_\sms{dust}(t) = \frac{\Sigma_\sms{gas}(t)}
                            {R_\sms{\snii}(t)\,\langle m_\sms{ISM}\rangle},
\end{equation}
where $R_\sms{\snii}(t)$ is the \snii\ rate per unit area, and 
$\langle m_\sms{ISM}\rangle$ is the average effective mass of gas, swept up 
by a single SN remnant, where the dust is returned back to the gas phase, 
by either 
thermal sputtering, or grain-grain collision \citep{jones96}.
The value of $\langle m_\sms{ISM}\rangle$ is largely unknown. 
We explore its effects, by varying it from $\langle m_\sms{ISM}\rangle=0\msun$
(no destruction), to $\langle m_\sms{ISM}\rangle=300\msun$ (typical 
destruction).
The latter value is typical of our Galaxy.
Indeed, if $M_\sms{gas}^\sms{Gal}\simeq5\E{9}\msun$ is the total mass of gas in 
our Galaxy, $R_\sms{SN}^\sms{Gal}\simeq1/30\;\rm yr^{-1}$, its average SN rate,
and $\tau_\sms{dust}^\sms{Gal}\simeq5\E{8}\;\rm yr$, the mean lifetime of an
ISM dust particle \citep{jones04}, then:
\begin{equation}
  \langle m_\sms{ISM}\rangle
    \simeq \frac{M_\sms{gas}^\sms{Gal}}
                {R_\sms{SN}^\sms{Gal}\,\tau_\sms{dust}^\sms{Gal}}
    \simeq 300\msun.
\end{equation}

An additional destruction mechanism that apllies only to PAHs and
very small grains is their photoevaporation in intense radiation
fields.
This destruction mechanism primarily affects the abundance of PAHs, and is 
taken into account in the SED model that calculates the dust abundances, by 
assuming that PAHs do not survive in \hii\ regions.

\begin{figure}[htbp]
  \centering
  \plotone{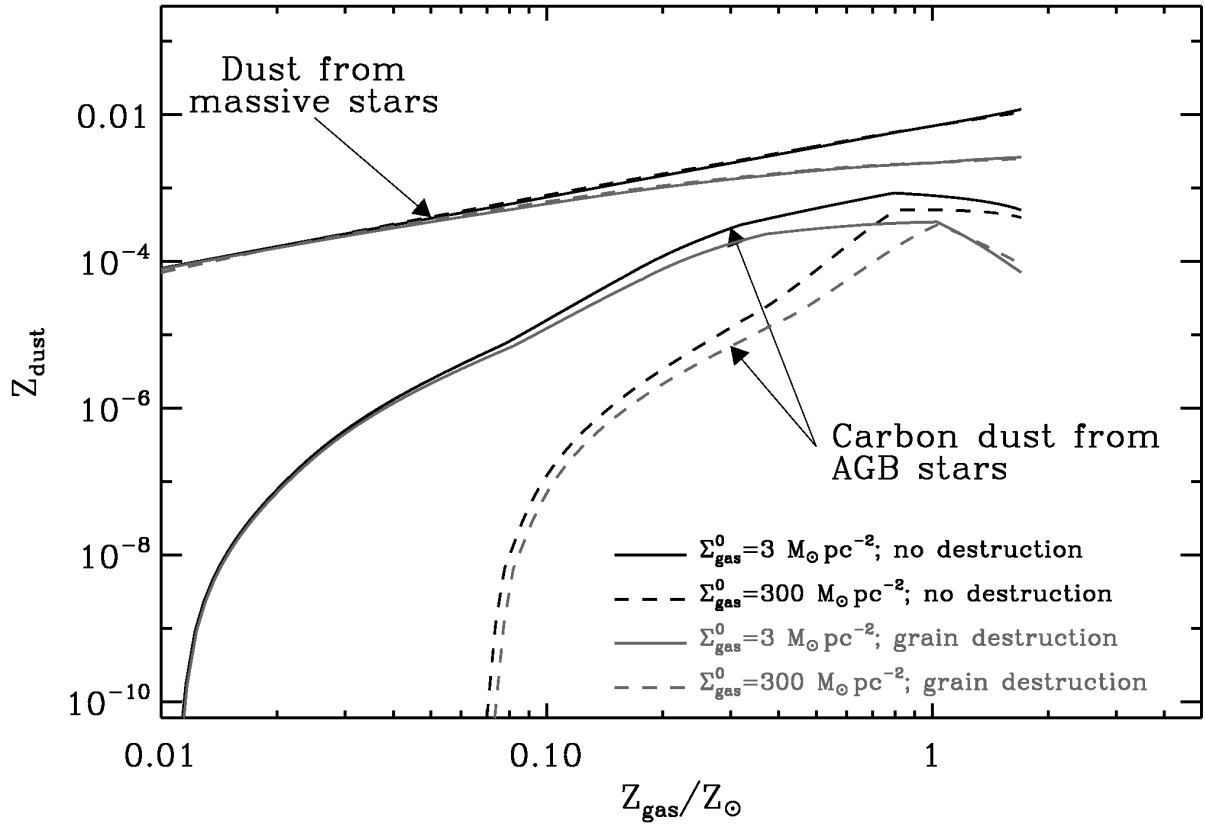}
  \caption{Evolution of the dust content with the metallicity of the ISM.
           We adopt $\langle m_\sms{ISM}\rangle=300\msun$,
           for the curves with grain destruction.}
  \label{fig:dustvol}
\end{figure}

\reffig{fig:dustvol} shows the evolution of the total dust content formed by massive stars, and the carbon dust formed by AGB stars, for various initial conditions, and destruction efficiencies.
The \snii\ dust evolves almost linearly with the metallicity, in absence of destruction, because the metal enrichment is dominated by massive stars.
In contrast, the AGB carbon dust starts rising when the metallicity of the
ISM is around $1/20\zsun$.
This value corresponds to a time of $\sim100$~Myr (\reffig{fig:gasvol}), 
which the lifetime of the most massive AGB stars.
Hence, the carbon dust produced by AGB stars is injected into the ISM, with a
delay which corresponds to the lifetime of the stars.
This evolutionary trend was previously noted by \citet{dwek98} and 
\citet{morgan03}.
The change in the slope of the AGB carbon dust, around $1\zsun$, is simply 
due to the fact that AGB stars of lifetime longer than $\sim1$~Gyr are 
oxygen rich.
The dust destruction effects the evolution for $Z\gtrsim0.1\zsun$.

  \subsection{Examining the Consistency Between the Stellar Populations and
              the Elemental Evolution}
  \label{sec:consistency}

\begin{figure}[htbp]
  \centering
  \plotone{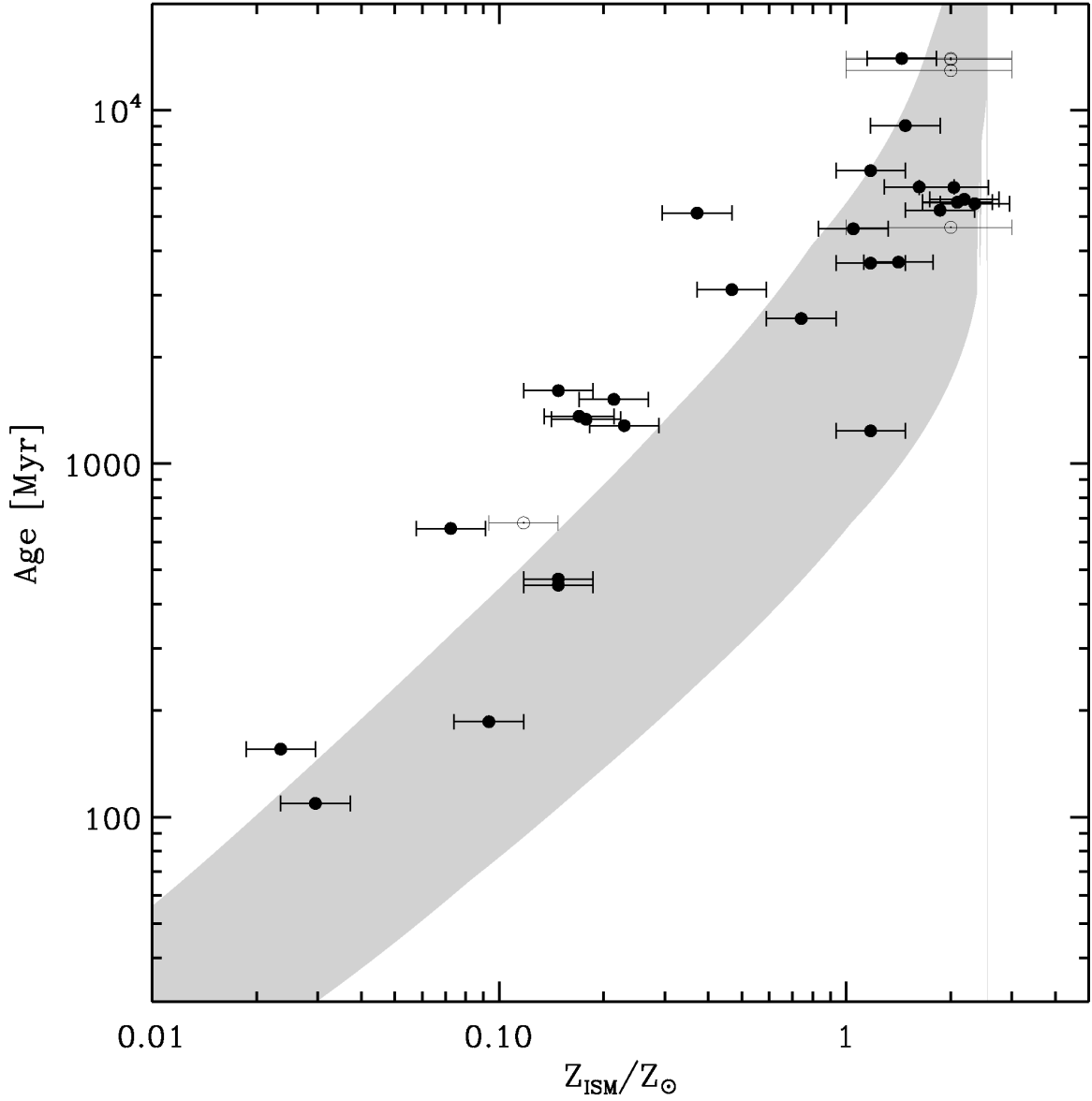}
  \caption{Consistency between the stellar populations and the elemental
           evolution.
           The error bars are the age of the galaxies derived from population
           synthesis modeling of the optical and near-IR broad-band observations
           (\reftab{tab:results}), as a function of the observed metallicity
           of the gas (\reftab{tab:source}).
           The grey stripe shows the range of values from the elemental 
           evolution model (\reffig{fig:gasvol}).}
  \label{fig:agevsZ}
\end{figure}

For consistency, we used the same prescription for the evolution of the star 
formation rate in both the population synthesis (\refS{sec:results})
and the elemental evolution modeling (\refS{sec:evZ}).
However, we added a late burst of star formation to calculate the galactic SED, which will only have a limited effect on the final elemental abundances of the
galaxy.
The two models differ in the stellar yields, with the latter using the more 
recent yields for AGB stars.
\reffig{fig:agevsZ} examines the effect of this different yields by plotting
the galactic age versus metallicity relation for our sample of galaxies.
The model calculations, shown as a grey stripe, are in general good agreement
with the data.
However, for sub-solar metallicities ($\lesssim0.3\zsun$), several galaxies appear
to have an older age, from the stellar point of view, than what would be 
inferred from their metal enrichment.
This discrepancy is probably the consequence of assuming a smooth star formation
history.
The differences will only manifest themselves as a moderate change in the stellar spectra, and will have no affect on the conclusions of the paper.

  \subsection{PAHs and the Delayed Injection of AGB Carbon Dust}
  \label{sec:pahvsZ}

From an observational point of view, PAHs are believed to form in the 
circumstellar envelopes of carbon rich AGB stars, and to be subsequently
ejected into the ISM through stellar winds.
The paucity of UV photons prevents the direct detection of these PAHs, during
their formation process in the post-AGB phase of the evolution of these stars
\citep{hony01,boersma06}.
However, PAHs are abundantly observed during the later planetary nebula phase
\citet{hony01}.
In what follows, we will assume that PAHs are only formed in the 
envelopes of AGB stars.

\begin{figure*}[htbp]
  \centering
  \plotone{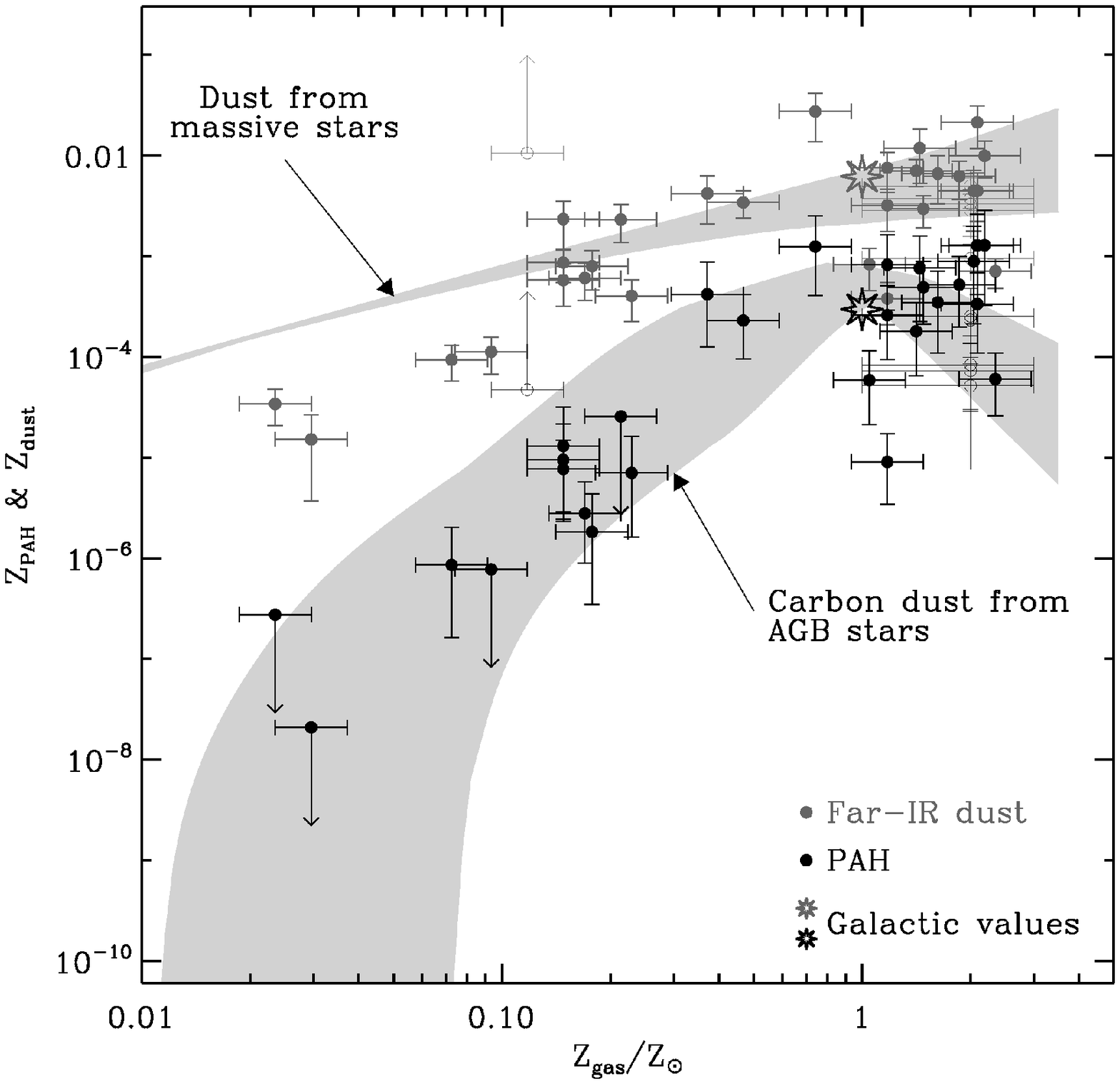}
  \caption{Comparison between the metallicity trends of the PAH abundance
           derived from the observed SED and those derived from the 
           chemical evolution model.
           The figure highlights the different evolutionary trend of \snii-
           and AGB-condensed dust.}
  \label{fig:pahvsZ}
\end{figure*}
\reffig{fig:pahvsZ} shows the comparison between the dust and PAH abundances
derived from the observations (\reffig{fig:pahvsOovH}), and the ones produced 
by the evolution model (\reffig{fig:dustvol}).
The agreement between the PAH-to-gas mass ratio, and the carbon dust 
production by AGB stars is very good, with the galaxies \hen\ and \IC{342} as
the only outliers.
\hen\ is a solar metallicity blue compact dwarf galaxy.
Its SED is similar to lower metallicity galaxies \citep{galliano05}, 
however its metallicity is very uncertain; for example \citet{vacca92} quoted 
$1/6\zsun$.
This uncertainty may reside in the fact that this galaxy has two cores,
resulting from a merger.
The properties of these nuclei are different in terms of dust absorption
\citep{phillips84}, molecular gas content \citep{baas94}, compact source
distribution \citep{cabanac05}, and mid-IR spectrum \citep{martin-hernandez06}.
Thus, our global approach may not apply to this object.
On the contrary the PAH-to-dust mass ratio of \IC{342} is consistent with 
the one of other galaxies with the same metallicity (\reffig{fig:pahvsOovH}).
Only its values of $Z_\sms{PAH}$ and $Z_\sms{dust}$ are systematically shifted.
As mentioned in \refS{sec:results} this galaxy suffers a lot of extinction 
and confusion since it is located at low Galactic latitude.
This could alter the estimate of the absolute dust-to-gas mass ratios of this 
object.
We conclude that the striking evolution of the PAH content in galaxies with metallicity can 
naturally be explained by the delayed injection of carbon dust into the ISM, by AGB stars.

In principle, PAHs can form by other processes, for example, by 
the hydrogenation of small carbon grains in the ISM.
Initially, their abundance will then follow the evolutionary trend of 
SN-condensed carbon dust. 
However, their absence in low metallicity systems suggests that they are 
efficiently destroyed, presumably by shocks or UV radiation, as envisioned in by 
\citet{madden06} or \citet{ohalloran06}. 
At later times, the rate of PAH production by these interstellar processes
will follow the evolutionary trend of AGB stars, since they will be the major source of carbon dust in the ISM.

The global trend of PAH abundance with metallicity does not preclude the 
possibility of local variations of PAH abundance in individual galaxies. 
For example, the SMC which has on the average a metallicity that is $1/6\zsun$
\citep{dufour82} has at least one region, the molecular cloud \smcb\ 
\citep{reach00}, with a PAH-to-dust ratio that is comparable to the Galactic
value \citep{li02}. 
Even in our Galaxy, there are regions with ``super Galactic'' PAH abundances
\citep[{e.g.}][]{ridderstad06}.
Such local abundance variations are a natural consequences of processes, such as 
mixing of stellar ejecta and cycling between the ISM phases, that when globally 
averaged, will follow the general trend of PAH abundances with metallicity 
depicted in \reffig{fig:pahvsZ}.

  \subsection{The Evolutionary Trend of SN-Condensed Dust}
  \label{sec:SNdust}

The dust content is in good agreement with the dust production by
\snii, down to $\sim0.1\zsun$ (\reffig{fig:pahvsZ}).
Below this value, the model systematically overestimates the observed 
dust content.
\citet{lisenfeld98} found a similar deficiency, from IRAS observations of blue
compact dwarf and dwarf irregular galaxies.
There may be several reasons for this discrepancy.

First, we may have overestimated the gas mass, in the lowest metallicity 
sources.
Indeed, the distribution of \hi\ of these galaxies extends farther out of 
the star forming region, as discussed in \refS{sec:results}.
We have attempted to correct this effect, for \izw, \sbs\ and \viizw, by 
considering only the gas content associated with the optical galaxy.
However, we could not correct for the gas located out of the star forming 
region and along the line of sight.

Second, we may have underestimated the dust mass by overlooking a cold 
dust component.
Indeed, \citet{galliano03,galliano05} showed that the millimetre excess 
observed in the SEDs of \ngc{1140}, \ngc{1569}, \iizw, and \hen, could be
consistently explained by the presence of very cold dust, accounting for
40 to $80\%$ of the total dust mass.
On \reffig{fig:pahvsZ}, the dust-to-gas mass ratio of these four galaxies
is indeed, below the \snii\ production rate by a factor of $\sim2$.
Assuming that the high clumpiness of the ISM is a general property of low-metallicity systems, and that the filling factor and/or contrast density of
these clumps rises when the metallicity drops, we have a natural explanation
for this deviation.
We can not address this issue, because of the lack of submillimetre data for these very low-metallicity galaxies.

Third, the discrepancy between the predicted and observed \snii\ dust at low 
metallicity could be due to several parameters or assumptions used in the 
chemical evolution modeling.
(1) The IMF could play an important role in the absolute value of the 
dust production by \snii.
It would indicate that the slope of the IMF is metallicity-dependent, and that
the contribution of massive stars is lower, at very low $Z_\sms{ISM}$, which
seems to be unlikely, both from a theoretical point of view (star formation),
and an observational point of view (star counts).
(2) A much more likely explanation could come from the fact that we assumed
that the condensation into dust of the elements ejected by the \snii\ and
their mixing in the ISM is instantaneous.
If the major part of the dust was to condense into the ISM, and not directly
into the \snii-ejecta, then the dust formation would be delayed after the
death of the massive stars.
(3) Finally, the dust production rates, computed from our dust evolution model, 
implicitely assume that the star formation rate of the galaxy is smooth and that
the mixing is instantaneous.
This hypothesis is certainly valid for evolved systems, but could be wrong for
very young objects.
For example, \citet{legrand00} suggested that the star formation history of 
\izw\ is not continuous.


\section{CONCLUSION AND SUMMARY}
\label{sec:conclusion}

The weakness of the mid-IR aromatic features in low-metallicity environments has 
been traditionally interpreted as the consequence of the increased selective
destruction efficiency of the PAHs in these environments. 
In this paper, we presented a new interpretation for the observed correlation 
of the intensity of the mid-IR emission from PAHs with galaxies' metallicity. 
In our model, this trend is a manifestation of the evolution of the abundance
of interstellar carbon that formed in AGB stars with time.
A fraction of the carbon dust formed in AGB stars in the form of PAH 
macromolecules offering a natural explanation of the evolutionary trend of 
PAH abundance with galactic metallicity (or time).

To ascertain this trend, we first converted the trend of observed mid-IR fluxes
to PAH and dust abundances, by modeling the UV-to-radio SEDs in a sample of 
35 nearby galaxies, with metallicities ranging 
from $1/50$ to $\sim3\zsun$. 
Our models represent the most detailed decomposition of the dust emission
into its gas phase components using a wide range of astrophyical constraints, including: the free-free and mid-IR emissions to constrain the gas and dust
radiation from \hii\ regions; the far-IR and optical
emission to constrain the ISRF that heats the dust in PDRs.
From these models, we determined the abundances of the PAHs and other dust
species associated to the neutral phase of the ISM. 
We then used a chemical evolution model to calculate the abundances of SN- 
and AGB-condensed dust as a function of time or metallicity. 
The model takes into account the delayed recycling of the ejecta from 
low-mass stars caused by their finite main-sequence lifetime. 

The main conclusions of this paper are the followings.
\begin{enumerate}
  \item From the decomposition of the IR emission into its various emission
    components and dust species, we discovered two distinct evolutionary trends 
    for PAHs and other dust species in galaxies.
  \item Comparison of these trends with the chemical evolution models showed 
    that the abundance of PAHs follows that of the carbon dust from AGB stars.
    The remaining dust species follows the evolutionary trend of SN-condensed
    dust.
  \item The delayed injection of PAHs into the ISM provides a natural
    explanation for the paucity of these large molecules in low metallicity
    systems.
    The subsequent rise in the PAH-to-gas mass ratio with metallicity 
    is then a natural consequence of the increasing contribution of AGB stars 
    to the chemical enrichment of the ISM as they evolve off the main sequence.
    The trend of the other dust species is a natural consequence of the
    evolution of SN-condensed dust which is 
    instantaneously injected into the ISM after the birth of the progenitor 
    star.
  \item While the model is very successful in observing the gross general 
    trend, there are some systematic deviations showing that  
    dust-to-gas mass ratios inferred from observations fall below the calculated 
    value for SN-condensed dust in the lowest metallicity galaxies. 
    This discrepancy may be due to one or more of the 
    following:\textlist{\thetextlist~an overestimate of the \hi\ gas mass used 
      to derive these dust-to-gas mass ratios; 
    \thetextlist~an underestimate of the dust mass due to the possible presence
      of a cold dust component; and
    \thetextlist~a more complex star formation history than used in the 
    model calculations.}
  \item As an aside, in our analysis of the mid-IR spectra of nearby galaxies, 
    we detected 
    the $7.7\mic$ aromatic feature at the $4\sigma$ level in \viizw, a dwarf 
    galaxy with $Z\simeq1/20\zsun$.
    This is the lowest metallicity galaxy for which PAH emission has been 
    detected to date.
\end{enumerate}

The success of our chemical evolution model, in reproducing the trend of PAH 
abundances with metallicity, strongly suggests the importance of stellar 
evolutionary effects in determining the abundances and composition of dust in 
galaxies. 
These will have important consequences for determining the opacity of galaxies 
and their reradiated thermal IR emission. 
Chemical evolution models for dust must therefore be an integral part of 
population synthesis models, providing a self-consistent link between the 
stellar and dust emission components of the SED of galaxies.


\acknowledgments

We thank Els Peeters for her expert advices on the IRS spectra extraction.
We are grateful to Sacha Hony for a useful discussion about AGB stars.
This work was performed while two of us (F.~G. \& P.~C.) held a National 
Research Council/Oak Ridge Associated Universities research associateship 
award at NASA Goddard Space Flight Center.
E.~D. acknowledges the support of NASA's LTSA03-0000-065. 
This study is based in part on
observations with \iso, an ESA project with instruments funded by ESA 
Member States (especially the PI countries: France, Germany, the 
Netherlands and the United Kingdom) and with the participation of ISAS 
and NASA.
This work is also based in part on observations made with the \spitzST, 
which is operated by the Jet Propulsion Laboratory, California 
Institute of Technology under a contract with NASA.
This research has made use of the HYPERLEDA database
(http://leda.univ-lyon1.fr) and of the NASA/IPAC Extragalactic Database (NED)
which is also operated by the Jet Propulsion Laboratory, California Institute 
of Technology, under contract with the National Aeronautics and Space 
Administration.
We acknowledge the extensive use of the Levenberg-Marquardt least-squares 
fitting procedure and the Adams-Bashford-Moulton Ordinary Differential Equation
solver written by Craig B.~Markwardt.


\bibliographystyle{/Users/Fred/Documents/Astro/TeXstyle/Packages_AAS/aas}
\bibliography{/Users/Fred/Documents/Astro/TeXstyle/references}

\end{document}